\documentclass[twocolumn]{aastex631}

\pdfoutput=1
\usepackage[utf8]{inputenc}
\usepackage{array}
\usepackage{xspace}
\usepackage{xstring}
\usepackage{acronym}
\usepackage{tabularx}
\usepackage{comment}
\usepackage{multirow}
\usepackage{rotating}
\usepackage{booktabs}
\usepackage{amsmath,amssymb,graphicx}
\usepackage{comment}
\usepackage{dcolumn}
\usepackage{makecell}
\usepackage{scrextend}
\usepackage{textgreek}
\usepackage{graphicx}
\usepackage[caption=false]{subfig}
\usepackage{makerobust}

\newcolumntype{Y}{>{\centering\arraybackslash}X}
\newcolumntype{Z}{>{\raggedright\arraybackslash}X}
\newcolumntype{K}{>{\raggedleft\arraybackslash}X}
\newcolumntype{U}{>{\hsize=1.01\hsize}Y}
\newcolumntype{V}{>{\hsize=1.2\hsize}Y}
\newcolumntype{W}{>{\hsize=0.71\hsize}Y}

\definecolor{notecolor}{rgb}{0.4, 0.2, 0.1}

\newcommand{\Mc}{\ensuremath{\mathcal{M}}\xspace}

\newcommand{\Msun}{\ensuremath{M_\odot}\xspace}

\newcommand{\chieff}{\ensuremath{\chi_\mathrm{eff}}\xspace}

\newcommand{\SNR}{\ensuremath{\mathrm{S}/\mathrm{N}}\xspace}

\newcommand\perGpcyr{\ensuremath{\text{Gpc}^{-3}\,\text{yr}^{-1}}}

\newcommand{\VT}{\ensuremath{\langle VT \rangle}}

\newcommand{\aoneminus}[1]{\IfEqCase{#1}{{GW230529ay_combined_imrphm_high_spin}{0.37}{GW230529ay_combined_imrphm_low_spin}{0.37}{GW230529ay_imrphen_nsbh}{0.09}{GW230529ay_imrseobv4_nsbh}{0.12}{GW230529ay_imrpv2_low_spin}{0.33}}}
\newcommand{\aonemed}[1]{\IfEqCase{#1}{{GW230529ay_combined_imrphm_high_spin}{0.44}{GW230529ay_combined_imrphm_low_spin}{0.46}{GW230529ay_imrphen_nsbh}{0.09}{GW230529ay_imrseobv4_nsbh}{0.12}{GW230529ay_imrpv2_low_spin}{0.42}}}
\newcommand{\aoneplus}[1]{\IfEqCase{#1}{{GW230529ay_combined_imrphm_high_spin}{0.40}{GW230529ay_combined_imrphm_low_spin}{0.37}{GW230529ay_imrphen_nsbh}{0.23}{GW230529ay_imrseobv4_nsbh}{0.29}{GW230529ay_imrpv2_low_spin}{0.43}}}
\newcommand{\aoneonepercent}[1]{\IfEqCase{#1}{{GW230529ay_combined_imrphm_high_spin}{0.01}{GW230529ay_combined_imrphm_low_spin}{0.02}{GW230529ay_imrphen_nsbh}{0.00}{GW230529ay_imrseobv4_nsbh}{0.00}{GW230529ay_imrpv2_low_spin}{0.02}}}
\newcommand{\aoneninetyninepercent}[1]{\IfEqCase{#1}{{GW230529ay_combined_imrphm_high_spin}{0.95}{GW230529ay_combined_imrphm_low_spin}{0.94}{GW230529ay_imrphen_nsbh}{0.38}{GW230529ay_imrseobv4_nsbh}{0.47}{GW230529ay_imrpv2_low_spin}{0.96}}}
\newcommand{\aonefivepercent}[1]{\IfEqCase{#1}{{GW230529ay_combined_imrphm_high_spin}{0.07}{GW230529ay_combined_imrphm_low_spin}{0.10}{GW230529ay_imrphen_nsbh}{0.00}{GW230529ay_imrseobv4_nsbh}{0.01}{GW230529ay_imrpv2_low_spin}{0.09}}}
\newcommand{\aoneninetyfivepercent}[1]{\IfEqCase{#1}{{GW230529ay_combined_imrphm_high_spin}{0.85}{GW230529ay_combined_imrphm_low_spin}{0.84}{GW230529ay_imrphen_nsbh}{0.32}{GW230529ay_imrseobv4_nsbh}{0.41}{GW230529ay_imrpv2_low_spin}{0.86}}}
\newcommand{\aoneninetypercent}[1]{\IfEqCase{#1}{{GW230529ay_combined_imrphm_high_spin}{0.77}{GW230529ay_combined_imrphm_low_spin}{0.76}{GW230529ay_imrphen_nsbh}{0.28}{GW230529ay_imrseobv4_nsbh}{0.36}{GW230529ay_imrpv2_low_spin}{0.76}}}
\newcommand{\atwominus}[1]{\IfEqCase{#1}{{GW230529ay_combined_imrphm_high_spin}{0.38}{GW230529ay_combined_imrphm_low_spin}{0.02}{GW230529ay_imrphen_nsbh}{0.01}{GW230529ay_imrseobv4_nsbh}{0.01}{GW230529ay_imrpv2_low_spin}{0.02}}}
\newcommand{\atwomed}[1]{\IfEqCase{#1}{{GW230529ay_combined_imrphm_high_spin}{0.43}{GW230529ay_combined_imrphm_low_spin}{0.02}{GW230529ay_imrphen_nsbh}{0.01}{GW230529ay_imrseobv4_nsbh}{0.01}{GW230529ay_imrpv2_low_spin}{0.02}}}
\newcommand{\atwoplus}[1]{\IfEqCase{#1}{{GW230529ay_combined_imrphm_high_spin}{0.48}{GW230529ay_combined_imrphm_low_spin}{0.02}{GW230529ay_imrphen_nsbh}{0.03}{GW230529ay_imrseobv4_nsbh}{0.03}{GW230529ay_imrpv2_low_spin}{0.02}}}
\newcommand{\atwoonepercent}[1]{\IfEqCase{#1}{{GW230529ay_combined_imrphm_high_spin}{0.01}{GW230529ay_combined_imrphm_low_spin}{0.00}{GW230529ay_imrphen_nsbh}{0.00}{GW230529ay_imrseobv4_nsbh}{0.00}{GW230529ay_imrpv2_low_spin}{0.00}}}
\newcommand{\atwoninetyninepercent}[1]{\IfEqCase{#1}{{GW230529ay_combined_imrphm_high_spin}{0.98}{GW230529ay_combined_imrphm_low_spin}{0.05}{GW230529ay_imrphen_nsbh}{0.04}{GW230529ay_imrseobv4_nsbh}{0.04}{GW230529ay_imrpv2_low_spin}{0.05}}}
\newcommand{\atwofivepercent}[1]{\IfEqCase{#1}{{GW230529ay_combined_imrphm_high_spin}{0.05}{GW230529ay_combined_imrphm_low_spin}{0.00}{GW230529ay_imrphen_nsbh}{0.00}{GW230529ay_imrseobv4_nsbh}{0.00}{GW230529ay_imrpv2_low_spin}{0.00}}}
\newcommand{\atwoninetyfivepercent}[1]{\IfEqCase{#1}{{GW230529ay_combined_imrphm_high_spin}{0.91}{GW230529ay_combined_imrphm_low_spin}{0.05}{GW230529ay_imrphen_nsbh}{0.03}{GW230529ay_imrseobv4_nsbh}{0.04}{GW230529ay_imrpv2_low_spin}{0.05}}}
\newcommand{\atwoninetypercent}[1]{\IfEqCase{#1}{{GW230529ay_combined_imrphm_high_spin}{0.84}{GW230529ay_combined_imrphm_low_spin}{0.04}{GW230529ay_imrphen_nsbh}{0.03}{GW230529ay_imrseobv4_nsbh}{0.03}{GW230529ay_imrpv2_low_spin}{0.05}}}
\newcommand{\baryonicmassoneminus}[1]{\IfEqCase{#1}{{GW230529ay_combined_imrphm_high_spin}{-}{GW230529ay_combined_imrphm_low_spin}{-}{GW230529ay_imrphen_nsbh}{1.4}{GW230529ay_imrseobv4_nsbh}{1.6}{GW230529ay_imrpv2_low_spin}{1.2}}}
\newcommand{\baryonicmassonemed}[1]{\IfEqCase{#1}{{GW230529ay_combined_imrphm_high_spin}{-}{GW230529ay_combined_imrphm_low_spin}{-}{GW230529ay_imrphen_nsbh}{5.4}{GW230529ay_imrseobv4_nsbh}{5.1}{GW230529ay_imrpv2_low_spin}{3.9}}}
\newcommand{\baryonicmassoneplus}[1]{\IfEqCase{#1}{{GW230529ay_combined_imrphm_high_spin}{-}{GW230529ay_combined_imrphm_low_spin}{-}{GW230529ay_imrphen_nsbh}{1.4}{GW230529ay_imrseobv4_nsbh}{1.6}{GW230529ay_imrpv2_low_spin}{1.2}}}
\newcommand{\baryonicmassoneonepercent}[1]{\IfEqCase{#1}{{GW230529ay_combined_imrphm_high_spin}{-}{GW230529ay_combined_imrphm_low_spin}{-}{GW230529ay_imrphen_nsbh}{3.6}{GW230529ay_imrseobv4_nsbh}{3.3}{GW230529ay_imrpv2_low_spin}{2.6}}}
\newcommand{\baryonicmassoneninetyninepercent}[1]{\IfEqCase{#1}{{GW230529ay_combined_imrphm_high_spin}{-}{GW230529ay_combined_imrphm_low_spin}{-}{GW230529ay_imrphen_nsbh}{7.6}{GW230529ay_imrseobv4_nsbh}{7.5}{GW230529ay_imrpv2_low_spin}{5.6}}}
\newcommand{\baryonicmassonefivepercent}[1]{\IfEqCase{#1}{{GW230529ay_combined_imrphm_high_spin}{-}{GW230529ay_combined_imrphm_low_spin}{-}{GW230529ay_imrphen_nsbh}{3.9}{GW230529ay_imrseobv4_nsbh}{3.6}{GW230529ay_imrpv2_low_spin}{2.7}}}
\newcommand{\baryonicmassoneninetyfivepercent}[1]{\IfEqCase{#1}{{GW230529ay_combined_imrphm_high_spin}{-}{GW230529ay_combined_imrphm_low_spin}{-}{GW230529ay_imrphen_nsbh}{6.8}{GW230529ay_imrseobv4_nsbh}{6.7}{GW230529ay_imrpv2_low_spin}{5.2}}}
\newcommand{\baryonicmassoneninetypercent}[1]{\IfEqCase{#1}{{GW230529ay_combined_imrphm_high_spin}{-}{GW230529ay_combined_imrphm_low_spin}{-}{GW230529ay_imrphen_nsbh}{6.4}{GW230529ay_imrseobv4_nsbh}{6.3}{GW230529ay_imrpv2_low_spin}{5.0}}}
\newcommand{\baryonicmassonesourceminus}[1]{\IfEqCase{#1}{{GW230529ay_combined_imrphm_high_spin}{-}{GW230529ay_combined_imrphm_low_spin}{-}{GW230529ay_imrphen_nsbh}{1.4}{GW230529ay_imrseobv4_nsbh}{1.5}{GW230529ay_imrpv2_low_spin}{1.2}}}
\newcommand{\baryonicmassonesourcemed}[1]{\IfEqCase{#1}{{GW230529ay_combined_imrphm_high_spin}{-}{GW230529ay_combined_imrphm_low_spin}{-}{GW230529ay_imrphen_nsbh}{5.1}{GW230529ay_imrseobv4_nsbh}{4.9}{GW230529ay_imrpv2_low_spin}{3.8}}}
\newcommand{\baryonicmassonesourceplus}[1]{\IfEqCase{#1}{{GW230529ay_combined_imrphm_high_spin}{-}{GW230529ay_combined_imrphm_low_spin}{-}{GW230529ay_imrphen_nsbh}{1.4}{GW230529ay_imrseobv4_nsbh}{1.5}{GW230529ay_imrpv2_low_spin}{1.2}}}
\newcommand{\baryonicmassonesourceonepercent}[1]{\IfEqCase{#1}{{GW230529ay_combined_imrphm_high_spin}{-}{GW230529ay_combined_imrphm_low_spin}{-}{GW230529ay_imrphen_nsbh}{3.5}{GW230529ay_imrseobv4_nsbh}{3.2}{GW230529ay_imrpv2_low_spin}{2.5}}}
\newcommand{\baryonicmassonesourceninetyninepercent}[1]{\IfEqCase{#1}{{GW230529ay_combined_imrphm_high_spin}{-}{GW230529ay_combined_imrphm_low_spin}{-}{GW230529ay_imrphen_nsbh}{7.3}{GW230529ay_imrseobv4_nsbh}{7.2}{GW230529ay_imrpv2_low_spin}{5.4}}}
\newcommand{\baryonicmassonesourcefivepercent}[1]{\IfEqCase{#1}{{GW230529ay_combined_imrphm_high_spin}{-}{GW230529ay_combined_imrphm_low_spin}{-}{GW230529ay_imrphen_nsbh}{3.8}{GW230529ay_imrseobv4_nsbh}{3.4}{GW230529ay_imrpv2_low_spin}{2.6}}}
\newcommand{\baryonicmassonesourceninetyfivepercent}[1]{\IfEqCase{#1}{{GW230529ay_combined_imrphm_high_spin}{-}{GW230529ay_combined_imrphm_low_spin}{-}{GW230529ay_imrphen_nsbh}{6.5}{GW230529ay_imrseobv4_nsbh}{6.4}{GW230529ay_imrpv2_low_spin}{5.0}}}
\newcommand{\baryonicmassonesourceninetypercent}[1]{\IfEqCase{#1}{{GW230529ay_combined_imrphm_high_spin}{-}{GW230529ay_combined_imrphm_low_spin}{-}{GW230529ay_imrphen_nsbh}{6.1}{GW230529ay_imrseobv4_nsbh}{6.0}{GW230529ay_imrpv2_low_spin}{4.8}}}
\newcommand{\baryonicmasstwominus}[1]{\IfEqCase{#1}{{GW230529ay_combined_imrphm_high_spin}{-}{GW230529ay_combined_imrphm_low_spin}{-}{GW230529ay_imrphen_nsbh}{0.3}{GW230529ay_imrseobv4_nsbh}{0.3}{GW230529ay_imrpv2_low_spin}{0.3}}}
\newcommand{\baryonicmasstwomed}[1]{\IfEqCase{#1}{{GW230529ay_combined_imrphm_high_spin}{-}{GW230529ay_combined_imrphm_low_spin}{-}{GW230529ay_imrphen_nsbh}{1.5}{GW230529ay_imrseobv4_nsbh}{1.6}{GW230529ay_imrpv2_low_spin}{1.7}}}
\newcommand{\baryonicmasstwoplus}[1]{\IfEqCase{#1}{{GW230529ay_combined_imrphm_high_spin}{-}{GW230529ay_combined_imrphm_low_spin}{-}{GW230529ay_imrphen_nsbh}{0.5}{GW230529ay_imrseobv4_nsbh}{0.6}{GW230529ay_imrpv2_low_spin}{0.7}}}
\newcommand{\baryonicmasstwoonepercent}[1]{\IfEqCase{#1}{{GW230529ay_combined_imrphm_high_spin}{-}{GW230529ay_combined_imrphm_low_spin}{-}{GW230529ay_imrphen_nsbh}{1.2}{GW230529ay_imrseobv4_nsbh}{1.2}{GW230529ay_imrpv2_low_spin}{1.3}}}
\newcommand{\baryonicmasstwoninetyninepercent}[1]{\IfEqCase{#1}{{GW230529ay_combined_imrphm_high_spin}{-}{GW230529ay_combined_imrphm_low_spin}{-}{GW230529ay_imrphen_nsbh}{2.2}{GW230529ay_imrseobv4_nsbh}{2.4}{GW230529ay_imrpv2_low_spin}{2.5}}}
\newcommand{\baryonicmasstwofivepercent}[1]{\IfEqCase{#1}{{GW230529ay_combined_imrphm_high_spin}{-}{GW230529ay_combined_imrphm_low_spin}{-}{GW230529ay_imrphen_nsbh}{1.3}{GW230529ay_imrseobv4_nsbh}{1.3}{GW230529ay_imrpv2_low_spin}{1.4}}}
\newcommand{\baryonicmasstwoninetyfivepercent}[1]{\IfEqCase{#1}{{GW230529ay_combined_imrphm_high_spin}{-}{GW230529ay_combined_imrphm_low_spin}{-}{GW230529ay_imrphen_nsbh}{2.0}{GW230529ay_imrseobv4_nsbh}{2.2}{GW230529ay_imrpv2_low_spin}{2.5}}}
\newcommand{\baryonicmasstwoninetypercent}[1]{\IfEqCase{#1}{{GW230529ay_combined_imrphm_high_spin}{-}{GW230529ay_combined_imrphm_low_spin}{-}{GW230529ay_imrphen_nsbh}{1.9}{GW230529ay_imrseobv4_nsbh}{2.1}{GW230529ay_imrpv2_low_spin}{2.4}}}
\newcommand{\baryonicmasstwosourceminus}[1]{\IfEqCase{#1}{{GW230529ay_combined_imrphm_high_spin}{-}{GW230529ay_combined_imrphm_low_spin}{-}{GW230529ay_imrphen_nsbh}{0.3}{GW230529ay_imrseobv4_nsbh}{0.3}{GW230529ay_imrpv2_low_spin}{0.3}}}
\newcommand{\baryonicmasstwosourcemed}[1]{\IfEqCase{#1}{{GW230529ay_combined_imrphm_high_spin}{-}{GW230529ay_combined_imrphm_low_spin}{-}{GW230529ay_imrphen_nsbh}{1.5}{GW230529ay_imrseobv4_nsbh}{1.5}{GW230529ay_imrpv2_low_spin}{1.7}}}
\newcommand{\baryonicmasstwosourceplus}[1]{\IfEqCase{#1}{{GW230529ay_combined_imrphm_high_spin}{-}{GW230529ay_combined_imrphm_low_spin}{-}{GW230529ay_imrphen_nsbh}{0.5}{GW230529ay_imrseobv4_nsbh}{0.6}{GW230529ay_imrpv2_low_spin}{0.7}}}
\newcommand{\baryonicmasstwosourceonepercent}[1]{\IfEqCase{#1}{{GW230529ay_combined_imrphm_high_spin}{-}{GW230529ay_combined_imrphm_low_spin}{-}{GW230529ay_imrphen_nsbh}{1.1}{GW230529ay_imrseobv4_nsbh}{1.1}{GW230529ay_imrpv2_low_spin}{1.3}}}
\newcommand{\baryonicmasstwosourceninetyninepercent}[1]{\IfEqCase{#1}{{GW230529ay_combined_imrphm_high_spin}{-}{GW230529ay_combined_imrphm_low_spin}{-}{GW230529ay_imrphen_nsbh}{2.1}{GW230529ay_imrseobv4_nsbh}{2.3}{GW230529ay_imrpv2_low_spin}{2.4}}}
\newcommand{\baryonicmasstwosourcefivepercent}[1]{\IfEqCase{#1}{{GW230529ay_combined_imrphm_high_spin}{-}{GW230529ay_combined_imrphm_low_spin}{-}{GW230529ay_imrphen_nsbh}{1.2}{GW230529ay_imrseobv4_nsbh}{1.2}{GW230529ay_imrpv2_low_spin}{1.3}}}
\newcommand{\baryonicmasstwosourceninetyfivepercent}[1]{\IfEqCase{#1}{{GW230529ay_combined_imrphm_high_spin}{-}{GW230529ay_combined_imrphm_low_spin}{-}{GW230529ay_imrphen_nsbh}{1.9}{GW230529ay_imrseobv4_nsbh}{2.1}{GW230529ay_imrpv2_low_spin}{2.3}}}
\newcommand{\baryonicmasstwosourceninetypercent}[1]{\IfEqCase{#1}{{GW230529ay_combined_imrphm_high_spin}{-}{GW230529ay_combined_imrphm_low_spin}{-}{GW230529ay_imrphen_nsbh}{1.8}{GW230529ay_imrseobv4_nsbh}{2.0}{GW230529ay_imrpv2_low_spin}{2.3}}}
\newcommand{\baryonictorusmassminus}[1]{\IfEqCase{#1}{{GW230529ay_combined_imrphm_high_spin}{-}{GW230529ay_combined_imrphm_low_spin}{-}{GW230529ay_imrphen_nsbh}{0.2}{GW230529ay_imrseobv4_nsbh}{-}{GW230529ay_imrpv2_low_spin}{-}}}
\newcommand{\baryonictorusmassmed}[1]{\IfEqCase{#1}{{GW230529ay_combined_imrphm_high_spin}{-}{GW230529ay_combined_imrphm_low_spin}{-}{GW230529ay_imrphen_nsbh}{0.2}{GW230529ay_imrseobv4_nsbh}{-}{GW230529ay_imrpv2_low_spin}{-}}}
\newcommand{\baryonictorusmassplus}[1]{\IfEqCase{#1}{{GW230529ay_combined_imrphm_high_spin}{-}{GW230529ay_combined_imrphm_low_spin}{-}{GW230529ay_imrphen_nsbh}{0.2}{GW230529ay_imrseobv4_nsbh}{-}{GW230529ay_imrpv2_low_spin}{-}}}
\newcommand{\baryonictorusmassonepercent}[1]{\IfEqCase{#1}{{GW230529ay_combined_imrphm_high_spin}{-}{GW230529ay_combined_imrphm_low_spin}{-}{GW230529ay_imrphen_nsbh}{0.0}{GW230529ay_imrseobv4_nsbh}{-}{GW230529ay_imrpv2_low_spin}{-}}}
\newcommand{\baryonictorusmassninetyninepercent}[1]{\IfEqCase{#1}{{GW230529ay_combined_imrphm_high_spin}{-}{GW230529ay_combined_imrphm_low_spin}{-}{GW230529ay_imrphen_nsbh}{0.4}{GW230529ay_imrseobv4_nsbh}{-}{GW230529ay_imrpv2_low_spin}{-}}}
\newcommand{\baryonictorusmassfivepercent}[1]{\IfEqCase{#1}{{GW230529ay_combined_imrphm_high_spin}{-}{GW230529ay_combined_imrphm_low_spin}{-}{GW230529ay_imrphen_nsbh}{0.0}{GW230529ay_imrseobv4_nsbh}{-}{GW230529ay_imrpv2_low_spin}{-}}}
\newcommand{\baryonictorusmassninetyfivepercent}[1]{\IfEqCase{#1}{{GW230529ay_combined_imrphm_high_spin}{-}{GW230529ay_combined_imrphm_low_spin}{-}{GW230529ay_imrphen_nsbh}{0.4}{GW230529ay_imrseobv4_nsbh}{-}{GW230529ay_imrpv2_low_spin}{-}}}
\newcommand{\baryonictorusmassninetypercent}[1]{\IfEqCase{#1}{{GW230529ay_combined_imrphm_high_spin}{-}{GW230529ay_combined_imrphm_low_spin}{-}{GW230529ay_imrphen_nsbh}{0.3}{GW230529ay_imrseobv4_nsbh}{-}{GW230529ay_imrpv2_low_spin}{-}}}
\newcommand{\baryonictorusmasssourceminus}[1]{\IfEqCase{#1}{{GW230529ay_combined_imrphm_high_spin}{-}{GW230529ay_combined_imrphm_low_spin}{-}{GW230529ay_imrphen_nsbh}{0.2}{GW230529ay_imrseobv4_nsbh}{-}{GW230529ay_imrpv2_low_spin}{-}}}
\newcommand{\baryonictorusmasssourcemed}[1]{\IfEqCase{#1}{{GW230529ay_combined_imrphm_high_spin}{-}{GW230529ay_combined_imrphm_low_spin}{-}{GW230529ay_imrphen_nsbh}{0.2}{GW230529ay_imrseobv4_nsbh}{-}{GW230529ay_imrpv2_low_spin}{-}}}
\newcommand{\baryonictorusmasssourceplus}[1]{\IfEqCase{#1}{{GW230529ay_combined_imrphm_high_spin}{-}{GW230529ay_combined_imrphm_low_spin}{-}{GW230529ay_imrphen_nsbh}{0.1}{GW230529ay_imrseobv4_nsbh}{-}{GW230529ay_imrpv2_low_spin}{-}}}
\newcommand{\baryonictorusmasssourceonepercent}[1]{\IfEqCase{#1}{{GW230529ay_combined_imrphm_high_spin}{-}{GW230529ay_combined_imrphm_low_spin}{-}{GW230529ay_imrphen_nsbh}{0.0}{GW230529ay_imrseobv4_nsbh}{-}{GW230529ay_imrpv2_low_spin}{-}}}
\newcommand{\baryonictorusmasssourceninetyninepercent}[1]{\IfEqCase{#1}{{GW230529ay_combined_imrphm_high_spin}{-}{GW230529ay_combined_imrphm_low_spin}{-}{GW230529ay_imrphen_nsbh}{0.4}{GW230529ay_imrseobv4_nsbh}{-}{GW230529ay_imrpv2_low_spin}{-}}}
\newcommand{\baryonictorusmasssourcefivepercent}[1]{\IfEqCase{#1}{{GW230529ay_combined_imrphm_high_spin}{-}{GW230529ay_combined_imrphm_low_spin}{-}{GW230529ay_imrphen_nsbh}{0.0}{GW230529ay_imrseobv4_nsbh}{-}{GW230529ay_imrpv2_low_spin}{-}}}
\newcommand{\baryonictorusmasssourceninetyfivepercent}[1]{\IfEqCase{#1}{{GW230529ay_combined_imrphm_high_spin}{-}{GW230529ay_combined_imrphm_low_spin}{-}{GW230529ay_imrphen_nsbh}{0.3}{GW230529ay_imrseobv4_nsbh}{-}{GW230529ay_imrpv2_low_spin}{-}}}
\newcommand{\baryonictorusmasssourceninetypercent}[1]{\IfEqCase{#1}{{GW230529ay_combined_imrphm_high_spin}{-}{GW230529ay_combined_imrphm_low_spin}{-}{GW230529ay_imrphen_nsbh}{0.3}{GW230529ay_imrseobv4_nsbh}{-}{GW230529ay_imrpv2_low_spin}{-}}}
\newcommand{\chieffminus}[1]{\IfEqCase{#1}{{GW230529ay_combined_imrphm_high_spin}{0.17}{GW230529ay_combined_imrphm_low_spin}{0.17}{GW230529ay_imrphen_nsbh}{0.15}{GW230529ay_imrseobv4_nsbh}{0.16}{GW230529ay_imrpv2_low_spin}{0.11}}}
\newcommand{\chieffmed}[1]{\IfEqCase{#1}{{GW230529ay_combined_imrphm_high_spin}{-0.10}{GW230529ay_combined_imrphm_low_spin}{-0.10}{GW230529ay_imrphen_nsbh}{-0.05}{GW230529ay_imrseobv4_nsbh}{-0.08}{GW230529ay_imrpv2_low_spin}{-0.11}}}
\newcommand{\chieffplus}[1]{\IfEqCase{#1}{{GW230529ay_combined_imrphm_high_spin}{0.12}{GW230529ay_combined_imrphm_low_spin}{0.10}{GW230529ay_imrphen_nsbh}{0.14}{GW230529ay_imrseobv4_nsbh}{0.16}{GW230529ay_imrpv2_low_spin}{0.13}}}
\newcommand{\chieffonepercent}[1]{\IfEqCase{#1}{{GW230529ay_combined_imrphm_high_spin}{-0.29}{GW230529ay_combined_imrphm_low_spin}{-0.29}{GW230529ay_imrphen_nsbh}{-0.22}{GW230529ay_imrseobv4_nsbh}{-0.25}{GW230529ay_imrpv2_low_spin}{-0.24}}}
\newcommand{\chieffninetyninepercent}[1]{\IfEqCase{#1}{{GW230529ay_combined_imrphm_high_spin}{0.10}{GW230529ay_combined_imrphm_low_spin}{0.07}{GW230529ay_imrphen_nsbh}{0.16}{GW230529ay_imrseobv4_nsbh}{0.15}{GW230529ay_imrpv2_low_spin}{0.06}}}
\newcommand{\chiefffivepercent}[1]{\IfEqCase{#1}{{GW230529ay_combined_imrphm_high_spin}{-0.27}{GW230529ay_combined_imrphm_low_spin}{-0.27}{GW230529ay_imrphen_nsbh}{-0.20}{GW230529ay_imrseobv4_nsbh}{-0.23}{GW230529ay_imrpv2_low_spin}{-0.22}}}
\newcommand{\chieffninetyfivepercent}[1]{\IfEqCase{#1}{{GW230529ay_combined_imrphm_high_spin}{0.02}{GW230529ay_combined_imrphm_low_spin}{0.00}{GW230529ay_imrphen_nsbh}{0.09}{GW230529ay_imrseobv4_nsbh}{0.08}{GW230529ay_imrpv2_low_spin}{0.02}}}
\newcommand{\chieffninetypercent}[1]{\IfEqCase{#1}{{GW230529ay_combined_imrphm_high_spin}{-0.01}{GW230529ay_combined_imrphm_low_spin}{-0.02}{GW230529ay_imrphen_nsbh}{0.05}{GW230529ay_imrseobv4_nsbh}{0.05}{GW230529ay_imrpv2_low_spin}{0.00}}}
\newcommand{\chipminus}[1]{\IfEqCase{#1}{{GW230529ay_combined_imrphm_high_spin}{0.30}{GW230529ay_combined_imrphm_low_spin}{0.34}{GW230529ay_imrphen_nsbh}{0.00}{GW230529ay_imrseobv4_nsbh}{0.00}{GW230529ay_imrpv2_low_spin}{0.29}}}
\newcommand{\chipmed}[1]{\IfEqCase{#1}{{GW230529ay_combined_imrphm_high_spin}{0.40}{GW230529ay_combined_imrphm_low_spin}{0.40}{GW230529ay_imrphen_nsbh}{0.00}{GW230529ay_imrseobv4_nsbh}{0.00}{GW230529ay_imrpv2_low_spin}{0.35}}}
\newcommand{\chipplus}[1]{\IfEqCase{#1}{{GW230529ay_combined_imrphm_high_spin}{0.39}{GW230529ay_combined_imrphm_low_spin}{0.37}{GW230529ay_imrphen_nsbh}{0.00}{GW230529ay_imrseobv4_nsbh}{0.00}{GW230529ay_imrpv2_low_spin}{0.44}}}
\newcommand{\chiponepercent}[1]{\IfEqCase{#1}{{GW230529ay_combined_imrphm_high_spin}{0.05}{GW230529ay_combined_imrphm_low_spin}{0.02}{GW230529ay_imrphen_nsbh}{0.00}{GW230529ay_imrseobv4_nsbh}{0.00}{GW230529ay_imrpv2_low_spin}{0.01}}}
\newcommand{\chipninetyninepercent}[1]{\IfEqCase{#1}{{GW230529ay_combined_imrphm_high_spin}{0.89}{GW230529ay_combined_imrphm_low_spin}{0.87}{GW230529ay_imrphen_nsbh}{0.00}{GW230529ay_imrseobv4_nsbh}{0.00}{GW230529ay_imrpv2_low_spin}{0.90}}}
\newcommand{\chipfivepercent}[1]{\IfEqCase{#1}{{GW230529ay_combined_imrphm_high_spin}{0.10}{GW230529ay_combined_imrphm_low_spin}{0.06}{GW230529ay_imrphen_nsbh}{0.00}{GW230529ay_imrseobv4_nsbh}{0.00}{GW230529ay_imrpv2_low_spin}{0.06}}}
\newcommand{\chipninetyfivepercent}[1]{\IfEqCase{#1}{{GW230529ay_combined_imrphm_high_spin}{0.79}{GW230529ay_combined_imrphm_low_spin}{0.77}{GW230529ay_imrphen_nsbh}{0.00}{GW230529ay_imrseobv4_nsbh}{0.00}{GW230529ay_imrpv2_low_spin}{0.79}}}
\newcommand{\chipninetypercent}[1]{\IfEqCase{#1}{{GW230529ay_combined_imrphm_high_spin}{0.71}{GW230529ay_combined_imrphm_low_spin}{0.70}{GW230529ay_imrphen_nsbh}{0.00}{GW230529ay_imrseobv4_nsbh}{0.00}{GW230529ay_imrpv2_low_spin}{0.69}}}
\newcommand{\chirpmassminus}[1]{\IfEqCase{#1}{{GW230529ay_combined_imrphm_high_spin}{0.002}{GW230529ay_combined_imrphm_low_spin}{0.002}{GW230529ay_imrphen_nsbh}{0.002}{GW230529ay_imrseobv4_nsbh}{0.002}{GW230529ay_imrpv2_low_spin}{0.002}}}
\newcommand{\chirpmassmed}[1]{\IfEqCase{#1}{{GW230529ay_combined_imrphm_high_spin}{2.026}{GW230529ay_combined_imrphm_low_spin}{2.026}{GW230529ay_imrphen_nsbh}{2.027}{GW230529ay_imrseobv4_nsbh}{2.027}{GW230529ay_imrpv2_low_spin}{2.027}}}
\newcommand{\chirpmassplus}[1]{\IfEqCase{#1}{{GW230529ay_combined_imrphm_high_spin}{0.002}{GW230529ay_combined_imrphm_low_spin}{0.002}{GW230529ay_imrphen_nsbh}{0.002}{GW230529ay_imrseobv4_nsbh}{0.002}{GW230529ay_imrpv2_low_spin}{0.002}}}
\newcommand{\chirpmassonepercent}[1]{\IfEqCase{#1}{{GW230529ay_combined_imrphm_high_spin}{2.024}{GW230529ay_combined_imrphm_low_spin}{2.024}{GW230529ay_imrphen_nsbh}{2.025}{GW230529ay_imrseobv4_nsbh}{2.024}{GW230529ay_imrpv2_low_spin}{2.025}}}
\newcommand{\chirpmassninetyninepercent}[1]{\IfEqCase{#1}{{GW230529ay_combined_imrphm_high_spin}{2.029}{GW230529ay_combined_imrphm_low_spin}{2.029}{GW230529ay_imrphen_nsbh}{2.030}{GW230529ay_imrseobv4_nsbh}{2.030}{GW230529ay_imrpv2_low_spin}{2.029}}}
\newcommand{\chirpmassfivepercent}[1]{\IfEqCase{#1}{{GW230529ay_combined_imrphm_high_spin}{2.024}{GW230529ay_combined_imrphm_low_spin}{2.024}{GW230529ay_imrphen_nsbh}{2.025}{GW230529ay_imrseobv4_nsbh}{2.025}{GW230529ay_imrpv2_low_spin}{2.025}}}
\newcommand{\chirpmassninetyfivepercent}[1]{\IfEqCase{#1}{{GW230529ay_combined_imrphm_high_spin}{2.028}{GW230529ay_combined_imrphm_low_spin}{2.028}{GW230529ay_imrphen_nsbh}{2.029}{GW230529ay_imrseobv4_nsbh}{2.029}{GW230529ay_imrpv2_low_spin}{2.029}}}
\newcommand{\chirpmassninetypercent}[1]{\IfEqCase{#1}{{GW230529ay_combined_imrphm_high_spin}{2.027}{GW230529ay_combined_imrphm_low_spin}{2.027}{GW230529ay_imrphen_nsbh}{2.028}{GW230529ay_imrseobv4_nsbh}{2.028}{GW230529ay_imrpv2_low_spin}{2.028}}}
\newcommand{\chirpmasssourceminus}[1]{\IfEqCase{#1}{{GW230529ay_combined_imrphm_high_spin}{0.04}{GW230529ay_combined_imrphm_low_spin}{0.04}{GW230529ay_imrphen_nsbh}{0.04}{GW230529ay_imrseobv4_nsbh}{0.04}{GW230529ay_imrpv2_low_spin}{0.04}}}
\newcommand{\chirpmasssourcemed}[1]{\IfEqCase{#1}{{GW230529ay_combined_imrphm_high_spin}{1.94}{GW230529ay_combined_imrphm_low_spin}{1.94}{GW230529ay_imrphen_nsbh}{1.94}{GW230529ay_imrseobv4_nsbh}{1.94}{GW230529ay_imrpv2_low_spin}{1.94}}}
\newcommand{\chirpmasssourceplus}[1]{\IfEqCase{#1}{{GW230529ay_combined_imrphm_high_spin}{0.04}{GW230529ay_combined_imrphm_low_spin}{0.04}{GW230529ay_imrphen_nsbh}{0.04}{GW230529ay_imrseobv4_nsbh}{0.04}{GW230529ay_imrpv2_low_spin}{0.04}}}
\newcommand{\chirpmasssourceonepercent}[1]{\IfEqCase{#1}{{GW230529ay_combined_imrphm_high_spin}{1.89}{GW230529ay_combined_imrphm_low_spin}{1.89}{GW230529ay_imrphen_nsbh}{1.89}{GW230529ay_imrseobv4_nsbh}{1.89}{GW230529ay_imrpv2_low_spin}{1.89}}}
\newcommand{\chirpmasssourceninetyninepercent}[1]{\IfEqCase{#1}{{GW230529ay_combined_imrphm_high_spin}{1.99}{GW230529ay_combined_imrphm_low_spin}{1.99}{GW230529ay_imrphen_nsbh}{2.00}{GW230529ay_imrseobv4_nsbh}{2.00}{GW230529ay_imrpv2_low_spin}{2.00}}}
\newcommand{\chirpmasssourcefivepercent}[1]{\IfEqCase{#1}{{GW230529ay_combined_imrphm_high_spin}{1.90}{GW230529ay_combined_imrphm_low_spin}{1.90}{GW230529ay_imrphen_nsbh}{1.90}{GW230529ay_imrseobv4_nsbh}{1.90}{GW230529ay_imrpv2_low_spin}{1.90}}}
\newcommand{\chirpmasssourceninetyfivepercent}[1]{\IfEqCase{#1}{{GW230529ay_combined_imrphm_high_spin}{1.98}{GW230529ay_combined_imrphm_low_spin}{1.98}{GW230529ay_imrphen_nsbh}{1.98}{GW230529ay_imrseobv4_nsbh}{1.99}{GW230529ay_imrpv2_low_spin}{1.98}}}
\newcommand{\chirpmasssourceninetypercent}[1]{\IfEqCase{#1}{{GW230529ay_combined_imrphm_high_spin}{1.97}{GW230529ay_combined_imrphm_low_spin}{1.97}{GW230529ay_imrphen_nsbh}{1.98}{GW230529ay_imrseobv4_nsbh}{1.98}{GW230529ay_imrpv2_low_spin}{1.98}}}
\newcommand{\comovingdistanceminus}[1]{\IfEqCase{#1}{{GW230529ay_combined_imrphm_high_spin}{90}{GW230529ay_combined_imrphm_low_spin}{90}{GW230529ay_imrphen_nsbh}{95}{GW230529ay_imrseobv4_nsbh}{95}{GW230529ay_imrpv2_low_spin}{92}}}
\newcommand{\comovingdistancemed}[1]{\IfEqCase{#1}{{GW230529ay_combined_imrphm_high_spin}{193}{GW230529ay_combined_imrphm_low_spin}{189}{GW230529ay_imrphen_nsbh}{193}{GW230529ay_imrseobv4_nsbh}{186}{GW230529ay_imrpv2_low_spin}{188}}}
\newcommand{\comovingdistanceplus}[1]{\IfEqCase{#1}{{GW230529ay_combined_imrphm_high_spin}{92}{GW230529ay_combined_imrphm_low_spin}{97}{GW230529ay_imrphen_nsbh}{99}{GW230529ay_imrseobv4_nsbh}{98}{GW230529ay_imrpv2_low_spin}{100}}}
\newcommand{\comovingdistanceonepercent}[1]{\IfEqCase{#1}{{GW230529ay_combined_imrphm_high_spin}{72}{GW230529ay_combined_imrphm_low_spin}{72}{GW230529ay_imrphen_nsbh}{68}{GW230529ay_imrseobv4_nsbh}{65}{GW230529ay_imrpv2_low_spin}{67}}}
\newcommand{\comovingdistanceninetyninepercent}[1]{\IfEqCase{#1}{{GW230529ay_combined_imrphm_high_spin}{319}{GW230529ay_combined_imrphm_low_spin}{321}{GW230529ay_imrphen_nsbh}{325}{GW230529ay_imrseobv4_nsbh}{318}{GW230529ay_imrpv2_low_spin}{322}}}
\newcommand{\comovingdistancefivepercent}[1]{\IfEqCase{#1}{{GW230529ay_combined_imrphm_high_spin}{102}{GW230529ay_combined_imrphm_low_spin}{99}{GW230529ay_imrphen_nsbh}{98}{GW230529ay_imrseobv4_nsbh}{91}{GW230529ay_imrpv2_low_spin}{95}}}
\newcommand{\comovingdistanceninetyfivepercent}[1]{\IfEqCase{#1}{{GW230529ay_combined_imrphm_high_spin}{285}{GW230529ay_combined_imrphm_low_spin}{286}{GW230529ay_imrphen_nsbh}{292}{GW230529ay_imrseobv4_nsbh}{285}{GW230529ay_imrpv2_low_spin}{287}}}
\newcommand{\comovingdistanceninetypercent}[1]{\IfEqCase{#1}{{GW230529ay_combined_imrphm_high_spin}{267}{GW230529ay_combined_imrphm_low_spin}{266}{GW230529ay_imrphen_nsbh}{274}{GW230529ay_imrseobv4_nsbh}{266}{GW230529ay_imrpv2_low_spin}{268}}}
\newcommand{\compactnessoneminus}[1]{\IfEqCase{#1}{{GW230529ay_combined_imrphm_high_spin}{-}{GW230529ay_combined_imrphm_low_spin}{-}{GW230529ay_imrphen_nsbh}{0.0}{GW230529ay_imrseobv4_nsbh}{0.0}{GW230529ay_imrpv2_low_spin}{0.1}}}
\newcommand{\compactnessonemed}[1]{\IfEqCase{#1}{{GW230529ay_combined_imrphm_high_spin}{-}{GW230529ay_combined_imrphm_low_spin}{-}{GW230529ay_imrphen_nsbh}{0.5}{GW230529ay_imrseobv4_nsbh}{0.5}{GW230529ay_imrpv2_low_spin}{0.2}}}
\newcommand{\compactnessoneplus}[1]{\IfEqCase{#1}{{GW230529ay_combined_imrphm_high_spin}{-}{GW230529ay_combined_imrphm_low_spin}{-}{GW230529ay_imrphen_nsbh}{0.0}{GW230529ay_imrseobv4_nsbh}{0.0}{GW230529ay_imrpv2_low_spin}{0.1}}}
\newcommand{\compactnessoneonepercent}[1]{\IfEqCase{#1}{{GW230529ay_combined_imrphm_high_spin}{-}{GW230529ay_combined_imrphm_low_spin}{-}{GW230529ay_imrphen_nsbh}{0.5}{GW230529ay_imrseobv4_nsbh}{0.5}{GW230529ay_imrpv2_low_spin}{0.1}}}
\newcommand{\compactnessoneninetyninepercent}[1]{\IfEqCase{#1}{{GW230529ay_combined_imrphm_high_spin}{-}{GW230529ay_combined_imrphm_low_spin}{-}{GW230529ay_imrphen_nsbh}{0.5}{GW230529ay_imrseobv4_nsbh}{0.5}{GW230529ay_imrpv2_low_spin}{0.3}}}
\newcommand{\compactnessonefivepercent}[1]{\IfEqCase{#1}{{GW230529ay_combined_imrphm_high_spin}{-}{GW230529ay_combined_imrphm_low_spin}{-}{GW230529ay_imrphen_nsbh}{0.5}{GW230529ay_imrseobv4_nsbh}{0.5}{GW230529ay_imrpv2_low_spin}{0.1}}}
\newcommand{\compactnessoneninetyfivepercent}[1]{\IfEqCase{#1}{{GW230529ay_combined_imrphm_high_spin}{-}{GW230529ay_combined_imrphm_low_spin}{-}{GW230529ay_imrphen_nsbh}{0.5}{GW230529ay_imrseobv4_nsbh}{0.5}{GW230529ay_imrpv2_low_spin}{0.3}}}
\newcommand{\compactnessoneninetypercent}[1]{\IfEqCase{#1}{{GW230529ay_combined_imrphm_high_spin}{-}{GW230529ay_combined_imrphm_low_spin}{-}{GW230529ay_imrphen_nsbh}{0.5}{GW230529ay_imrseobv4_nsbh}{0.5}{GW230529ay_imrpv2_low_spin}{0.2}}}
\newcommand{\compactnesstwominus}[1]{\IfEqCase{#1}{{GW230529ay_combined_imrphm_high_spin}{-}{GW230529ay_combined_imrphm_low_spin}{-}{GW230529ay_imrphen_nsbh}{0.0}{GW230529ay_imrseobv4_nsbh}{0.0}{GW230529ay_imrpv2_low_spin}{0.0}}}
\newcommand{\compactnesstwomed}[1]{\IfEqCase{#1}{{GW230529ay_combined_imrphm_high_spin}{-}{GW230529ay_combined_imrphm_low_spin}{-}{GW230529ay_imrphen_nsbh}{0.1}{GW230529ay_imrseobv4_nsbh}{0.1}{GW230529ay_imrpv2_low_spin}{0.1}}}
\newcommand{\compactnesstwoplus}[1]{\IfEqCase{#1}{{GW230529ay_combined_imrphm_high_spin}{-}{GW230529ay_combined_imrphm_low_spin}{-}{GW230529ay_imrphen_nsbh}{0.1}{GW230529ay_imrseobv4_nsbh}{0.1}{GW230529ay_imrpv2_low_spin}{0.1}}}
\newcommand{\compactnesstwoonepercent}[1]{\IfEqCase{#1}{{GW230529ay_combined_imrphm_high_spin}{-}{GW230529ay_combined_imrphm_low_spin}{-}{GW230529ay_imrphen_nsbh}{0.1}{GW230529ay_imrseobv4_nsbh}{0.1}{GW230529ay_imrpv2_low_spin}{0.1}}}
\newcommand{\compactnesstwoninetyninepercent}[1]{\IfEqCase{#1}{{GW230529ay_combined_imrphm_high_spin}{-}{GW230529ay_combined_imrphm_low_spin}{-}{GW230529ay_imrphen_nsbh}{0.2}{GW230529ay_imrseobv4_nsbh}{0.2}{GW230529ay_imrpv2_low_spin}{0.2}}}
\newcommand{\compactnesstwofivepercent}[1]{\IfEqCase{#1}{{GW230529ay_combined_imrphm_high_spin}{-}{GW230529ay_combined_imrphm_low_spin}{-}{GW230529ay_imrphen_nsbh}{0.1}{GW230529ay_imrseobv4_nsbh}{0.1}{GW230529ay_imrpv2_low_spin}{0.1}}}
\newcommand{\compactnesstwoninetyfivepercent}[1]{\IfEqCase{#1}{{GW230529ay_combined_imrphm_high_spin}{-}{GW230529ay_combined_imrphm_low_spin}{-}{GW230529ay_imrphen_nsbh}{0.2}{GW230529ay_imrseobv4_nsbh}{0.2}{GW230529ay_imrpv2_low_spin}{0.2}}}
\newcommand{\compactnesstwoninetypercent}[1]{\IfEqCase{#1}{{GW230529ay_combined_imrphm_high_spin}{-}{GW230529ay_combined_imrphm_low_spin}{-}{GW230529ay_imrphen_nsbh}{0.2}{GW230529ay_imrseobv4_nsbh}{0.2}{GW230529ay_imrpv2_low_spin}{0.2}}}
\newcommand{\cosiotaminus}[1]{\IfEqCase{#1}{{GW230529ay_combined_imrphm_high_spin}{0.90}{GW230529ay_combined_imrphm_low_spin}{0.92}{GW230529ay_imrphen_nsbh}{0.99}{GW230529ay_imrseobv4_nsbh}{0.97}{GW230529ay_imrpv2_low_spin}{0.96}}}
\newcommand{\cosiotamed}[1]{\IfEqCase{#1}{{GW230529ay_combined_imrphm_high_spin}{-0.05}{GW230529ay_combined_imrphm_low_spin}{-0.04}{GW230529ay_imrphen_nsbh}{0.03}{GW230529ay_imrseobv4_nsbh}{0.01}{GW230529ay_imrpv2_low_spin}{0.01}}}
\newcommand{\cosiotaplus}[1]{\IfEqCase{#1}{{GW230529ay_combined_imrphm_high_spin}{1.01}{GW230529ay_combined_imrphm_low_spin}{0.99}{GW230529ay_imrphen_nsbh}{0.94}{GW230529ay_imrseobv4_nsbh}{0.96}{GW230529ay_imrpv2_low_spin}{0.95}}}
\newcommand{\cosiotaonepercent}[1]{\IfEqCase{#1}{{GW230529ay_combined_imrphm_high_spin}{-0.99}{GW230529ay_combined_imrphm_low_spin}{-0.99}{GW230529ay_imrphen_nsbh}{-0.99}{GW230529ay_imrseobv4_nsbh}{-0.99}{GW230529ay_imrpv2_low_spin}{-0.99}}}
\newcommand{\cosiotaninetyninepercent}[1]{\IfEqCase{#1}{{GW230529ay_combined_imrphm_high_spin}{0.99}{GW230529ay_combined_imrphm_low_spin}{0.99}{GW230529ay_imrphen_nsbh}{0.99}{GW230529ay_imrseobv4_nsbh}{0.99}{GW230529ay_imrpv2_low_spin}{0.99}}}
\newcommand{\cosiotafivepercent}[1]{\IfEqCase{#1}{{GW230529ay_combined_imrphm_high_spin}{-0.96}{GW230529ay_combined_imrphm_low_spin}{-0.96}{GW230529ay_imrphen_nsbh}{-0.97}{GW230529ay_imrseobv4_nsbh}{-0.97}{GW230529ay_imrpv2_low_spin}{-0.95}}}
\newcommand{\cosiotaninetyfivepercent}[1]{\IfEqCase{#1}{{GW230529ay_combined_imrphm_high_spin}{0.95}{GW230529ay_combined_imrphm_low_spin}{0.95}{GW230529ay_imrphen_nsbh}{0.97}{GW230529ay_imrseobv4_nsbh}{0.97}{GW230529ay_imrpv2_low_spin}{0.96}}}
\newcommand{\cosiotaninetypercent}[1]{\IfEqCase{#1}{{GW230529ay_combined_imrphm_high_spin}{0.91}{GW230529ay_combined_imrphm_low_spin}{0.91}{GW230529ay_imrphen_nsbh}{0.93}{GW230529ay_imrseobv4_nsbh}{0.93}{GW230529ay_imrpv2_low_spin}{0.92}}}
\newcommand{\costhetajnminus}[1]{\IfEqCase{#1}{{GW230529ay_combined_imrphm_high_spin}{0.89}{GW230529ay_combined_imrphm_low_spin}{0.92}{GW230529ay_imrphen_nsbh}{0.99}{GW230529ay_imrseobv4_nsbh}{0.97}{GW230529ay_imrpv2_low_spin}{0.97}}}
\newcommand{\costhetajnmed}[1]{\IfEqCase{#1}{{GW230529ay_combined_imrphm_high_spin}{-0.07}{GW230529ay_combined_imrphm_low_spin}{-0.04}{GW230529ay_imrphen_nsbh}{0.03}{GW230529ay_imrseobv4_nsbh}{0.01}{GW230529ay_imrpv2_low_spin}{0.01}}}
\newcommand{\costhetajnplus}[1]{\IfEqCase{#1}{{GW230529ay_combined_imrphm_high_spin}{1.02}{GW230529ay_combined_imrphm_low_spin}{0.99}{GW230529ay_imrphen_nsbh}{0.94}{GW230529ay_imrseobv4_nsbh}{0.96}{GW230529ay_imrpv2_low_spin}{0.95}}}
\newcommand{\costhetajnonepercent}[1]{\IfEqCase{#1}{{GW230529ay_combined_imrphm_high_spin}{-0.99}{GW230529ay_combined_imrphm_low_spin}{-0.99}{GW230529ay_imrphen_nsbh}{-0.99}{GW230529ay_imrseobv4_nsbh}{-0.99}{GW230529ay_imrpv2_low_spin}{-0.99}}}
\newcommand{\costhetajnninetyninepercent}[1]{\IfEqCase{#1}{{GW230529ay_combined_imrphm_high_spin}{0.99}{GW230529ay_combined_imrphm_low_spin}{0.99}{GW230529ay_imrphen_nsbh}{0.99}{GW230529ay_imrseobv4_nsbh}{0.99}{GW230529ay_imrpv2_low_spin}{0.99}}}
\newcommand{\costhetajnfivepercent}[1]{\IfEqCase{#1}{{GW230529ay_combined_imrphm_high_spin}{-0.96}{GW230529ay_combined_imrphm_low_spin}{-0.95}{GW230529ay_imrphen_nsbh}{-0.97}{GW230529ay_imrseobv4_nsbh}{-0.97}{GW230529ay_imrpv2_low_spin}{-0.96}}}
\newcommand{\costhetajnninetyfivepercent}[1]{\IfEqCase{#1}{{GW230529ay_combined_imrphm_high_spin}{0.95}{GW230529ay_combined_imrphm_low_spin}{0.95}{GW230529ay_imrphen_nsbh}{0.97}{GW230529ay_imrseobv4_nsbh}{0.97}{GW230529ay_imrpv2_low_spin}{0.96}}}
\newcommand{\costhetajnninetypercent}[1]{\IfEqCase{#1}{{GW230529ay_combined_imrphm_high_spin}{0.91}{GW230529ay_combined_imrphm_low_spin}{0.91}{GW230529ay_imrphen_nsbh}{0.93}{GW230529ay_imrseobv4_nsbh}{0.93}{GW230529ay_imrpv2_low_spin}{0.92}}}
\newcommand{\costiltoneminus}[1]{\IfEqCase{#1}{{GW230529ay_combined_imrphm_high_spin}{0.57}{GW230529ay_combined_imrphm_low_spin}{0.53}{GW230529ay_imrphen_nsbh}{0.00}{GW230529ay_imrseobv4_nsbh}{0.00}{GW230529ay_imrpv2_low_spin}{0.50}}}
\newcommand{\costiltonemed}[1]{\IfEqCase{#1}{{GW230529ay_combined_imrphm_high_spin}{-0.30}{GW230529ay_combined_imrphm_low_spin}{-0.35}{GW230529ay_imrphen_nsbh}{-1.00}{GW230529ay_imrseobv4_nsbh}{-1.00}{GW230529ay_imrpv2_low_spin}{-0.42}}}
\newcommand{\costiltoneplus}[1]{\IfEqCase{#1}{{GW230529ay_combined_imrphm_high_spin}{0.63}{GW230529ay_combined_imrphm_low_spin}{0.38}{GW230529ay_imrphen_nsbh}{2.00}{GW230529ay_imrseobv4_nsbh}{2.00}{GW230529ay_imrpv2_low_spin}{0.52}}}
\newcommand{\costiltoneonepercent}[1]{\IfEqCase{#1}{{GW230529ay_combined_imrphm_high_spin}{-0.97}{GW230529ay_combined_imrphm_low_spin}{-0.98}{GW230529ay_imrphen_nsbh}{-1.00}{GW230529ay_imrseobv4_nsbh}{-1.00}{GW230529ay_imrpv2_low_spin}{-0.98}}}
\newcommand{\costiltoneninetyninepercent}[1]{\IfEqCase{#1}{{GW230529ay_combined_imrphm_high_spin}{0.73}{GW230529ay_combined_imrphm_low_spin}{0.58}{GW230529ay_imrphen_nsbh}{1.00}{GW230529ay_imrseobv4_nsbh}{1.00}{GW230529ay_imrpv2_low_spin}{0.41}}}
\newcommand{\costiltonefivepercent}[1]{\IfEqCase{#1}{{GW230529ay_combined_imrphm_high_spin}{-0.87}{GW230529ay_combined_imrphm_low_spin}{-0.88}{GW230529ay_imrphen_nsbh}{-1.00}{GW230529ay_imrseobv4_nsbh}{-1.00}{GW230529ay_imrpv2_low_spin}{-0.91}}}
\newcommand{\costiltoneninetyfivepercent}[1]{\IfEqCase{#1}{{GW230529ay_combined_imrphm_high_spin}{0.34}{GW230529ay_combined_imrphm_low_spin}{0.02}{GW230529ay_imrphen_nsbh}{1.00}{GW230529ay_imrseobv4_nsbh}{1.00}{GW230529ay_imrpv2_low_spin}{0.11}}}
\newcommand{\costiltoneninetypercent}[1]{\IfEqCase{#1}{{GW230529ay_combined_imrphm_high_spin}{0.15}{GW230529ay_combined_imrphm_low_spin}{-0.08}{GW230529ay_imrphen_nsbh}{1.00}{GW230529ay_imrseobv4_nsbh}{1.00}{GW230529ay_imrpv2_low_spin}{0.01}}}
\newcommand{\costilttwominus}[1]{\IfEqCase{#1}{{GW230529ay_combined_imrphm_high_spin}{0.76}{GW230529ay_combined_imrphm_low_spin}{0.90}{GW230529ay_imrphen_nsbh}{0.00}{GW230529ay_imrseobv4_nsbh}{0.00}{GW230529ay_imrpv2_low_spin}{0.87}}}
\newcommand{\costilttwomed}[1]{\IfEqCase{#1}{{GW230529ay_combined_imrphm_high_spin}{-0.14}{GW230529ay_combined_imrphm_low_spin}{-0.00}{GW230529ay_imrphen_nsbh}{-1.00}{GW230529ay_imrseobv4_nsbh}{-1.00}{GW230529ay_imrpv2_low_spin}{-0.04}}}
\newcommand{\costilttwoplus}[1]{\IfEqCase{#1}{{GW230529ay_combined_imrphm_high_spin}{0.97}{GW230529ay_combined_imrphm_low_spin}{0.90}{GW230529ay_imrphen_nsbh}{2.00}{GW230529ay_imrseobv4_nsbh}{2.00}{GW230529ay_imrpv2_low_spin}{0.93}}}
\newcommand{\costilttwoonepercent}[1]{\IfEqCase{#1}{{GW230529ay_combined_imrphm_high_spin}{-0.98}{GW230529ay_combined_imrphm_low_spin}{-0.98}{GW230529ay_imrphen_nsbh}{-1.00}{GW230529ay_imrseobv4_nsbh}{-1.00}{GW230529ay_imrpv2_low_spin}{-0.98}}}
\newcommand{\costilttwoninetyninepercent}[1]{\IfEqCase{#1}{{GW230529ay_combined_imrphm_high_spin}{0.96}{GW230529ay_combined_imrphm_low_spin}{0.98}{GW230529ay_imrphen_nsbh}{1.00}{GW230529ay_imrseobv4_nsbh}{1.00}{GW230529ay_imrpv2_low_spin}{0.98}}}
\newcommand{\costilttwofivepercent}[1]{\IfEqCase{#1}{{GW230529ay_combined_imrphm_high_spin}{-0.90}{GW230529ay_combined_imrphm_low_spin}{-0.90}{GW230529ay_imrphen_nsbh}{-1.00}{GW230529ay_imrseobv4_nsbh}{-1.00}{GW230529ay_imrpv2_low_spin}{-0.91}}}
\newcommand{\costilttwoninetyfivepercent}[1]{\IfEqCase{#1}{{GW230529ay_combined_imrphm_high_spin}{0.83}{GW230529ay_combined_imrphm_low_spin}{0.90}{GW230529ay_imrphen_nsbh}{1.00}{GW230529ay_imrseobv4_nsbh}{1.00}{GW230529ay_imrpv2_low_spin}{0.89}}}
\newcommand{\costilttwoninetypercent}[1]{\IfEqCase{#1}{{GW230529ay_combined_imrphm_high_spin}{0.68}{GW230529ay_combined_imrphm_low_spin}{0.80}{GW230529ay_imrphen_nsbh}{1.00}{GW230529ay_imrseobv4_nsbh}{1.00}{GW230529ay_imrpv2_low_spin}{0.79}}}
\newcommand{\decminus}[1]{\IfEqCase{#1}{{GW230529ay_combined_imrphm_high_spin}{1.01722}{GW230529ay_combined_imrphm_low_spin}{1.07278}{GW230529ay_imrphen_nsbh}{1.07111}{GW230529ay_imrseobv4_nsbh}{1.08237}{GW230529ay_imrpv2_low_spin}{1.04604}}}
\newcommand{\decmed}[1]{\IfEqCase{#1}{{GW230529ay_combined_imrphm_high_spin}{-0.02585}{GW230529ay_combined_imrphm_low_spin}{0.00375}{GW230529ay_imrphen_nsbh}{0.01140}{GW230529ay_imrseobv4_nsbh}{0.01095}{GW230529ay_imrpv2_low_spin}{-0.02655}}}
\newcommand{\decplus}[1]{\IfEqCase{#1}{{GW230529ay_combined_imrphm_high_spin}{1.04884}{GW230529ay_combined_imrphm_low_spin}{1.04498}{GW230529ay_imrphen_nsbh}{1.04077}{GW230529ay_imrseobv4_nsbh}{1.03897}{GW230529ay_imrpv2_low_spin}{1.07937}}}
\newcommand{\deconepercent}[1]{\IfEqCase{#1}{{GW230529ay_combined_imrphm_high_spin}{-1.31720}{GW230529ay_combined_imrphm_low_spin}{-1.33254}{GW230529ay_imrphen_nsbh}{-1.33499}{GW230529ay_imrseobv4_nsbh}{-1.35571}{GW230529ay_imrpv2_low_spin}{-1.35189}}}
\newcommand{\decninetyninepercent}[1]{\IfEqCase{#1}{{GW230529ay_combined_imrphm_high_spin}{1.30985}{GW230529ay_combined_imrphm_low_spin}{1.32964}{GW230529ay_imrphen_nsbh}{1.32838}{GW230529ay_imrseobv4_nsbh}{1.32271}{GW230529ay_imrpv2_low_spin}{1.35890}}}
\newcommand{\decfivepercent}[1]{\IfEqCase{#1}{{GW230529ay_combined_imrphm_high_spin}{-1.04307}{GW230529ay_combined_imrphm_low_spin}{-1.06904}{GW230529ay_imrphen_nsbh}{-1.05971}{GW230529ay_imrseobv4_nsbh}{-1.07141}{GW230529ay_imrpv2_low_spin}{-1.07260}}}
\newcommand{\decninetyfivepercent}[1]{\IfEqCase{#1}{{GW230529ay_combined_imrphm_high_spin}{1.02299}{GW230529ay_combined_imrphm_low_spin}{1.04873}{GW230529ay_imrphen_nsbh}{1.05217}{GW230529ay_imrseobv4_nsbh}{1.04992}{GW230529ay_imrpv2_low_spin}{1.05282}}}
\newcommand{\decninetypercent}[1]{\IfEqCase{#1}{{GW230529ay_combined_imrphm_high_spin}{0.82808}{GW230529ay_combined_imrphm_low_spin}{0.85607}{GW230529ay_imrphen_nsbh}{0.87266}{GW230529ay_imrseobv4_nsbh}{0.86521}{GW230529ay_imrpv2_low_spin}{0.85056}}}
\newcommand{\deltalambdaminus}[1]{\IfEqCase{#1}{{GW230529ay_combined_imrphm_high_spin}{-}{GW230529ay_combined_imrphm_low_spin}{-}{GW230529ay_imrphen_nsbh}{65}{GW230529ay_imrseobv4_nsbh}{72}{GW230529ay_imrpv2_low_spin}{471}}}
\newcommand{\deltalambdamed}[1]{\IfEqCase{#1}{{GW230529ay_combined_imrphm_high_spin}{-}{GW230529ay_combined_imrphm_low_spin}{-}{GW230529ay_imrphen_nsbh}{72}{GW230529ay_imrseobv4_nsbh}{79}{GW230529ay_imrpv2_low_spin}{-7}}}
\newcommand{\deltalambdaplus}[1]{\IfEqCase{#1}{{GW230529ay_combined_imrphm_high_spin}{-}{GW230529ay_combined_imrphm_low_spin}{-}{GW230529ay_imrphen_nsbh}{248}{GW230529ay_imrseobv4_nsbh}{266}{GW230529ay_imrpv2_low_spin}{265}}}
\newcommand{\deltalambdaonepercent}[1]{\IfEqCase{#1}{{GW230529ay_combined_imrphm_high_spin}{-}{GW230529ay_combined_imrphm_low_spin}{-}{GW230529ay_imrphen_nsbh}{1}{GW230529ay_imrseobv4_nsbh}{2}{GW230529ay_imrpv2_low_spin}{-886}}}
\newcommand{\deltalambdaninetyninepercent}[1]{\IfEqCase{#1}{{GW230529ay_combined_imrphm_high_spin}{-}{GW230529ay_combined_imrphm_low_spin}{-}{GW230529ay_imrphen_nsbh}{457}{GW230529ay_imrseobv4_nsbh}{498}{GW230529ay_imrpv2_low_spin}{406}}}
\newcommand{\deltalambdafivepercent}[1]{\IfEqCase{#1}{{GW230529ay_combined_imrphm_high_spin}{-}{GW230529ay_combined_imrphm_low_spin}{-}{GW230529ay_imrphen_nsbh}{7}{GW230529ay_imrseobv4_nsbh}{7}{GW230529ay_imrpv2_low_spin}{-478}}}
\newcommand{\deltalambdaninetyfivepercent}[1]{\IfEqCase{#1}{{GW230529ay_combined_imrphm_high_spin}{-}{GW230529ay_combined_imrphm_low_spin}{-}{GW230529ay_imrphen_nsbh}{320}{GW230529ay_imrseobv4_nsbh}{345}{GW230529ay_imrpv2_low_spin}{258}}}
\newcommand{\deltalambdaninetypercent}[1]{\IfEqCase{#1}{{GW230529ay_combined_imrphm_high_spin}{-}{GW230529ay_combined_imrphm_low_spin}{-}{GW230529ay_imrphen_nsbh}{247}{GW230529ay_imrseobv4_nsbh}{273}{GW230529ay_imrpv2_low_spin}{167}}}
\newcommand{\finalmasssourceminus}[1]{\IfEqCase{#1}{{GW230529ay_combined_imrphm_high_spin}{-}{GW230529ay_combined_imrphm_low_spin}{-}{GW230529ay_imrphen_nsbh}{0.6}{GW230529ay_imrseobv4_nsbh}{0.6}{GW230529ay_imrpv2_low_spin}{-}}}
\newcommand{\finalmasssourcemed}[1]{\IfEqCase{#1}{{GW230529ay_combined_imrphm_high_spin}{-}{GW230529ay_combined_imrphm_low_spin}{-}{GW230529ay_imrphen_nsbh}{5.0}{GW230529ay_imrseobv4_nsbh}{4.9}{GW230529ay_imrpv2_low_spin}{-}}}
\newcommand{\finalmasssourceplus}[1]{\IfEqCase{#1}{{GW230529ay_combined_imrphm_high_spin}{-}{GW230529ay_combined_imrphm_low_spin}{-}{GW230529ay_imrphen_nsbh}{0.8}{GW230529ay_imrseobv4_nsbh}{0.9}{GW230529ay_imrpv2_low_spin}{-}}}
\newcommand{\finalmasssourceonepercent}[1]{\IfEqCase{#1}{{GW230529ay_combined_imrphm_high_spin}{-}{GW230529ay_combined_imrphm_low_spin}{-}{GW230529ay_imrphen_nsbh}{4.3}{GW230529ay_imrseobv4_nsbh}{4.2}{GW230529ay_imrpv2_low_spin}{-}}}
\newcommand{\finalmasssourceninetyninepercent}[1]{\IfEqCase{#1}{{GW230529ay_combined_imrphm_high_spin}{-}{GW230529ay_combined_imrphm_low_spin}{-}{GW230529ay_imrphen_nsbh}{6.4}{GW230529ay_imrseobv4_nsbh}{6.3}{GW230529ay_imrpv2_low_spin}{-}}}
\newcommand{\finalmasssourcefivepercent}[1]{\IfEqCase{#1}{{GW230529ay_combined_imrphm_high_spin}{-}{GW230529ay_combined_imrphm_low_spin}{-}{GW230529ay_imrphen_nsbh}{4.4}{GW230529ay_imrseobv4_nsbh}{4.3}{GW230529ay_imrpv2_low_spin}{-}}}
\newcommand{\finalmasssourceninetyfivepercent}[1]{\IfEqCase{#1}{{GW230529ay_combined_imrphm_high_spin}{-}{GW230529ay_combined_imrphm_low_spin}{-}{GW230529ay_imrphen_nsbh}{5.9}{GW230529ay_imrseobv4_nsbh}{5.8}{GW230529ay_imrpv2_low_spin}{-}}}
\newcommand{\finalmasssourceninetypercent}[1]{\IfEqCase{#1}{{GW230529ay_combined_imrphm_high_spin}{-}{GW230529ay_combined_imrphm_low_spin}{-}{GW230529ay_imrphen_nsbh}{5.6}{GW230529ay_imrseobv4_nsbh}{5.6}{GW230529ay_imrpv2_low_spin}{-}}}
\newcommand{\finalspinminus}[1]{\IfEqCase{#1}{{GW230529ay_combined_imrphm_high_spin}{-}{GW230529ay_combined_imrphm_low_spin}{-}{GW230529ay_imrphen_nsbh}{0.02}{GW230529ay_imrseobv4_nsbh}{0.02}{GW230529ay_imrpv2_low_spin}{-}}}
\newcommand{\finalspinmed}[1]{\IfEqCase{#1}{{GW230529ay_combined_imrphm_high_spin}{-}{GW230529ay_combined_imrphm_low_spin}{-}{GW230529ay_imrphen_nsbh}{0.53}{GW230529ay_imrseobv4_nsbh}{0.53}{GW230529ay_imrpv2_low_spin}{-}}}
\newcommand{\finalspinplus}[1]{\IfEqCase{#1}{{GW230529ay_combined_imrphm_high_spin}{-}{GW230529ay_combined_imrphm_low_spin}{-}{GW230529ay_imrphen_nsbh}{0.05}{GW230529ay_imrseobv4_nsbh}{0.06}{GW230529ay_imrpv2_low_spin}{-}}}
\newcommand{\finalspinonepercent}[1]{\IfEqCase{#1}{{GW230529ay_combined_imrphm_high_spin}{-}{GW230529ay_combined_imrphm_low_spin}{-}{GW230529ay_imrphen_nsbh}{0.51}{GW230529ay_imrseobv4_nsbh}{0.51}{GW230529ay_imrpv2_low_spin}{-}}}
\newcommand{\finalspinninetyninepercent}[1]{\IfEqCase{#1}{{GW230529ay_combined_imrphm_high_spin}{-}{GW230529ay_combined_imrphm_low_spin}{-}{GW230529ay_imrphen_nsbh}{0.59}{GW230529ay_imrseobv4_nsbh}{0.61}{GW230529ay_imrpv2_low_spin}{-}}}
\newcommand{\finalspinfivepercent}[1]{\IfEqCase{#1}{{GW230529ay_combined_imrphm_high_spin}{-}{GW230529ay_combined_imrphm_low_spin}{-}{GW230529ay_imrphen_nsbh}{0.51}{GW230529ay_imrseobv4_nsbh}{0.51}{GW230529ay_imrpv2_low_spin}{-}}}
\newcommand{\finalspinninetyfivepercent}[1]{\IfEqCase{#1}{{GW230529ay_combined_imrphm_high_spin}{-}{GW230529ay_combined_imrphm_low_spin}{-}{GW230529ay_imrphen_nsbh}{0.58}{GW230529ay_imrseobv4_nsbh}{0.59}{GW230529ay_imrpv2_low_spin}{-}}}
\newcommand{\finalspinninetypercent}[1]{\IfEqCase{#1}{{GW230529ay_combined_imrphm_high_spin}{-}{GW230529ay_combined_imrphm_low_spin}{-}{GW230529ay_imrphen_nsbh}{0.57}{GW230529ay_imrseobv4_nsbh}{0.58}{GW230529ay_imrpv2_low_spin}{-}}}
\newcommand{\geocenttimeminus}[1]{\IfEqCase{#1}{{GW230529ay_combined_imrphm_high_spin}{0.0}{GW230529ay_combined_imrphm_low_spin}{0.0}{GW230529ay_imrphen_nsbh}{0.0}{GW230529ay_imrseobv4_nsbh}{0.0}{GW230529ay_imrpv2_low_spin}{0.0}}}
\newcommand{\geocenttimemed}[1]{\IfEqCase{#1}{{GW230529ay_combined_imrphm_high_spin}{1369419318.7}{GW230529ay_combined_imrphm_low_spin}{1369419318.7}{GW230529ay_imrphen_nsbh}{1369419318.7}{GW230529ay_imrseobv4_nsbh}{1369419318.8}{GW230529ay_imrpv2_low_spin}{1369419318.7}}}
\newcommand{\geocenttimeplus}[1]{\IfEqCase{#1}{{GW230529ay_combined_imrphm_high_spin}{0.0}{GW230529ay_combined_imrphm_low_spin}{0.0}{GW230529ay_imrphen_nsbh}{0.0}{GW230529ay_imrseobv4_nsbh}{0.0}{GW230529ay_imrpv2_low_spin}{0.0}}}
\newcommand{\geocenttimeonepercent}[1]{\IfEqCase{#1}{{GW230529ay_combined_imrphm_high_spin}{1369419318.7}{GW230529ay_combined_imrphm_low_spin}{1369419318.7}{GW230529ay_imrphen_nsbh}{1369419318.7}{GW230529ay_imrseobv4_nsbh}{1369419318.7}{GW230529ay_imrpv2_low_spin}{1369419318.7}}}
\newcommand{\geocenttimeninetyninepercent}[1]{\IfEqCase{#1}{{GW230529ay_combined_imrphm_high_spin}{1369419318.8}{GW230529ay_combined_imrphm_low_spin}{1369419318.8}{GW230529ay_imrphen_nsbh}{1369419318.8}{GW230529ay_imrseobv4_nsbh}{1369419318.8}{GW230529ay_imrpv2_low_spin}{1369419318.8}}}
\newcommand{\geocenttimefivepercent}[1]{\IfEqCase{#1}{{GW230529ay_combined_imrphm_high_spin}{1369419318.7}{GW230529ay_combined_imrphm_low_spin}{1369419318.7}{GW230529ay_imrphen_nsbh}{1369419318.7}{GW230529ay_imrseobv4_nsbh}{1369419318.7}{GW230529ay_imrpv2_low_spin}{1369419318.7}}}
\newcommand{\geocenttimeninetyfivepercent}[1]{\IfEqCase{#1}{{GW230529ay_combined_imrphm_high_spin}{1369419318.8}{GW230529ay_combined_imrphm_low_spin}{1369419318.8}{GW230529ay_imrphen_nsbh}{1369419318.8}{GW230529ay_imrseobv4_nsbh}{1369419318.8}{GW230529ay_imrpv2_low_spin}{1369419318.8}}}
\newcommand{\geocenttimeninetypercent}[1]{\IfEqCase{#1}{{GW230529ay_combined_imrphm_high_spin}{1369419318.8}{GW230529ay_combined_imrphm_low_spin}{1369419318.8}{GW230529ay_imrphen_nsbh}{1369419318.8}{GW230529ay_imrseobv4_nsbh}{1369419318.8}{GW230529ay_imrpv2_low_spin}{1369419318.8}}}
\newcommand{\iotaminus}[1]{\IfEqCase{#1}{{GW230529ay_combined_imrphm_high_spin}{1.32}{GW230529ay_combined_imrphm_low_spin}{1.30}{GW230529ay_imrphen_nsbh}{1.28}{GW230529ay_imrseobv4_nsbh}{1.31}{GW230529ay_imrpv2_low_spin}{1.28}}}
\newcommand{\iotamed}[1]{\IfEqCase{#1}{{GW230529ay_combined_imrphm_high_spin}{1.63}{GW230529ay_combined_imrphm_low_spin}{1.61}{GW230529ay_imrphen_nsbh}{1.54}{GW230529ay_imrseobv4_nsbh}{1.56}{GW230529ay_imrpv2_low_spin}{1.56}}}
\newcommand{\iotaplus}[1]{\IfEqCase{#1}{{GW230529ay_combined_imrphm_high_spin}{1.23}{GW230529ay_combined_imrphm_low_spin}{1.23}{GW230529ay_imrphen_nsbh}{1.34}{GW230529ay_imrseobv4_nsbh}{1.32}{GW230529ay_imrpv2_low_spin}{1.27}}}
\newcommand{\iotaonepercent}[1]{\IfEqCase{#1}{{GW230529ay_combined_imrphm_high_spin}{0.13}{GW230529ay_combined_imrphm_low_spin}{0.14}{GW230529ay_imrphen_nsbh}{0.11}{GW230529ay_imrseobv4_nsbh}{0.12}{GW230529ay_imrpv2_low_spin}{0.12}}}
\newcommand{\iotaninetyninepercent}[1]{\IfEqCase{#1}{{GW230529ay_combined_imrphm_high_spin}{3.01}{GW230529ay_combined_imrphm_low_spin}{3.00}{GW230529ay_imrphen_nsbh}{3.02}{GW230529ay_imrseobv4_nsbh}{3.03}{GW230529ay_imrpv2_low_spin}{3.00}}}
\newcommand{\iotafivepercent}[1]{\IfEqCase{#1}{{GW230529ay_combined_imrphm_high_spin}{0.30}{GW230529ay_combined_imrphm_low_spin}{0.31}{GW230529ay_imrphen_nsbh}{0.26}{GW230529ay_imrseobv4_nsbh}{0.26}{GW230529ay_imrpv2_low_spin}{0.28}}}
\newcommand{\iotaninetyfivepercent}[1]{\IfEqCase{#1}{{GW230529ay_combined_imrphm_high_spin}{2.85}{GW230529ay_combined_imrphm_low_spin}{2.84}{GW230529ay_imrphen_nsbh}{2.88}{GW230529ay_imrseobv4_nsbh}{2.89}{GW230529ay_imrpv2_low_spin}{2.84}}}
\newcommand{\iotaninetypercent}[1]{\IfEqCase{#1}{{GW230529ay_combined_imrphm_high_spin}{2.73}{GW230529ay_combined_imrphm_low_spin}{2.71}{GW230529ay_imrphen_nsbh}{2.75}{GW230529ay_imrseobv4_nsbh}{2.77}{GW230529ay_imrpv2_low_spin}{2.71}}}
\newcommand{\lambdaoneminus}[1]{\IfEqCase{#1}{{GW230529ay_combined_imrphm_high_spin}{-}{GW230529ay_combined_imrphm_low_spin}{-}{GW230529ay_imrphen_nsbh}{0}{GW230529ay_imrseobv4_nsbh}{0}{GW230529ay_imrpv2_low_spin}{238}}}
\newcommand{\lambdaonemed}[1]{\IfEqCase{#1}{{GW230529ay_combined_imrphm_high_spin}{-}{GW230529ay_combined_imrphm_low_spin}{-}{GW230529ay_imrphen_nsbh}{0}{GW230529ay_imrseobv4_nsbh}{0}{GW230529ay_imrpv2_low_spin}{259}}}
\newcommand{\lambdaoneplus}[1]{\IfEqCase{#1}{{GW230529ay_combined_imrphm_high_spin}{-}{GW230529ay_combined_imrphm_low_spin}{-}{GW230529ay_imrphen_nsbh}{0}{GW230529ay_imrseobv4_nsbh}{0}{GW230529ay_imrpv2_low_spin}{1969}}}
\newcommand{\lambdaoneonepercent}[1]{\IfEqCase{#1}{{GW230529ay_combined_imrphm_high_spin}{-}{GW230529ay_combined_imrphm_low_spin}{-}{GW230529ay_imrphen_nsbh}{0}{GW230529ay_imrseobv4_nsbh}{0}{GW230529ay_imrpv2_low_spin}{4}}}
\newcommand{\lambdaoneninetyninepercent}[1]{\IfEqCase{#1}{{GW230529ay_combined_imrphm_high_spin}{-}{GW230529ay_combined_imrphm_low_spin}{-}{GW230529ay_imrphen_nsbh}{0}{GW230529ay_imrseobv4_nsbh}{0}{GW230529ay_imrpv2_low_spin}{3638}}}
\newcommand{\lambdaonefivepercent}[1]{\IfEqCase{#1}{{GW230529ay_combined_imrphm_high_spin}{-}{GW230529ay_combined_imrphm_low_spin}{-}{GW230529ay_imrphen_nsbh}{0}{GW230529ay_imrseobv4_nsbh}{0}{GW230529ay_imrpv2_low_spin}{20}}}
\newcommand{\lambdaoneninetyfivepercent}[1]{\IfEqCase{#1}{{GW230529ay_combined_imrphm_high_spin}{-}{GW230529ay_combined_imrphm_low_spin}{-}{GW230529ay_imrphen_nsbh}{0}{GW230529ay_imrseobv4_nsbh}{0}{GW230529ay_imrpv2_low_spin}{2228}}}
\newcommand{\lambdaoneninetypercent}[1]{\IfEqCase{#1}{{GW230529ay_combined_imrphm_high_spin}{-}{GW230529ay_combined_imrphm_low_spin}{-}{GW230529ay_imrphen_nsbh}{0}{GW230529ay_imrseobv4_nsbh}{0}{GW230529ay_imrpv2_low_spin}{1462}}}
\newcommand{\lambdatwominus}[1]{\IfEqCase{#1}{{GW230529ay_combined_imrphm_high_spin}{-}{GW230529ay_combined_imrphm_low_spin}{-}{GW230529ay_imrphen_nsbh}{2497}{GW230529ay_imrseobv4_nsbh}{2144}{GW230529ay_imrpv2_low_spin}{1610}}}
\newcommand{\lambdatwomed}[1]{\IfEqCase{#1}{{GW230529ay_combined_imrphm_high_spin}{-}{GW230529ay_combined_imrphm_low_spin}{-}{GW230529ay_imrphen_nsbh}{2866}{GW230529ay_imrseobv4_nsbh}{2457}{GW230529ay_imrpv2_low_spin}{1778}}}
\newcommand{\lambdatwoplus}[1]{\IfEqCase{#1}{{GW230529ay_combined_imrphm_high_spin}{-}{GW230529ay_combined_imrphm_low_spin}{-}{GW230529ay_imrphen_nsbh}{1894}{GW230529ay_imrseobv4_nsbh}{2207}{GW230529ay_imrpv2_low_spin}{2742}}}
\newcommand{\lambdatwoonepercent}[1]{\IfEqCase{#1}{{GW230529ay_combined_imrphm_high_spin}{-}{GW230529ay_combined_imrphm_low_spin}{-}{GW230529ay_imrphen_nsbh}{75}{GW230529ay_imrseobv4_nsbh}{67}{GW230529ay_imrpv2_low_spin}{34}}}
\newcommand{\lambdatwoninetyninepercent}[1]{\IfEqCase{#1}{{GW230529ay_combined_imrphm_high_spin}{-}{GW230529ay_combined_imrphm_low_spin}{-}{GW230529ay_imrphen_nsbh}{4953}{GW230529ay_imrseobv4_nsbh}{4926}{GW230529ay_imrpv2_low_spin}{4884}}}
\newcommand{\lambdatwofivepercent}[1]{\IfEqCase{#1}{{GW230529ay_combined_imrphm_high_spin}{-}{GW230529ay_combined_imrphm_low_spin}{-}{GW230529ay_imrphen_nsbh}{369}{GW230529ay_imrseobv4_nsbh}{314}{GW230529ay_imrpv2_low_spin}{168}}}
\newcommand{\lambdatwoninetyfivepercent}[1]{\IfEqCase{#1}{{GW230529ay_combined_imrphm_high_spin}{-}{GW230529ay_combined_imrphm_low_spin}{-}{GW230529ay_imrphen_nsbh}{4760}{GW230529ay_imrseobv4_nsbh}{4664}{GW230529ay_imrpv2_low_spin}{4520}}}
\newcommand{\lambdatwoninetypercent}[1]{\IfEqCase{#1}{{GW230529ay_combined_imrphm_high_spin}{-}{GW230529ay_combined_imrphm_low_spin}{-}{GW230529ay_imrphen_nsbh}{4519}{GW230529ay_imrseobv4_nsbh}{4365}{GW230529ay_imrpv2_low_spin}{4087}}}
\newcommand{\loglikelihoodminus}[1]{\IfEqCase{#1}{{GW230529ay_combined_imrphm_high_spin}{4.4}{GW230529ay_combined_imrphm_low_spin}{4.3}{GW230529ay_imrphen_nsbh}{3.4}{GW230529ay_imrseobv4_nsbh}{3.5}{GW230529ay_imrpv2_low_spin}{4.7}}}
\newcommand{\loglikelihoodmed}[1]{\IfEqCase{#1}{{GW230529ay_combined_imrphm_high_spin}{64.3}{GW230529ay_combined_imrphm_low_spin}{64.2}{GW230529ay_imrphen_nsbh}{59.6}{GW230529ay_imrseobv4_nsbh}{59.7}{GW230529ay_imrpv2_low_spin}{60.0}}}
\newcommand{\loglikelihoodplus}[1]{\IfEqCase{#1}{{GW230529ay_combined_imrphm_high_spin}{3.6}{GW230529ay_combined_imrphm_low_spin}{3.4}{GW230529ay_imrphen_nsbh}{2.2}{GW230529ay_imrseobv4_nsbh}{2.1}{GW230529ay_imrpv2_low_spin}{3.1}}}
\newcommand{\loglikelihoodonepercent}[1]{\IfEqCase{#1}{{GW230529ay_combined_imrphm_high_spin}{57.6}{GW230529ay_combined_imrphm_low_spin}{57.6}{GW230529ay_imrphen_nsbh}{54.3}{GW230529ay_imrseobv4_nsbh}{54.0}{GW230529ay_imrpv2_low_spin}{52.8}}}
\newcommand{\loglikelihoodninetyninepercent}[1]{\IfEqCase{#1}{{GW230529ay_combined_imrphm_high_spin}{68.9}{GW230529ay_combined_imrphm_low_spin}{68.6}{GW230529ay_imrphen_nsbh}{62.5}{GW230529ay_imrseobv4_nsbh}{62.4}{GW230529ay_imrpv2_low_spin}{64.2}}}
\newcommand{\loglikelihoodfivepercent}[1]{\IfEqCase{#1}{{GW230529ay_combined_imrphm_high_spin}{59.8}{GW230529ay_combined_imrphm_low_spin}{60.0}{GW230529ay_imrphen_nsbh}{56.3}{GW230529ay_imrseobv4_nsbh}{56.2}{GW230529ay_imrpv2_low_spin}{55.3}}}
\newcommand{\loglikelihoodninetyfivepercent}[1]{\IfEqCase{#1}{{GW230529ay_combined_imrphm_high_spin}{67.9}{GW230529ay_combined_imrphm_low_spin}{67.6}{GW230529ay_imrphen_nsbh}{61.9}{GW230529ay_imrseobv4_nsbh}{61.8}{GW230529ay_imrpv2_low_spin}{63.1}}}
\newcommand{\loglikelihoodninetypercent}[1]{\IfEqCase{#1}{{GW230529ay_combined_imrphm_high_spin}{67.2}{GW230529ay_combined_imrphm_low_spin}{67.0}{GW230529ay_imrphen_nsbh}{61.4}{GW230529ay_imrseobv4_nsbh}{61.4}{GW230529ay_imrpv2_low_spin}{62.5}}}
\newcommand{\logpriorminus}[1]{\IfEqCase{#1}{{GW230529ay_combined_imrphm_high_spin}{6.4}{GW230529ay_combined_imrphm_low_spin}{6.4}{GW230529ay_imrphen_nsbh}{6.5}{GW230529ay_imrseobv4_nsbh}{6.5}{GW230529ay_imrpv2_low_spin}{6.5}}}
\newcommand{\logpriormed}[1]{\IfEqCase{#1}{{GW230529ay_combined_imrphm_high_spin}{44.5}{GW230529ay_combined_imrphm_low_spin}{47.5}{GW230529ay_imrphen_nsbh}{46.5}{GW230529ay_imrseobv4_nsbh}{46.0}{GW230529ay_imrpv2_low_spin}{32.1}}}
\newcommand{\logpriorplus}[1]{\IfEqCase{#1}{{GW230529ay_combined_imrphm_high_spin}{4.6}{GW230529ay_combined_imrphm_low_spin}{4.5}{GW230529ay_imrphen_nsbh}{4.8}{GW230529ay_imrseobv4_nsbh}{4.9}{GW230529ay_imrpv2_low_spin}{4.6}}}
\newcommand{\logprioronepercent}[1]{\IfEqCase{#1}{{GW230529ay_combined_imrphm_high_spin}{34.7}{GW230529ay_combined_imrphm_low_spin}{37.7}{GW230529ay_imrphen_nsbh}{37.2}{GW230529ay_imrseobv4_nsbh}{36.1}{GW230529ay_imrpv2_low_spin}{22.4}}}
\newcommand{\logpriorninetyninepercent}[1]{\IfEqCase{#1}{{GW230529ay_combined_imrphm_high_spin}{50.6}{GW230529ay_combined_imrphm_low_spin}{53.4}{GW230529ay_imrphen_nsbh}{52.9}{GW230529ay_imrseobv4_nsbh}{52.5}{GW230529ay_imrpv2_low_spin}{38.1}}}
\newcommand{\logpriorfivepercent}[1]{\IfEqCase{#1}{{GW230529ay_combined_imrphm_high_spin}{38.2}{GW230529ay_combined_imrphm_low_spin}{41.1}{GW230529ay_imrphen_nsbh}{40.0}{GW230529ay_imrseobv4_nsbh}{39.5}{GW230529ay_imrpv2_low_spin}{25.6}}}
\newcommand{\logpriorninetyfivepercent}[1]{\IfEqCase{#1}{{GW230529ay_combined_imrphm_high_spin}{49.2}{GW230529ay_combined_imrphm_low_spin}{52.0}{GW230529ay_imrphen_nsbh}{51.3}{GW230529ay_imrseobv4_nsbh}{50.9}{GW230529ay_imrpv2_low_spin}{36.7}}}
\newcommand{\logpriorninetypercent}[1]{\IfEqCase{#1}{{GW230529ay_combined_imrphm_high_spin}{48.3}{GW230529ay_combined_imrphm_low_spin}{51.1}{GW230529ay_imrphen_nsbh}{50.4}{GW230529ay_imrseobv4_nsbh}{49.9}{GW230529ay_imrpv2_low_spin}{35.8}}}
\newcommand{\luminositydistanceminus}[1]{\IfEqCase{#1}{{GW230529ay_combined_imrphm_high_spin}{96}{GW230529ay_combined_imrphm_low_spin}{96}{GW230529ay_imrphen_nsbh}{101}{GW230529ay_imrseobv4_nsbh}{101}{GW230529ay_imrpv2_low_spin}{98}}}
\newcommand{\luminositydistancemed}[1]{\IfEqCase{#1}{{GW230529ay_combined_imrphm_high_spin}{201}{GW230529ay_combined_imrphm_low_spin}{197}{GW230529ay_imrphen_nsbh}{202}{GW230529ay_imrseobv4_nsbh}{194}{GW230529ay_imrpv2_low_spin}{196}}}
\newcommand{\luminositydistanceplus}[1]{\IfEqCase{#1}{{GW230529ay_combined_imrphm_high_spin}{102}{GW230529ay_combined_imrphm_low_spin}{108}{GW230529ay_imrphen_nsbh}{110}{GW230529ay_imrseobv4_nsbh}{109}{GW230529ay_imrpv2_low_spin}{111}}}
\newcommand{\luminositydistanceonepercent}[1]{\IfEqCase{#1}{{GW230529ay_combined_imrphm_high_spin}{73}{GW230529ay_combined_imrphm_low_spin}{73}{GW230529ay_imrphen_nsbh}{69}{GW230529ay_imrseobv4_nsbh}{66}{GW230529ay_imrpv2_low_spin}{68}}}
\newcommand{\luminositydistanceninetyninepercent}[1]{\IfEqCase{#1}{{GW230529ay_combined_imrphm_high_spin}{343}{GW230529ay_combined_imrphm_low_spin}{345}{GW230529ay_imrphen_nsbh}{349}{GW230529ay_imrseobv4_nsbh}{341}{GW230529ay_imrpv2_low_spin}{346}}}
\newcommand{\luminositydistancefivepercent}[1]{\IfEqCase{#1}{{GW230529ay_combined_imrphm_high_spin}{105}{GW230529ay_combined_imrphm_low_spin}{101}{GW230529ay_imrphen_nsbh}{101}{GW230529ay_imrseobv4_nsbh}{93}{GW230529ay_imrpv2_low_spin}{97}}}
\newcommand{\luminositydistanceninetyfivepercent}[1]{\IfEqCase{#1}{{GW230529ay_combined_imrphm_high_spin}{303}{GW230529ay_combined_imrphm_low_spin}{305}{GW230529ay_imrphen_nsbh}{312}{GW230529ay_imrseobv4_nsbh}{303}{GW230529ay_imrpv2_low_spin}{306}}}
\newcommand{\luminositydistanceninetypercent}[1]{\IfEqCase{#1}{{GW230529ay_combined_imrphm_high_spin}{283}{GW230529ay_combined_imrphm_low_spin}{282}{GW230529ay_imrphen_nsbh}{292}{GW230529ay_imrseobv4_nsbh}{282}{GW230529ay_imrpv2_low_spin}{284}}}
\newcommand{\massoneminus}[1]{\IfEqCase{#1}{{GW230529ay_combined_imrphm_high_spin}{1.2}{GW230529ay_combined_imrphm_low_spin}{1.2}{GW230529ay_imrphen_nsbh}{1.1}{GW230529ay_imrseobv4_nsbh}{1.2}{GW230529ay_imrpv2_low_spin}{1.1}}}
\newcommand{\massonemed}[1]{\IfEqCase{#1}{{GW230529ay_combined_imrphm_high_spin}{3.8}{GW230529ay_combined_imrphm_low_spin}{3.8}{GW230529ay_imrphen_nsbh}{4.0}{GW230529ay_imrseobv4_nsbh}{3.8}{GW230529ay_imrpv2_low_spin}{3.5}}}
\newcommand{\massoneplus}[1]{\IfEqCase{#1}{{GW230529ay_combined_imrphm_high_spin}{0.9}{GW230529ay_combined_imrphm_low_spin}{0.7}{GW230529ay_imrphen_nsbh}{1.0}{GW230529ay_imrseobv4_nsbh}{1.2}{GW230529ay_imrpv2_low_spin}{1.0}}}
\newcommand{\massoneonepercent}[1]{\IfEqCase{#1}{{GW230529ay_combined_imrphm_high_spin}{2.4}{GW230529ay_combined_imrphm_low_spin}{2.4}{GW230529ay_imrphen_nsbh}{2.7}{GW230529ay_imrseobv4_nsbh}{2.4}{GW230529ay_imrpv2_low_spin}{2.4}}}
\newcommand{\massoneninetyninepercent}[1]{\IfEqCase{#1}{{GW230529ay_combined_imrphm_high_spin}{5.2}{GW230529ay_combined_imrphm_low_spin}{5.0}{GW230529ay_imrphen_nsbh}{5.7}{GW230529ay_imrseobv4_nsbh}{5.5}{GW230529ay_imrpv2_low_spin}{4.8}}}
\newcommand{\massonefivepercent}[1]{\IfEqCase{#1}{{GW230529ay_combined_imrphm_high_spin}{2.6}{GW230529ay_combined_imrphm_low_spin}{2.6}{GW230529ay_imrphen_nsbh}{2.9}{GW230529ay_imrseobv4_nsbh}{2.6}{GW230529ay_imrpv2_low_spin}{2.4}}}
\newcommand{\massoneninetyfivepercent}[1]{\IfEqCase{#1}{{GW230529ay_combined_imrphm_high_spin}{4.7}{GW230529ay_combined_imrphm_low_spin}{4.5}{GW230529ay_imrphen_nsbh}{5.0}{GW230529ay_imrseobv4_nsbh}{5.0}{GW230529ay_imrpv2_low_spin}{4.5}}}
\newcommand{\massoneninetypercent}[1]{\IfEqCase{#1}{{GW230529ay_combined_imrphm_high_spin}{4.5}{GW230529ay_combined_imrphm_low_spin}{4.3}{GW230529ay_imrphen_nsbh}{4.8}{GW230529ay_imrseobv4_nsbh}{4.7}{GW230529ay_imrpv2_low_spin}{4.4}}}
\newcommand{\massonesourceminus}[1]{\IfEqCase{#1}{{GW230529ay_combined_imrphm_high_spin}{1.2}{GW230529ay_combined_imrphm_low_spin}{1.2}{GW230529ay_imrphen_nsbh}{1.0}{GW230529ay_imrseobv4_nsbh}{1.1}{GW230529ay_imrpv2_low_spin}{1.0}}}
\newcommand{\massonesourcemed}[1]{\IfEqCase{#1}{{GW230529ay_combined_imrphm_high_spin}{3.6}{GW230529ay_combined_imrphm_low_spin}{3.6}{GW230529ay_imrphen_nsbh}{3.8}{GW230529ay_imrseobv4_nsbh}{3.7}{GW230529ay_imrpv2_low_spin}{3.4}}}
\newcommand{\massonesourceplus}[1]{\IfEqCase{#1}{{GW230529ay_combined_imrphm_high_spin}{0.8}{GW230529ay_combined_imrphm_low_spin}{0.7}{GW230529ay_imrphen_nsbh}{1.0}{GW230529ay_imrseobv4_nsbh}{1.1}{GW230529ay_imrpv2_low_spin}{1.0}}}
\newcommand{\massonesourceonepercent}[1]{\IfEqCase{#1}{{GW230529ay_combined_imrphm_high_spin}{2.3}{GW230529ay_combined_imrphm_low_spin}{2.3}{GW230529ay_imrphen_nsbh}{2.6}{GW230529ay_imrseobv4_nsbh}{2.4}{GW230529ay_imrpv2_low_spin}{2.2}}}
\newcommand{\massonesourceninetyninepercent}[1]{\IfEqCase{#1}{{GW230529ay_combined_imrphm_high_spin}{5.0}{GW230529ay_combined_imrphm_low_spin}{4.8}{GW230529ay_imrphen_nsbh}{5.4}{GW230529ay_imrseobv4_nsbh}{5.3}{GW230529ay_imrpv2_low_spin}{4.6}}}
\newcommand{\massonesourcefivepercent}[1]{\IfEqCase{#1}{{GW230529ay_combined_imrphm_high_spin}{2.5}{GW230529ay_combined_imrphm_low_spin}{2.5}{GW230529ay_imrphen_nsbh}{2.8}{GW230529ay_imrseobv4_nsbh}{2.5}{GW230529ay_imrpv2_low_spin}{2.3}}}
\newcommand{\massonesourceninetyfivepercent}[1]{\IfEqCase{#1}{{GW230529ay_combined_imrphm_high_spin}{4.5}{GW230529ay_combined_imrphm_low_spin}{4.3}{GW230529ay_imrphen_nsbh}{4.8}{GW230529ay_imrseobv4_nsbh}{4.8}{GW230529ay_imrpv2_low_spin}{4.3}}}
\newcommand{\massonesourceninetypercent}[1]{\IfEqCase{#1}{{GW230529ay_combined_imrphm_high_spin}{4.3}{GW230529ay_combined_imrphm_low_spin}{4.2}{GW230529ay_imrphen_nsbh}{4.6}{GW230529ay_imrseobv4_nsbh}{4.5}{GW230529ay_imrpv2_low_spin}{4.2}}}
\newcommand{\masstwominus}[1]{\IfEqCase{#1}{{GW230529ay_combined_imrphm_high_spin}{0.2}{GW230529ay_combined_imrphm_low_spin}{0.2}{GW230529ay_imrphen_nsbh}{0.2}{GW230529ay_imrseobv4_nsbh}{0.3}{GW230529ay_imrpv2_low_spin}{0.3}}}
\newcommand{\masstwomed}[1]{\IfEqCase{#1}{{GW230529ay_combined_imrphm_high_spin}{1.5}{GW230529ay_combined_imrphm_low_spin}{1.5}{GW230529ay_imrphen_nsbh}{1.4}{GW230529ay_imrseobv4_nsbh}{1.5}{GW230529ay_imrpv2_low_spin}{1.6}}}
\newcommand{\masstwoplus}[1]{\IfEqCase{#1}{{GW230529ay_combined_imrphm_high_spin}{0.6}{GW230529ay_combined_imrphm_low_spin}{0.6}{GW230529ay_imrphen_nsbh}{0.4}{GW230529ay_imrseobv4_nsbh}{0.6}{GW230529ay_imrpv2_low_spin}{0.6}}}
\newcommand{\masstwoonepercent}[1]{\IfEqCase{#1}{{GW230529ay_combined_imrphm_high_spin}{1.2}{GW230529ay_combined_imrphm_low_spin}{1.2}{GW230529ay_imrphen_nsbh}{1.1}{GW230529ay_imrseobv4_nsbh}{1.1}{GW230529ay_imrpv2_low_spin}{1.2}}}
\newcommand{\masstwoninetyninepercent}[1]{\IfEqCase{#1}{{GW230529ay_combined_imrphm_high_spin}{2.3}{GW230529ay_combined_imrphm_low_spin}{2.3}{GW230529ay_imrphen_nsbh}{2.0}{GW230529ay_imrseobv4_nsbh}{2.2}{GW230529ay_imrpv2_low_spin}{2.3}}}
\newcommand{\masstwofivepercent}[1]{\IfEqCase{#1}{{GW230529ay_combined_imrphm_high_spin}{1.3}{GW230529ay_combined_imrphm_low_spin}{1.3}{GW230529ay_imrphen_nsbh}{1.2}{GW230529ay_imrseobv4_nsbh}{1.2}{GW230529ay_imrpv2_low_spin}{1.3}}}
\newcommand{\masstwoninetyfivepercent}[1]{\IfEqCase{#1}{{GW230529ay_combined_imrphm_high_spin}{2.1}{GW230529ay_combined_imrphm_low_spin}{2.1}{GW230529ay_imrphen_nsbh}{1.9}{GW230529ay_imrseobv4_nsbh}{2.1}{GW230529ay_imrpv2_low_spin}{2.2}}}
\newcommand{\masstwoninetypercent}[1]{\IfEqCase{#1}{{GW230529ay_combined_imrphm_high_spin}{1.9}{GW230529ay_combined_imrphm_low_spin}{2.0}{GW230529ay_imrphen_nsbh}{1.8}{GW230529ay_imrseobv4_nsbh}{1.9}{GW230529ay_imrpv2_low_spin}{2.1}}}
\newcommand{\masstwosourceminus}[1]{\IfEqCase{#1}{{GW230529ay_combined_imrphm_high_spin}{0.2}{GW230529ay_combined_imrphm_low_spin}{0.2}{GW230529ay_imrphen_nsbh}{0.2}{GW230529ay_imrseobv4_nsbh}{0.3}{GW230529ay_imrpv2_low_spin}{0.3}}}
\newcommand{\masstwosourcemed}[1]{\IfEqCase{#1}{{GW230529ay_combined_imrphm_high_spin}{1.4}{GW230529ay_combined_imrphm_low_spin}{1.4}{GW230529ay_imrphen_nsbh}{1.4}{GW230529ay_imrseobv4_nsbh}{1.4}{GW230529ay_imrpv2_low_spin}{1.5}}}
\newcommand{\masstwosourceplus}[1]{\IfEqCase{#1}{{GW230529ay_combined_imrphm_high_spin}{0.6}{GW230529ay_combined_imrphm_low_spin}{0.6}{GW230529ay_imrphen_nsbh}{0.4}{GW230529ay_imrseobv4_nsbh}{0.5}{GW230529ay_imrpv2_low_spin}{0.6}}}
\newcommand{\masstwosourceonepercent}[1]{\IfEqCase{#1}{{GW230529ay_combined_imrphm_high_spin}{1.1}{GW230529ay_combined_imrphm_low_spin}{1.1}{GW230529ay_imrphen_nsbh}{1.0}{GW230529ay_imrseobv4_nsbh}{1.1}{GW230529ay_imrpv2_low_spin}{1.2}}}
\newcommand{\masstwosourceninetyninepercent}[1]{\IfEqCase{#1}{{GW230529ay_combined_imrphm_high_spin}{2.2}{GW230529ay_combined_imrphm_low_spin}{2.2}{GW230529ay_imrphen_nsbh}{1.9}{GW230529ay_imrseobv4_nsbh}{2.1}{GW230529ay_imrpv2_low_spin}{2.2}}}
\newcommand{\masstwosourcefivepercent}[1]{\IfEqCase{#1}{{GW230529ay_combined_imrphm_high_spin}{1.2}{GW230529ay_combined_imrphm_low_spin}{1.2}{GW230529ay_imrphen_nsbh}{1.1}{GW230529ay_imrseobv4_nsbh}{1.1}{GW230529ay_imrpv2_low_spin}{1.2}}}
\newcommand{\masstwosourceninetyfivepercent}[1]{\IfEqCase{#1}{{GW230529ay_combined_imrphm_high_spin}{2.0}{GW230529ay_combined_imrphm_low_spin}{2.0}{GW230529ay_imrphen_nsbh}{1.8}{GW230529ay_imrseobv4_nsbh}{2.0}{GW230529ay_imrpv2_low_spin}{2.1}}}
\newcommand{\masstwosourceninetypercent}[1]{\IfEqCase{#1}{{GW230529ay_combined_imrphm_high_spin}{1.9}{GW230529ay_combined_imrphm_low_spin}{1.9}{GW230529ay_imrphen_nsbh}{1.7}{GW230529ay_imrseobv4_nsbh}{1.9}{GW230529ay_imrpv2_low_spin}{2.1}}}
\newcommand{\massratiominus}[1]{\IfEqCase{#1}{{GW230529ay_combined_imrphm_high_spin}{0.12}{GW230529ay_combined_imrphm_low_spin}{0.10}{GW230529ay_imrphen_nsbh}{0.12}{GW230529ay_imrseobv4_nsbh}{0.15}{GW230529ay_imrpv2_low_spin}{0.17}}}
\newcommand{\massratiomed}[1]{\IfEqCase{#1}{{GW230529ay_combined_imrphm_high_spin}{0.39}{GW230529ay_combined_imrphm_low_spin}{0.39}{GW230529ay_imrphen_nsbh}{0.36}{GW230529ay_imrseobv4_nsbh}{0.39}{GW230529ay_imrpv2_low_spin}{0.45}}}
\newcommand{\massratioplus}[1]{\IfEqCase{#1}{{GW230529ay_combined_imrphm_high_spin}{0.41}{GW230529ay_combined_imrphm_low_spin}{0.43}{GW230529ay_imrphen_nsbh}{0.28}{GW230529ay_imrseobv4_nsbh}{0.38}{GW230529ay_imrpv2_low_spin}{0.46}}}
\newcommand{\massratioonepercent}[1]{\IfEqCase{#1}{{GW230529ay_combined_imrphm_high_spin}{0.22}{GW230529ay_combined_imrphm_low_spin}{0.24}{GW230529ay_imrphen_nsbh}{0.19}{GW230529ay_imrseobv4_nsbh}{0.20}{GW230529ay_imrpv2_low_spin}{0.25}}}
\newcommand{\massrationinetyninepercent}[1]{\IfEqCase{#1}{{GW230529ay_combined_imrphm_high_spin}{0.95}{GW230529ay_combined_imrphm_low_spin}{0.95}{GW230529ay_imrphen_nsbh}{0.74}{GW230529ay_imrseobv4_nsbh}{0.91}{GW230529ay_imrpv2_low_spin}{0.98}}}
\newcommand{\massratiofivepercent}[1]{\IfEqCase{#1}{{GW230529ay_combined_imrphm_high_spin}{0.27}{GW230529ay_combined_imrphm_low_spin}{0.29}{GW230529ay_imrphen_nsbh}{0.24}{GW230529ay_imrseobv4_nsbh}{0.24}{GW230529ay_imrpv2_low_spin}{0.29}}}
\newcommand{\massrationinetyfivepercent}[1]{\IfEqCase{#1}{{GW230529ay_combined_imrphm_high_spin}{0.80}{GW230529ay_combined_imrphm_low_spin}{0.82}{GW230529ay_imrphen_nsbh}{0.64}{GW230529ay_imrseobv4_nsbh}{0.77}{GW230529ay_imrpv2_low_spin}{0.92}}}
\newcommand{\massrationinetypercent}[1]{\IfEqCase{#1}{{GW230529ay_combined_imrphm_high_spin}{0.68}{GW230529ay_combined_imrphm_low_spin}{0.71}{GW230529ay_imrphen_nsbh}{0.57}{GW230529ay_imrseobv4_nsbh}{0.69}{GW230529ay_imrpv2_low_spin}{0.85}}}
\newcommand{\networkmatchedfiltersnrminus}[1]{\IfEqCase{#1}{{GW230529ay_combined_imrphm_high_spin}{0.4}{GW230529ay_combined_imrphm_low_spin}{0.4}{GW230529ay_imrphen_nsbh}{0.4}{GW230529ay_imrseobv4_nsbh}{0.4}{GW230529ay_imrpv2_low_spin}{0.5}}}
\newcommand{\networkmatchedfiltersnrmed}[1]{\IfEqCase{#1}{{GW230529ay_combined_imrphm_high_spin}{11.6}{GW230529ay_combined_imrphm_low_spin}{11.6}{GW230529ay_imrphen_nsbh}{11.4}{GW230529ay_imrseobv4_nsbh}{11.5}{GW230529ay_imrpv2_low_spin}{11.5}}}
\newcommand{\networkmatchedfiltersnrplus}[1]{\IfEqCase{#1}{{GW230529ay_combined_imrphm_high_spin}{0.3}{GW230529ay_combined_imrphm_low_spin}{0.3}{GW230529ay_imrphen_nsbh}{0.2}{GW230529ay_imrseobv4_nsbh}{0.2}{GW230529ay_imrpv2_low_spin}{0.3}}}
\newcommand{\networkmatchedfiltersnronepercent}[1]{\IfEqCase{#1}{{GW230529ay_combined_imrphm_high_spin}{11.0}{GW230529ay_combined_imrphm_low_spin}{11.0}{GW230529ay_imrphen_nsbh}{10.9}{GW230529ay_imrseobv4_nsbh}{10.9}{GW230529ay_imrpv2_low_spin}{10.8}}}
\newcommand{\networkmatchedfiltersnrninetyninepercent}[1]{\IfEqCase{#1}{{GW230529ay_combined_imrphm_high_spin}{12.0}{GW230529ay_combined_imrphm_low_spin}{12.0}{GW230529ay_imrphen_nsbh}{11.7}{GW230529ay_imrseobv4_nsbh}{11.7}{GW230529ay_imrpv2_low_spin}{11.9}}}
\newcommand{\networkmatchedfiltersnrfivepercent}[1]{\IfEqCase{#1}{{GW230529ay_combined_imrphm_high_spin}{11.2}{GW230529ay_combined_imrphm_low_spin}{11.2}{GW230529ay_imrphen_nsbh}{11.1}{GW230529ay_imrseobv4_nsbh}{11.1}{GW230529ay_imrpv2_low_spin}{11.0}}}
\newcommand{\networkmatchedfiltersnrninetyfivepercent}[1]{\IfEqCase{#1}{{GW230529ay_combined_imrphm_high_spin}{11.9}{GW230529ay_combined_imrphm_low_spin}{11.9}{GW230529ay_imrphen_nsbh}{11.6}{GW230529ay_imrseobv4_nsbh}{11.6}{GW230529ay_imrpv2_low_spin}{11.8}}}
\newcommand{\networkmatchedfiltersnrninetypercent}[1]{\IfEqCase{#1}{{GW230529ay_combined_imrphm_high_spin}{11.9}{GW230529ay_combined_imrphm_low_spin}{11.9}{GW230529ay_imrphen_nsbh}{11.6}{GW230529ay_imrseobv4_nsbh}{11.6}{GW230529ay_imrpv2_low_spin}{11.7}}}
\newcommand{\networkoptimalsnrminus}[1]{\IfEqCase{#1}{{GW230529ay_combined_imrphm_high_spin}{1.7}{GW230529ay_combined_imrphm_low_spin}{1.7}{GW230529ay_imrphen_nsbh}{1.7}{GW230529ay_imrseobv4_nsbh}{1.7}{GW230529ay_imrpv2_low_spin}{1.7}}}
\newcommand{\networkoptimalsnrmed}[1]{\IfEqCase{#1}{{GW230529ay_combined_imrphm_high_spin}{11.3}{GW230529ay_combined_imrphm_low_spin}{11.3}{GW230529ay_imrphen_nsbh}{11.1}{GW230529ay_imrseobv4_nsbh}{11.1}{GW230529ay_imrpv2_low_spin}{11.1}}}
\newcommand{\networkoptimalsnrplus}[1]{\IfEqCase{#1}{{GW230529ay_combined_imrphm_high_spin}{1.7}{GW230529ay_combined_imrphm_low_spin}{1.7}{GW230529ay_imrphen_nsbh}{1.6}{GW230529ay_imrseobv4_nsbh}{1.7}{GW230529ay_imrpv2_low_spin}{1.7}}}
\newcommand{\networkoptimalsnronepercent}[1]{\IfEqCase{#1}{{GW230529ay_combined_imrphm_high_spin}{8.8}{GW230529ay_combined_imrphm_low_spin}{8.9}{GW230529ay_imrphen_nsbh}{8.7}{GW230529ay_imrseobv4_nsbh}{8.7}{GW230529ay_imrpv2_low_spin}{8.7}}}
\newcommand{\networkoptimalsnrninetyninepercent}[1]{\IfEqCase{#1}{{GW230529ay_combined_imrphm_high_spin}{13.7}{GW230529ay_combined_imrphm_low_spin}{13.7}{GW230529ay_imrphen_nsbh}{13.4}{GW230529ay_imrseobv4_nsbh}{13.5}{GW230529ay_imrpv2_low_spin}{13.5}}}
\newcommand{\networkoptimalsnrfivepercent}[1]{\IfEqCase{#1}{{GW230529ay_combined_imrphm_high_spin}{9.5}{GW230529ay_combined_imrphm_low_spin}{9.6}{GW230529ay_imrphen_nsbh}{9.4}{GW230529ay_imrseobv4_nsbh}{9.4}{GW230529ay_imrpv2_low_spin}{9.4}}}
\newcommand{\networkoptimalsnrninetyfivepercent}[1]{\IfEqCase{#1}{{GW230529ay_combined_imrphm_high_spin}{13.0}{GW230529ay_combined_imrphm_low_spin}{13.0}{GW230529ay_imrphen_nsbh}{12.7}{GW230529ay_imrseobv4_nsbh}{12.8}{GW230529ay_imrpv2_low_spin}{12.8}}}
\newcommand{\networkoptimalsnrninetypercent}[1]{\IfEqCase{#1}{{GW230529ay_combined_imrphm_high_spin}{12.6}{GW230529ay_combined_imrphm_low_spin}{12.6}{GW230529ay_imrphen_nsbh}{12.4}{GW230529ay_imrseobv4_nsbh}{12.4}{GW230529ay_imrpv2_low_spin}{12.5}}}
\newcommand{\phaseminus}[1]{\IfEqCase{#1}{{GW230529ay_combined_imrphm_high_spin}{2.32}{GW230529ay_combined_imrphm_low_spin}{2.37}{GW230529ay_imrphen_nsbh}{2.88}{GW230529ay_imrseobv4_nsbh}{2.83}{GW230529ay_imrpv2_low_spin}{2.78}}}
\newcommand{\phasemed}[1]{\IfEqCase{#1}{{GW230529ay_combined_imrphm_high_spin}{2.69}{GW230529ay_combined_imrphm_low_spin}{2.72}{GW230529ay_imrphen_nsbh}{3.16}{GW230529ay_imrseobv4_nsbh}{3.12}{GW230529ay_imrpv2_low_spin}{3.10}}}
\newcommand{\phaseplus}[1]{\IfEqCase{#1}{{GW230529ay_combined_imrphm_high_spin}{3.19}{GW230529ay_combined_imrphm_low_spin}{3.17}{GW230529ay_imrphen_nsbh}{2.83}{GW230529ay_imrseobv4_nsbh}{2.84}{GW230529ay_imrpv2_low_spin}{2.86}}}
\newcommand{\phaseonepercent}[1]{\IfEqCase{#1}{{GW230529ay_combined_imrphm_high_spin}{0.08}{GW230529ay_combined_imrphm_low_spin}{0.07}{GW230529ay_imrphen_nsbh}{0.05}{GW230529ay_imrseobv4_nsbh}{0.07}{GW230529ay_imrpv2_low_spin}{0.06}}}
\newcommand{\phaseninetyninepercent}[1]{\IfEqCase{#1}{{GW230529ay_combined_imrphm_high_spin}{6.20}{GW230529ay_combined_imrphm_low_spin}{6.21}{GW230529ay_imrphen_nsbh}{6.22}{GW230529ay_imrseobv4_nsbh}{6.21}{GW230529ay_imrpv2_low_spin}{6.22}}}
\newcommand{\phasefivepercent}[1]{\IfEqCase{#1}{{GW230529ay_combined_imrphm_high_spin}{0.37}{GW230529ay_combined_imrphm_low_spin}{0.35}{GW230529ay_imrphen_nsbh}{0.28}{GW230529ay_imrseobv4_nsbh}{0.29}{GW230529ay_imrpv2_low_spin}{0.32}}}
\newcommand{\phaseninetyfivepercent}[1]{\IfEqCase{#1}{{GW230529ay_combined_imrphm_high_spin}{5.88}{GW230529ay_combined_imrphm_low_spin}{5.90}{GW230529ay_imrphen_nsbh}{5.99}{GW230529ay_imrseobv4_nsbh}{5.96}{GW230529ay_imrpv2_low_spin}{5.97}}}
\newcommand{\phaseninetypercent}[1]{\IfEqCase{#1}{{GW230529ay_combined_imrphm_high_spin}{5.48}{GW230529ay_combined_imrphm_low_spin}{5.51}{GW230529ay_imrphen_nsbh}{5.64}{GW230529ay_imrseobv4_nsbh}{5.65}{GW230529ay_imrpv2_low_spin}{5.63}}}
\newcommand{\phioneminus}[1]{\IfEqCase{#1}{{GW230529ay_combined_imrphm_high_spin}{2.81}{GW230529ay_combined_imrphm_low_spin}{2.86}{GW230529ay_imrphen_nsbh}{0.00}{GW230529ay_imrseobv4_nsbh}{0.00}{GW230529ay_imrpv2_low_spin}{2.80}}}
\newcommand{\phionemed}[1]{\IfEqCase{#1}{{GW230529ay_combined_imrphm_high_spin}{3.11}{GW230529ay_combined_imrphm_low_spin}{3.18}{GW230529ay_imrphen_nsbh}{0.00}{GW230529ay_imrseobv4_nsbh}{0.00}{GW230529ay_imrpv2_low_spin}{3.10}}}
\newcommand{\phioneplus}[1]{\IfEqCase{#1}{{GW230529ay_combined_imrphm_high_spin}{2.87}{GW230529ay_combined_imrphm_low_spin}{2.80}{GW230529ay_imrphen_nsbh}{0.00}{GW230529ay_imrseobv4_nsbh}{0.00}{GW230529ay_imrpv2_low_spin}{2.87}}}
\newcommand{\phioneonepercent}[1]{\IfEqCase{#1}{{GW230529ay_combined_imrphm_high_spin}{0.06}{GW230529ay_combined_imrphm_low_spin}{0.07}{GW230529ay_imrphen_nsbh}{0.00}{GW230529ay_imrseobv4_nsbh}{0.00}{GW230529ay_imrpv2_low_spin}{0.06}}}
\newcommand{\phioneninetyninepercent}[1]{\IfEqCase{#1}{{GW230529ay_combined_imrphm_high_spin}{6.23}{GW230529ay_combined_imrphm_low_spin}{6.22}{GW230529ay_imrphen_nsbh}{0.00}{GW230529ay_imrseobv4_nsbh}{0.00}{GW230529ay_imrpv2_low_spin}{6.23}}}
\newcommand{\phionefivepercent}[1]{\IfEqCase{#1}{{GW230529ay_combined_imrphm_high_spin}{0.30}{GW230529ay_combined_imrphm_low_spin}{0.31}{GW230529ay_imrphen_nsbh}{0.00}{GW230529ay_imrseobv4_nsbh}{0.00}{GW230529ay_imrpv2_low_spin}{0.31}}}
\newcommand{\phioneninetyfivepercent}[1]{\IfEqCase{#1}{{GW230529ay_combined_imrphm_high_spin}{5.99}{GW230529ay_combined_imrphm_low_spin}{5.97}{GW230529ay_imrphen_nsbh}{0.00}{GW230529ay_imrseobv4_nsbh}{0.00}{GW230529ay_imrpv2_low_spin}{5.97}}}
\newcommand{\phioneninetypercent}[1]{\IfEqCase{#1}{{GW230529ay_combined_imrphm_high_spin}{5.67}{GW230529ay_combined_imrphm_low_spin}{5.66}{GW230529ay_imrphen_nsbh}{0.00}{GW230529ay_imrseobv4_nsbh}{0.00}{GW230529ay_imrpv2_low_spin}{5.68}}}
\newcommand{\phionetwominus}[1]{\IfEqCase{#1}{{GW230529ay_combined_imrphm_high_spin}{2.83}{GW230529ay_combined_imrphm_low_spin}{2.87}{GW230529ay_imrphen_nsbh}{0.00}{GW230529ay_imrseobv4_nsbh}{0.00}{GW230529ay_imrpv2_low_spin}{2.86}}}
\newcommand{\phionetwomed}[1]{\IfEqCase{#1}{{GW230529ay_combined_imrphm_high_spin}{3.16}{GW230529ay_combined_imrphm_low_spin}{3.17}{GW230529ay_imrphen_nsbh}{0.00}{GW230529ay_imrseobv4_nsbh}{0.00}{GW230529ay_imrpv2_low_spin}{3.17}}}
\newcommand{\phionetwoplus}[1]{\IfEqCase{#1}{{GW230529ay_combined_imrphm_high_spin}{2.81}{GW230529ay_combined_imrphm_low_spin}{2.80}{GW230529ay_imrphen_nsbh}{0.00}{GW230529ay_imrseobv4_nsbh}{0.00}{GW230529ay_imrpv2_low_spin}{2.79}}}
\newcommand{\phionetwoonepercent}[1]{\IfEqCase{#1}{{GW230529ay_combined_imrphm_high_spin}{0.06}{GW230529ay_combined_imrphm_low_spin}{0.06}{GW230529ay_imrphen_nsbh}{0.00}{GW230529ay_imrseobv4_nsbh}{0.00}{GW230529ay_imrpv2_low_spin}{0.07}}}
\newcommand{\phionetwoninetyninepercent}[1]{\IfEqCase{#1}{{GW230529ay_combined_imrphm_high_spin}{6.22}{GW230529ay_combined_imrphm_low_spin}{6.23}{GW230529ay_imrphen_nsbh}{0.00}{GW230529ay_imrseobv4_nsbh}{0.00}{GW230529ay_imrpv2_low_spin}{6.22}}}
\newcommand{\phionetwofivepercent}[1]{\IfEqCase{#1}{{GW230529ay_combined_imrphm_high_spin}{0.33}{GW230529ay_combined_imrphm_low_spin}{0.30}{GW230529ay_imrphen_nsbh}{0.00}{GW230529ay_imrseobv4_nsbh}{0.00}{GW230529ay_imrpv2_low_spin}{0.30}}}
\newcommand{\phionetwoninetyfivepercent}[1]{\IfEqCase{#1}{{GW230529ay_combined_imrphm_high_spin}{5.97}{GW230529ay_combined_imrphm_low_spin}{5.96}{GW230529ay_imrphen_nsbh}{0.00}{GW230529ay_imrseobv4_nsbh}{0.00}{GW230529ay_imrpv2_low_spin}{5.96}}}
\newcommand{\phionetwoninetypercent}[1]{\IfEqCase{#1}{{GW230529ay_combined_imrphm_high_spin}{5.66}{GW230529ay_combined_imrphm_low_spin}{5.66}{GW230529ay_imrphen_nsbh}{0.00}{GW230529ay_imrseobv4_nsbh}{0.00}{GW230529ay_imrpv2_low_spin}{5.63}}}
\newcommand{\phitwominus}[1]{\IfEqCase{#1}{{GW230529ay_combined_imrphm_high_spin}{2.81}{GW230529ay_combined_imrphm_low_spin}{2.84}{GW230529ay_imrphen_nsbh}{0.00}{GW230529ay_imrseobv4_nsbh}{0.00}{GW230529ay_imrpv2_low_spin}{2.80}}}
\newcommand{\phitwomed}[1]{\IfEqCase{#1}{{GW230529ay_combined_imrphm_high_spin}{3.11}{GW230529ay_combined_imrphm_low_spin}{3.16}{GW230529ay_imrphen_nsbh}{0.00}{GW230529ay_imrseobv4_nsbh}{0.00}{GW230529ay_imrpv2_low_spin}{3.14}}}
\newcommand{\phitwoplus}[1]{\IfEqCase{#1}{{GW230529ay_combined_imrphm_high_spin}{2.86}{GW230529ay_combined_imrphm_low_spin}{2.82}{GW230529ay_imrphen_nsbh}{0.00}{GW230529ay_imrseobv4_nsbh}{0.00}{GW230529ay_imrpv2_low_spin}{2.83}}}
\newcommand{\phitwoonepercent}[1]{\IfEqCase{#1}{{GW230529ay_combined_imrphm_high_spin}{0.07}{GW230529ay_combined_imrphm_low_spin}{0.06}{GW230529ay_imrphen_nsbh}{0.00}{GW230529ay_imrseobv4_nsbh}{0.00}{GW230529ay_imrpv2_low_spin}{0.07}}}
\newcommand{\phitwoninetyninepercent}[1]{\IfEqCase{#1}{{GW230529ay_combined_imrphm_high_spin}{6.22}{GW230529ay_combined_imrphm_low_spin}{6.22}{GW230529ay_imrphen_nsbh}{0.00}{GW230529ay_imrseobv4_nsbh}{0.00}{GW230529ay_imrpv2_low_spin}{6.22}}}
\newcommand{\phitwofivepercent}[1]{\IfEqCase{#1}{{GW230529ay_combined_imrphm_high_spin}{0.30}{GW230529ay_combined_imrphm_low_spin}{0.32}{GW230529ay_imrphen_nsbh}{0.00}{GW230529ay_imrseobv4_nsbh}{0.00}{GW230529ay_imrpv2_low_spin}{0.34}}}
\newcommand{\phitwoninetyfivepercent}[1]{\IfEqCase{#1}{{GW230529ay_combined_imrphm_high_spin}{5.97}{GW230529ay_combined_imrphm_low_spin}{5.98}{GW230529ay_imrphen_nsbh}{0.00}{GW230529ay_imrseobv4_nsbh}{0.00}{GW230529ay_imrpv2_low_spin}{5.97}}}
\newcommand{\phitwoninetypercent}[1]{\IfEqCase{#1}{{GW230529ay_combined_imrphm_high_spin}{5.66}{GW230529ay_combined_imrphm_low_spin}{5.66}{GW230529ay_imrphen_nsbh}{0.00}{GW230529ay_imrseobv4_nsbh}{0.00}{GW230529ay_imrpv2_low_spin}{5.64}}}
\newcommand{\phijlminus}[1]{\IfEqCase{#1}{{GW230529ay_combined_imrphm_high_spin}{2.84}{GW230529ay_combined_imrphm_low_spin}{2.88}{GW230529ay_imrphen_nsbh}{0.00}{GW230529ay_imrseobv4_nsbh}{0.00}{GW230529ay_imrpv2_low_spin}{2.84}}}
\newcommand{\phijlmed}[1]{\IfEqCase{#1}{{GW230529ay_combined_imrphm_high_spin}{3.15}{GW230529ay_combined_imrphm_low_spin}{3.18}{GW230529ay_imrphen_nsbh}{0.00}{GW230529ay_imrseobv4_nsbh}{0.00}{GW230529ay_imrpv2_low_spin}{3.18}}}
\newcommand{\phijlplus}[1]{\IfEqCase{#1}{{GW230529ay_combined_imrphm_high_spin}{2.82}{GW230529ay_combined_imrphm_low_spin}{2.79}{GW230529ay_imrphen_nsbh}{0.00}{GW230529ay_imrseobv4_nsbh}{0.00}{GW230529ay_imrpv2_low_spin}{2.77}}}
\newcommand{\phijlonepercent}[1]{\IfEqCase{#1}{{GW230529ay_combined_imrphm_high_spin}{0.06}{GW230529ay_combined_imrphm_low_spin}{0.06}{GW230529ay_imrphen_nsbh}{0.00}{GW230529ay_imrseobv4_nsbh}{0.00}{GW230529ay_imrpv2_low_spin}{0.08}}}
\newcommand{\phijlninetyninepercent}[1]{\IfEqCase{#1}{{GW230529ay_combined_imrphm_high_spin}{6.22}{GW230529ay_combined_imrphm_low_spin}{6.22}{GW230529ay_imrphen_nsbh}{0.00}{GW230529ay_imrseobv4_nsbh}{0.00}{GW230529ay_imrpv2_low_spin}{6.22}}}
\newcommand{\phijlfivepercent}[1]{\IfEqCase{#1}{{GW230529ay_combined_imrphm_high_spin}{0.31}{GW230529ay_combined_imrphm_low_spin}{0.30}{GW230529ay_imrphen_nsbh}{0.00}{GW230529ay_imrseobv4_nsbh}{0.00}{GW230529ay_imrpv2_low_spin}{0.34}}}
\newcommand{\phijlninetyfivepercent}[1]{\IfEqCase{#1}{{GW230529ay_combined_imrphm_high_spin}{5.97}{GW230529ay_combined_imrphm_low_spin}{5.97}{GW230529ay_imrphen_nsbh}{0.00}{GW230529ay_imrseobv4_nsbh}{0.00}{GW230529ay_imrpv2_low_spin}{5.95}}}
\newcommand{\phijlninetypercent}[1]{\IfEqCase{#1}{{GW230529ay_combined_imrphm_high_spin}{5.66}{GW230529ay_combined_imrphm_low_spin}{5.65}{GW230529ay_imrphen_nsbh}{0.00}{GW230529ay_imrseobv4_nsbh}{0.00}{GW230529ay_imrpv2_low_spin}{5.66}}}
\newcommand{\psiminus}[1]{\IfEqCase{#1}{{GW230529ay_combined_imrphm_high_spin}{1.42}{GW230529ay_combined_imrphm_low_spin}{1.43}{GW230529ay_imrphen_nsbh}{1.40}{GW230529ay_imrseobv4_nsbh}{1.41}{GW230529ay_imrpv2_low_spin}{1.41}}}
\newcommand{\psimed}[1]{\IfEqCase{#1}{{GW230529ay_combined_imrphm_high_spin}{1.57}{GW230529ay_combined_imrphm_low_spin}{1.58}{GW230529ay_imrphen_nsbh}{1.55}{GW230529ay_imrseobv4_nsbh}{1.57}{GW230529ay_imrpv2_low_spin}{1.58}}}
\newcommand{\psiplus}[1]{\IfEqCase{#1}{{GW230529ay_combined_imrphm_high_spin}{1.42}{GW230529ay_combined_imrphm_low_spin}{1.42}{GW230529ay_imrphen_nsbh}{1.44}{GW230529ay_imrseobv4_nsbh}{1.41}{GW230529ay_imrpv2_low_spin}{1.40}}}
\newcommand{\psionepercent}[1]{\IfEqCase{#1}{{GW230529ay_combined_imrphm_high_spin}{0.03}{GW230529ay_combined_imrphm_low_spin}{0.03}{GW230529ay_imrphen_nsbh}{0.03}{GW230529ay_imrseobv4_nsbh}{0.03}{GW230529ay_imrpv2_low_spin}{0.04}}}
\newcommand{\psininetyninepercent}[1]{\IfEqCase{#1}{{GW230529ay_combined_imrphm_high_spin}{3.11}{GW230529ay_combined_imrphm_low_spin}{3.11}{GW230529ay_imrphen_nsbh}{3.11}{GW230529ay_imrseobv4_nsbh}{3.12}{GW230529ay_imrpv2_low_spin}{3.11}}}
\newcommand{\psifivepercent}[1]{\IfEqCase{#1}{{GW230529ay_combined_imrphm_high_spin}{0.15}{GW230529ay_combined_imrphm_low_spin}{0.15}{GW230529ay_imrphen_nsbh}{0.15}{GW230529ay_imrseobv4_nsbh}{0.16}{GW230529ay_imrpv2_low_spin}{0.16}}}
\newcommand{\psininetyfivepercent}[1]{\IfEqCase{#1}{{GW230529ay_combined_imrphm_high_spin}{2.99}{GW230529ay_combined_imrphm_low_spin}{2.99}{GW230529ay_imrphen_nsbh}{2.98}{GW230529ay_imrseobv4_nsbh}{2.99}{GW230529ay_imrpv2_low_spin}{2.97}}}
\newcommand{\psininetypercent}[1]{\IfEqCase{#1}{{GW230529ay_combined_imrphm_high_spin}{2.85}{GW230529ay_combined_imrphm_low_spin}{2.84}{GW230529ay_imrphen_nsbh}{2.81}{GW230529ay_imrseobv4_nsbh}{2.84}{GW230529ay_imrpv2_low_spin}{2.82}}}
\newcommand{\raminus}[1]{\IfEqCase{#1}{{GW230529ay_combined_imrphm_high_spin}{2.75293}{GW230529ay_combined_imrphm_low_spin}{2.60395}{GW230529ay_imrphen_nsbh}{2.64119}{GW230529ay_imrseobv4_nsbh}{2.66536}{GW230529ay_imrpv2_low_spin}{2.81228}}}
\newcommand{\ramed}[1]{\IfEqCase{#1}{{GW230529ay_combined_imrphm_high_spin}{3.22060}{GW230529ay_combined_imrphm_low_spin}{3.06261}{GW230529ay_imrphen_nsbh}{3.08973}{GW230529ay_imrseobv4_nsbh}{3.11260}{GW230529ay_imrpv2_low_spin}{3.29576}}}
\newcommand{\raplus}[1]{\IfEqCase{#1}{{GW230529ay_combined_imrphm_high_spin}{2.20896}{GW230529ay_combined_imrphm_low_spin}{2.39008}{GW230529ay_imrphen_nsbh}{2.39369}{GW230529ay_imrseobv4_nsbh}{2.36920}{GW230529ay_imrpv2_low_spin}{2.17581}}}
\newcommand{\raonepercent}[1]{\IfEqCase{#1}{{GW230529ay_combined_imrphm_high_spin}{0.11869}{GW230529ay_combined_imrphm_low_spin}{0.11478}{GW230529ay_imrphen_nsbh}{0.12863}{GW230529ay_imrseobv4_nsbh}{0.13196}{GW230529ay_imrpv2_low_spin}{0.13058}}}
\newcommand{\raninetyninepercent}[1]{\IfEqCase{#1}{{GW230529ay_combined_imrphm_high_spin}{6.12148}{GW230529ay_combined_imrphm_low_spin}{6.13437}{GW230529ay_imrphen_nsbh}{6.13993}{GW230529ay_imrseobv4_nsbh}{6.12821}{GW230529ay_imrpv2_low_spin}{6.14771}}}
\newcommand{\rafivepercent}[1]{\IfEqCase{#1}{{GW230529ay_combined_imrphm_high_spin}{0.46767}{GW230529ay_combined_imrphm_low_spin}{0.45866}{GW230529ay_imrphen_nsbh}{0.44853}{GW230529ay_imrseobv4_nsbh}{0.44724}{GW230529ay_imrpv2_low_spin}{0.48348}}}
\newcommand{\raninetyfivepercent}[1]{\IfEqCase{#1}{{GW230529ay_combined_imrphm_high_spin}{5.42956}{GW230529ay_combined_imrphm_low_spin}{5.45269}{GW230529ay_imrphen_nsbh}{5.48342}{GW230529ay_imrseobv4_nsbh}{5.48180}{GW230529ay_imrpv2_low_spin}{5.47157}}}
\newcommand{\raninetypercent}[1]{\IfEqCase{#1}{{GW230529ay_combined_imrphm_high_spin}{5.03843}{GW230529ay_combined_imrphm_low_spin}{5.05539}{GW230529ay_imrphen_nsbh}{5.07293}{GW230529ay_imrseobv4_nsbh}{5.07216}{GW230529ay_imrpv2_low_spin}{5.09506}}}
\newcommand{\radiatedenergyminus}[1]{\IfEqCase{#1}{{GW230529ay_combined_imrphm_high_spin}{-}{GW230529ay_combined_imrphm_low_spin}{-}{GW230529ay_imrphen_nsbh}{0.0}{GW230529ay_imrseobv4_nsbh}{0.0}{GW230529ay_imrpv2_low_spin}{-}}}
\newcommand{\radiatedenergymed}[1]{\IfEqCase{#1}{{GW230529ay_combined_imrphm_high_spin}{-}{GW230529ay_combined_imrphm_low_spin}{-}{GW230529ay_imrphen_nsbh}{0.2}{GW230529ay_imrseobv4_nsbh}{0.2}{GW230529ay_imrpv2_low_spin}{-}}}
\newcommand{\radiatedenergyplus}[1]{\IfEqCase{#1}{{GW230529ay_combined_imrphm_high_spin}{-}{GW230529ay_combined_imrphm_low_spin}{-}{GW230529ay_imrphen_nsbh}{0.0}{GW230529ay_imrseobv4_nsbh}{0.0}{GW230529ay_imrpv2_low_spin}{-}}}
\newcommand{\radiatedenergyonepercent}[1]{\IfEqCase{#1}{{GW230529ay_combined_imrphm_high_spin}{-}{GW230529ay_combined_imrphm_low_spin}{-}{GW230529ay_imrphen_nsbh}{0.1}{GW230529ay_imrseobv4_nsbh}{0.1}{GW230529ay_imrpv2_low_spin}{-}}}
\newcommand{\radiatedenergyninetyninepercent}[1]{\IfEqCase{#1}{{GW230529ay_combined_imrphm_high_spin}{-}{GW230529ay_combined_imrphm_low_spin}{-}{GW230529ay_imrphen_nsbh}{0.2}{GW230529ay_imrseobv4_nsbh}{0.2}{GW230529ay_imrpv2_low_spin}{-}}}
\newcommand{\radiatedenergyfivepercent}[1]{\IfEqCase{#1}{{GW230529ay_combined_imrphm_high_spin}{-}{GW230529ay_combined_imrphm_low_spin}{-}{GW230529ay_imrphen_nsbh}{0.1}{GW230529ay_imrseobv4_nsbh}{0.1}{GW230529ay_imrpv2_low_spin}{-}}}
\newcommand{\radiatedenergyninetyfivepercent}[1]{\IfEqCase{#1}{{GW230529ay_combined_imrphm_high_spin}{-}{GW230529ay_combined_imrphm_low_spin}{-}{GW230529ay_imrphen_nsbh}{0.2}{GW230529ay_imrseobv4_nsbh}{0.2}{GW230529ay_imrpv2_low_spin}{-}}}
\newcommand{\radiatedenergyninetypercent}[1]{\IfEqCase{#1}{{GW230529ay_combined_imrphm_high_spin}{-}{GW230529ay_combined_imrphm_low_spin}{-}{GW230529ay_imrphen_nsbh}{0.2}{GW230529ay_imrseobv4_nsbh}{0.2}{GW230529ay_imrpv2_low_spin}{-}}}
\newcommand{\redshiftminus}[1]{\IfEqCase{#1}{{GW230529ay_combined_imrphm_high_spin}{0.02}{GW230529ay_combined_imrphm_low_spin}{0.02}{GW230529ay_imrphen_nsbh}{0.02}{GW230529ay_imrseobv4_nsbh}{0.02}{GW230529ay_imrpv2_low_spin}{0.02}}}
\newcommand{\redshiftmed}[1]{\IfEqCase{#1}{{GW230529ay_combined_imrphm_high_spin}{0.04}{GW230529ay_combined_imrphm_low_spin}{0.04}{GW230529ay_imrphen_nsbh}{0.04}{GW230529ay_imrseobv4_nsbh}{0.04}{GW230529ay_imrpv2_low_spin}{0.04}}}
\newcommand{\redshiftplus}[1]{\IfEqCase{#1}{{GW230529ay_combined_imrphm_high_spin}{0.02}{GW230529ay_combined_imrphm_low_spin}{0.02}{GW230529ay_imrphen_nsbh}{0.02}{GW230529ay_imrseobv4_nsbh}{0.02}{GW230529ay_imrpv2_low_spin}{0.02}}}
\newcommand{\redshiftonepercent}[1]{\IfEqCase{#1}{{GW230529ay_combined_imrphm_high_spin}{0.02}{GW230529ay_combined_imrphm_low_spin}{0.02}{GW230529ay_imrphen_nsbh}{0.02}{GW230529ay_imrseobv4_nsbh}{0.01}{GW230529ay_imrpv2_low_spin}{0.02}}}
\newcommand{\redshiftninetyninepercent}[1]{\IfEqCase{#1}{{GW230529ay_combined_imrphm_high_spin}{0.07}{GW230529ay_combined_imrphm_low_spin}{0.07}{GW230529ay_imrphen_nsbh}{0.07}{GW230529ay_imrseobv4_nsbh}{0.07}{GW230529ay_imrpv2_low_spin}{0.07}}}
\newcommand{\redshiftfivepercent}[1]{\IfEqCase{#1}{{GW230529ay_combined_imrphm_high_spin}{0.02}{GW230529ay_combined_imrphm_low_spin}{0.02}{GW230529ay_imrphen_nsbh}{0.02}{GW230529ay_imrseobv4_nsbh}{0.02}{GW230529ay_imrpv2_low_spin}{0.02}}}
\newcommand{\redshiftninetyfivepercent}[1]{\IfEqCase{#1}{{GW230529ay_combined_imrphm_high_spin}{0.07}{GW230529ay_combined_imrphm_low_spin}{0.07}{GW230529ay_imrphen_nsbh}{0.07}{GW230529ay_imrseobv4_nsbh}{0.07}{GW230529ay_imrpv2_low_spin}{0.07}}}
\newcommand{\redshiftninetypercent}[1]{\IfEqCase{#1}{{GW230529ay_combined_imrphm_high_spin}{0.06}{GW230529ay_combined_imrphm_low_spin}{0.06}{GW230529ay_imrphen_nsbh}{0.06}{GW230529ay_imrseobv4_nsbh}{0.06}{GW230529ay_imrpv2_low_spin}{0.06}}}
\newcommand{\spinonexminus}[1]{\IfEqCase{#1}{{GW230529ay_combined_imrphm_high_spin}{0.55}{GW230529ay_combined_imrphm_low_spin}{0.55}{GW230529ay_imrphen_nsbh}{0.00}{GW230529ay_imrseobv4_nsbh}{0.00}{GW230529ay_imrpv2_low_spin}{0.52}}}
\newcommand{\spinonexmed}[1]{\IfEqCase{#1}{{GW230529ay_combined_imrphm_high_spin}{0.00}{GW230529ay_combined_imrphm_low_spin}{-0.00}{GW230529ay_imrphen_nsbh}{0.00}{GW230529ay_imrseobv4_nsbh}{0.00}{GW230529ay_imrpv2_low_spin}{-0.00}}}
\newcommand{\spinonexplus}[1]{\IfEqCase{#1}{{GW230529ay_combined_imrphm_high_spin}{0.56}{GW230529ay_combined_imrphm_low_spin}{0.55}{GW230529ay_imrphen_nsbh}{0.00}{GW230529ay_imrseobv4_nsbh}{0.00}{GW230529ay_imrpv2_low_spin}{0.52}}}
\newcommand{\spinonexonepercent}[1]{\IfEqCase{#1}{{GW230529ay_combined_imrphm_high_spin}{-0.74}{GW230529ay_combined_imrphm_low_spin}{-0.73}{GW230529ay_imrphen_nsbh}{0.00}{GW230529ay_imrseobv4_nsbh}{0.00}{GW230529ay_imrpv2_low_spin}{-0.74}}}
\newcommand{\spinonexninetyninepercent}[1]{\IfEqCase{#1}{{GW230529ay_combined_imrphm_high_spin}{0.75}{GW230529ay_combined_imrphm_low_spin}{0.73}{GW230529ay_imrphen_nsbh}{0.00}{GW230529ay_imrseobv4_nsbh}{0.00}{GW230529ay_imrpv2_low_spin}{0.74}}}
\newcommand{\spinonexfivepercent}[1]{\IfEqCase{#1}{{GW230529ay_combined_imrphm_high_spin}{-0.55}{GW230529ay_combined_imrphm_low_spin}{-0.55}{GW230529ay_imrphen_nsbh}{0.00}{GW230529ay_imrseobv4_nsbh}{0.00}{GW230529ay_imrpv2_low_spin}{-0.52}}}
\newcommand{\spinonexninetyfivepercent}[1]{\IfEqCase{#1}{{GW230529ay_combined_imrphm_high_spin}{0.56}{GW230529ay_combined_imrphm_low_spin}{0.55}{GW230529ay_imrphen_nsbh}{0.00}{GW230529ay_imrseobv4_nsbh}{0.00}{GW230529ay_imrpv2_low_spin}{0.52}}}
\newcommand{\spinonexninetypercent}[1]{\IfEqCase{#1}{{GW230529ay_combined_imrphm_high_spin}{0.43}{GW230529ay_combined_imrphm_low_spin}{0.43}{GW230529ay_imrphen_nsbh}{0.00}{GW230529ay_imrseobv4_nsbh}{0.00}{GW230529ay_imrpv2_low_spin}{0.40}}}
\newcommand{\spinoneyminus}[1]{\IfEqCase{#1}{{GW230529ay_combined_imrphm_high_spin}{0.54}{GW230529ay_combined_imrphm_low_spin}{0.55}{GW230529ay_imrphen_nsbh}{0.00}{GW230529ay_imrseobv4_nsbh}{0.00}{GW230529ay_imrpv2_low_spin}{0.52}}}
\newcommand{\spinoneymed}[1]{\IfEqCase{#1}{{GW230529ay_combined_imrphm_high_spin}{0.00}{GW230529ay_combined_imrphm_low_spin}{-0.00}{GW230529ay_imrphen_nsbh}{0.00}{GW230529ay_imrseobv4_nsbh}{0.00}{GW230529ay_imrpv2_low_spin}{0.00}}}
\newcommand{\spinoneyplus}[1]{\IfEqCase{#1}{{GW230529ay_combined_imrphm_high_spin}{0.54}{GW230529ay_combined_imrphm_low_spin}{0.55}{GW230529ay_imrphen_nsbh}{0.00}{GW230529ay_imrseobv4_nsbh}{0.00}{GW230529ay_imrpv2_low_spin}{0.52}}}
\newcommand{\spinoneyonepercent}[1]{\IfEqCase{#1}{{GW230529ay_combined_imrphm_high_spin}{-0.73}{GW230529ay_combined_imrphm_low_spin}{-0.73}{GW230529ay_imrphen_nsbh}{0.00}{GW230529ay_imrseobv4_nsbh}{0.00}{GW230529ay_imrpv2_low_spin}{-0.74}}}
\newcommand{\spinoneyninetyninepercent}[1]{\IfEqCase{#1}{{GW230529ay_combined_imrphm_high_spin}{0.73}{GW230529ay_combined_imrphm_low_spin}{0.73}{GW230529ay_imrphen_nsbh}{0.00}{GW230529ay_imrseobv4_nsbh}{0.00}{GW230529ay_imrpv2_low_spin}{0.74}}}
\newcommand{\spinoneyfivepercent}[1]{\IfEqCase{#1}{{GW230529ay_combined_imrphm_high_spin}{-0.54}{GW230529ay_combined_imrphm_low_spin}{-0.55}{GW230529ay_imrphen_nsbh}{0.00}{GW230529ay_imrseobv4_nsbh}{0.00}{GW230529ay_imrpv2_low_spin}{-0.51}}}
\newcommand{\spinoneyninetyfivepercent}[1]{\IfEqCase{#1}{{GW230529ay_combined_imrphm_high_spin}{0.55}{GW230529ay_combined_imrphm_low_spin}{0.55}{GW230529ay_imrphen_nsbh}{0.00}{GW230529ay_imrseobv4_nsbh}{0.00}{GW230529ay_imrpv2_low_spin}{0.52}}}
\newcommand{\spinoneyninetypercent}[1]{\IfEqCase{#1}{{GW230529ay_combined_imrphm_high_spin}{0.42}{GW230529ay_combined_imrphm_low_spin}{0.43}{GW230529ay_imrphen_nsbh}{0.00}{GW230529ay_imrseobv4_nsbh}{0.00}{GW230529ay_imrpv2_low_spin}{0.40}}}
\newcommand{\spinonezminus}[1]{\IfEqCase{#1}{{GW230529ay_combined_imrphm_high_spin}{0.35}{GW230529ay_combined_imrphm_low_spin}{0.35}{GW230529ay_imrphen_nsbh}{0.25}{GW230529ay_imrseobv4_nsbh}{0.30}{GW230529ay_imrpv2_low_spin}{0.25}}}
\newcommand{\spinonezmed}[1]{\IfEqCase{#1}{{GW230529ay_combined_imrphm_high_spin}{-0.11}{GW230529ay_combined_imrphm_low_spin}{-0.14}{GW230529ay_imrphen_nsbh}{-0.07}{GW230529ay_imrseobv4_nsbh}{-0.11}{GW230529ay_imrpv2_low_spin}{-0.16}}}
\newcommand{\spinonezplus}[1]{\IfEqCase{#1}{{GW230529ay_combined_imrphm_high_spin}{0.19}{GW230529ay_combined_imrphm_low_spin}{0.15}{GW230529ay_imrphen_nsbh}{0.18}{GW230529ay_imrseobv4_nsbh}{0.21}{GW230529ay_imrpv2_low_spin}{0.19}}}
\newcommand{\spinonezonepercent}[1]{\IfEqCase{#1}{{GW230529ay_combined_imrphm_high_spin}{-0.62}{GW230529ay_combined_imrphm_low_spin}{-0.55}{GW230529ay_imrphen_nsbh}{-0.38}{GW230529ay_imrseobv4_nsbh}{-0.47}{GW230529ay_imrpv2_low_spin}{-0.46}}}
\newcommand{\spinonezninetyninepercent}[1]{\IfEqCase{#1}{{GW230529ay_combined_imrphm_high_spin}{0.16}{GW230529ay_combined_imrphm_low_spin}{0.08}{GW230529ay_imrphen_nsbh}{0.19}{GW230529ay_imrseobv4_nsbh}{0.18}{GW230529ay_imrpv2_low_spin}{0.07}}}
\newcommand{\spinonezfivepercent}[1]{\IfEqCase{#1}{{GW230529ay_combined_imrphm_high_spin}{-0.47}{GW230529ay_combined_imrphm_low_spin}{-0.49}{GW230529ay_imrphen_nsbh}{-0.32}{GW230529ay_imrseobv4_nsbh}{-0.41}{GW230529ay_imrpv2_low_spin}{-0.41}}}
\newcommand{\spinonezninetyfivepercent}[1]{\IfEqCase{#1}{{GW230529ay_combined_imrphm_high_spin}{0.08}{GW230529ay_combined_imrphm_low_spin}{0.00}{GW230529ay_imrphen_nsbh}{0.11}{GW230529ay_imrseobv4_nsbh}{0.10}{GW230529ay_imrpv2_low_spin}{0.03}}}
\newcommand{\spinonezninetypercent}[1]{\IfEqCase{#1}{{GW230529ay_combined_imrphm_high_spin}{0.03}{GW230529ay_combined_imrphm_low_spin}{-0.02}{GW230529ay_imrphen_nsbh}{0.07}{GW230529ay_imrseobv4_nsbh}{0.06}{GW230529ay_imrpv2_low_spin}{0.00}}}
\newcommand{\spintwoxminus}[1]{\IfEqCase{#1}{{GW230529ay_combined_imrphm_high_spin}{0.56}{GW230529ay_combined_imrphm_low_spin}{0.03}{GW230529ay_imrphen_nsbh}{0.00}{GW230529ay_imrseobv4_nsbh}{0.00}{GW230529ay_imrpv2_low_spin}{0.03}}}
\newcommand{\spintwoxmed}[1]{\IfEqCase{#1}{{GW230529ay_combined_imrphm_high_spin}{0.00}{GW230529ay_combined_imrphm_low_spin}{-0.00}{GW230529ay_imrphen_nsbh}{0.00}{GW230529ay_imrseobv4_nsbh}{0.00}{GW230529ay_imrpv2_low_spin}{-0.00}}}
\newcommand{\spintwoxplus}[1]{\IfEqCase{#1}{{GW230529ay_combined_imrphm_high_spin}{0.55}{GW230529ay_combined_imrphm_low_spin}{0.03}{GW230529ay_imrphen_nsbh}{0.00}{GW230529ay_imrseobv4_nsbh}{0.00}{GW230529ay_imrpv2_low_spin}{0.03}}}
\newcommand{\spintwoxonepercent}[1]{\IfEqCase{#1}{{GW230529ay_combined_imrphm_high_spin}{-0.78}{GW230529ay_combined_imrphm_low_spin}{-0.04}{GW230529ay_imrphen_nsbh}{0.00}{GW230529ay_imrseobv4_nsbh}{0.00}{GW230529ay_imrpv2_low_spin}{-0.04}}}
\newcommand{\spintwoxninetyninepercent}[1]{\IfEqCase{#1}{{GW230529ay_combined_imrphm_high_spin}{0.78}{GW230529ay_combined_imrphm_low_spin}{0.04}{GW230529ay_imrphen_nsbh}{0.00}{GW230529ay_imrseobv4_nsbh}{0.00}{GW230529ay_imrpv2_low_spin}{0.04}}}
\newcommand{\spintwoxfivepercent}[1]{\IfEqCase{#1}{{GW230529ay_combined_imrphm_high_spin}{-0.56}{GW230529ay_combined_imrphm_low_spin}{-0.03}{GW230529ay_imrphen_nsbh}{0.00}{GW230529ay_imrseobv4_nsbh}{0.00}{GW230529ay_imrpv2_low_spin}{-0.03}}}
\newcommand{\spintwoxninetyfivepercent}[1]{\IfEqCase{#1}{{GW230529ay_combined_imrphm_high_spin}{0.55}{GW230529ay_combined_imrphm_low_spin}{0.03}{GW230529ay_imrphen_nsbh}{0.00}{GW230529ay_imrseobv4_nsbh}{0.00}{GW230529ay_imrpv2_low_spin}{0.03}}}
\newcommand{\spintwoxninetypercent}[1]{\IfEqCase{#1}{{GW230529ay_combined_imrphm_high_spin}{0.41}{GW230529ay_combined_imrphm_low_spin}{0.02}{GW230529ay_imrphen_nsbh}{0.00}{GW230529ay_imrseobv4_nsbh}{0.00}{GW230529ay_imrpv2_low_spin}{0.02}}}
\newcommand{\spintwoyminus}[1]{\IfEqCase{#1}{{GW230529ay_combined_imrphm_high_spin}{0.55}{GW230529ay_combined_imrphm_low_spin}{0.03}{GW230529ay_imrphen_nsbh}{0.00}{GW230529ay_imrseobv4_nsbh}{0.00}{GW230529ay_imrpv2_low_spin}{0.03}}}
\newcommand{\spintwoymed}[1]{\IfEqCase{#1}{{GW230529ay_combined_imrphm_high_spin}{0.00}{GW230529ay_combined_imrphm_low_spin}{-0.00}{GW230529ay_imrphen_nsbh}{0.00}{GW230529ay_imrseobv4_nsbh}{0.00}{GW230529ay_imrpv2_low_spin}{-0.00}}}
\newcommand{\spintwoyplus}[1]{\IfEqCase{#1}{{GW230529ay_combined_imrphm_high_spin}{0.55}{GW230529ay_combined_imrphm_low_spin}{0.03}{GW230529ay_imrphen_nsbh}{0.00}{GW230529ay_imrseobv4_nsbh}{0.00}{GW230529ay_imrpv2_low_spin}{0.03}}}
\newcommand{\spintwoyonepercent}[1]{\IfEqCase{#1}{{GW230529ay_combined_imrphm_high_spin}{-0.78}{GW230529ay_combined_imrphm_low_spin}{-0.04}{GW230529ay_imrphen_nsbh}{0.00}{GW230529ay_imrseobv4_nsbh}{0.00}{GW230529ay_imrpv2_low_spin}{-0.04}}}
\newcommand{\spintwoyninetyninepercent}[1]{\IfEqCase{#1}{{GW230529ay_combined_imrphm_high_spin}{0.79}{GW230529ay_combined_imrphm_low_spin}{0.04}{GW230529ay_imrphen_nsbh}{0.00}{GW230529ay_imrseobv4_nsbh}{0.00}{GW230529ay_imrpv2_low_spin}{0.04}}}
\newcommand{\spintwoyfivepercent}[1]{\IfEqCase{#1}{{GW230529ay_combined_imrphm_high_spin}{-0.55}{GW230529ay_combined_imrphm_low_spin}{-0.03}{GW230529ay_imrphen_nsbh}{0.00}{GW230529ay_imrseobv4_nsbh}{0.00}{GW230529ay_imrpv2_low_spin}{-0.03}}}
\newcommand{\spintwoyninetyfivepercent}[1]{\IfEqCase{#1}{{GW230529ay_combined_imrphm_high_spin}{0.55}{GW230529ay_combined_imrphm_low_spin}{0.03}{GW230529ay_imrphen_nsbh}{0.00}{GW230529ay_imrseobv4_nsbh}{0.00}{GW230529ay_imrpv2_low_spin}{0.03}}}
\newcommand{\spintwoyninetypercent}[1]{\IfEqCase{#1}{{GW230529ay_combined_imrphm_high_spin}{0.40}{GW230529ay_combined_imrphm_low_spin}{0.02}{GW230529ay_imrphen_nsbh}{0.00}{GW230529ay_imrseobv4_nsbh}{0.00}{GW230529ay_imrpv2_low_spin}{0.02}}}
\newcommand{\spintwozminus}[1]{\IfEqCase{#1}{{GW230529ay_combined_imrphm_high_spin}{0.52}{GW230529ay_combined_imrphm_low_spin}{0.03}{GW230529ay_imrphen_nsbh}{0.03}{GW230529ay_imrseobv4_nsbh}{0.03}{GW230529ay_imrpv2_low_spin}{0.03}}}
\newcommand{\spintwozmed}[1]{\IfEqCase{#1}{{GW230529ay_combined_imrphm_high_spin}{-0.03}{GW230529ay_combined_imrphm_low_spin}{-0.00}{GW230529ay_imrphen_nsbh}{-0.00}{GW230529ay_imrseobv4_nsbh}{-0.00}{GW230529ay_imrpv2_low_spin}{-0.00}}}
\newcommand{\spintwozplus}[1]{\IfEqCase{#1}{{GW230529ay_combined_imrphm_high_spin}{0.43}{GW230529ay_combined_imrphm_low_spin}{0.03}{GW230529ay_imrphen_nsbh}{0.03}{GW230529ay_imrseobv4_nsbh}{0.03}{GW230529ay_imrpv2_low_spin}{0.03}}}
\newcommand{\spintwozonepercent}[1]{\IfEqCase{#1}{{GW230529ay_combined_imrphm_high_spin}{-0.71}{GW230529ay_combined_imrphm_low_spin}{-0.04}{GW230529ay_imrphen_nsbh}{-0.04}{GW230529ay_imrseobv4_nsbh}{-0.04}{GW230529ay_imrpv2_low_spin}{-0.04}}}
\newcommand{\spintwozninetyninepercent}[1]{\IfEqCase{#1}{{GW230529ay_combined_imrphm_high_spin}{0.65}{GW230529ay_combined_imrphm_low_spin}{0.04}{GW230529ay_imrphen_nsbh}{0.04}{GW230529ay_imrseobv4_nsbh}{0.04}{GW230529ay_imrpv2_low_spin}{0.04}}}
\newcommand{\spintwozfivepercent}[1]{\IfEqCase{#1}{{GW230529ay_combined_imrphm_high_spin}{-0.55}{GW230529ay_combined_imrphm_low_spin}{-0.03}{GW230529ay_imrphen_nsbh}{-0.03}{GW230529ay_imrseobv4_nsbh}{-0.03}{GW230529ay_imrpv2_low_spin}{-0.03}}}
\newcommand{\spintwozninetyfivepercent}[1]{\IfEqCase{#1}{{GW230529ay_combined_imrphm_high_spin}{0.40}{GW230529ay_combined_imrphm_low_spin}{0.03}{GW230529ay_imrphen_nsbh}{0.03}{GW230529ay_imrseobv4_nsbh}{0.03}{GW230529ay_imrpv2_low_spin}{0.03}}}
\newcommand{\spintwozninetypercent}[1]{\IfEqCase{#1}{{GW230529ay_combined_imrphm_high_spin}{0.27}{GW230529ay_combined_imrphm_low_spin}{0.02}{GW230529ay_imrphen_nsbh}{0.02}{GW230529ay_imrseobv4_nsbh}{0.02}{GW230529ay_imrpv2_low_spin}{0.02}}}
\newcommand{\symmetricmassratiominus}[1]{\IfEqCase{#1}{{GW230529ay_combined_imrphm_high_spin}{0.03}{GW230529ay_combined_imrphm_low_spin}{0.03}{GW230529ay_imrphen_nsbh}{0.04}{GW230529ay_imrseobv4_nsbh}{0.05}{GW230529ay_imrpv2_low_spin}{0.04}}}
\newcommand{\symmetricmassratiomed}[1]{\IfEqCase{#1}{{GW230529ay_combined_imrphm_high_spin}{0.20}{GW230529ay_combined_imrphm_low_spin}{0.20}{GW230529ay_imrphen_nsbh}{0.19}{GW230529ay_imrseobv4_nsbh}{0.20}{GW230529ay_imrpv2_low_spin}{0.21}}}
\newcommand{\symmetricmassratioplus}[1]{\IfEqCase{#1}{{GW230529ay_combined_imrphm_high_spin}{0.04}{GW230529ay_combined_imrphm_low_spin}{0.05}{GW230529ay_imrphen_nsbh}{0.04}{GW230529ay_imrseobv4_nsbh}{0.04}{GW230529ay_imrpv2_low_spin}{0.03}}}
\newcommand{\symmetricmassratioonepercent}[1]{\IfEqCase{#1}{{GW230529ay_combined_imrphm_high_spin}{0.15}{GW230529ay_combined_imrphm_low_spin}{0.16}{GW230529ay_imrphen_nsbh}{0.13}{GW230529ay_imrseobv4_nsbh}{0.14}{GW230529ay_imrpv2_low_spin}{0.16}}}
\newcommand{\symmetricmassrationinetyninepercent}[1]{\IfEqCase{#1}{{GW230529ay_combined_imrphm_high_spin}{0.25}{GW230529ay_combined_imrphm_low_spin}{0.25}{GW230529ay_imrphen_nsbh}{0.24}{GW230529ay_imrseobv4_nsbh}{0.25}{GW230529ay_imrpv2_low_spin}{0.25}}}
\newcommand{\symmetricmassratiofivepercent}[1]{\IfEqCase{#1}{{GW230529ay_combined_imrphm_high_spin}{0.17}{GW230529ay_combined_imrphm_low_spin}{0.17}{GW230529ay_imrphen_nsbh}{0.15}{GW230529ay_imrseobv4_nsbh}{0.16}{GW230529ay_imrpv2_low_spin}{0.17}}}
\newcommand{\symmetricmassrationinetyfivepercent}[1]{\IfEqCase{#1}{{GW230529ay_combined_imrphm_high_spin}{0.25}{GW230529ay_combined_imrphm_low_spin}{0.25}{GW230529ay_imrphen_nsbh}{0.24}{GW230529ay_imrseobv4_nsbh}{0.25}{GW230529ay_imrpv2_low_spin}{0.25}}}
\newcommand{\symmetricmassrationinetypercent}[1]{\IfEqCase{#1}{{GW230529ay_combined_imrphm_high_spin}{0.24}{GW230529ay_combined_imrphm_low_spin}{0.24}{GW230529ay_imrphen_nsbh}{0.23}{GW230529ay_imrseobv4_nsbh}{0.24}{GW230529ay_imrpv2_low_spin}{0.25}}}
\newcommand{\thetajnminus}[1]{\IfEqCase{#1}{{GW230529ay_combined_imrphm_high_spin}{1.33}{GW230529ay_combined_imrphm_low_spin}{1.30}{GW230529ay_imrphen_nsbh}{1.28}{GW230529ay_imrseobv4_nsbh}{1.31}{GW230529ay_imrpv2_low_spin}{1.28}}}
\newcommand{\thetajnmed}[1]{\IfEqCase{#1}{{GW230529ay_combined_imrphm_high_spin}{1.64}{GW230529ay_combined_imrphm_low_spin}{1.61}{GW230529ay_imrphen_nsbh}{1.54}{GW230529ay_imrseobv4_nsbh}{1.56}{GW230529ay_imrpv2_low_spin}{1.56}}}
\newcommand{\thetajnplus}[1]{\IfEqCase{#1}{{GW230529ay_combined_imrphm_high_spin}{1.21}{GW230529ay_combined_imrphm_low_spin}{1.23}{GW230529ay_imrphen_nsbh}{1.34}{GW230529ay_imrseobv4_nsbh}{1.32}{GW230529ay_imrpv2_low_spin}{1.28}}}
\newcommand{\thetajnonepercent}[1]{\IfEqCase{#1}{{GW230529ay_combined_imrphm_high_spin}{0.14}{GW230529ay_combined_imrphm_low_spin}{0.15}{GW230529ay_imrphen_nsbh}{0.11}{GW230529ay_imrseobv4_nsbh}{0.12}{GW230529ay_imrpv2_low_spin}{0.13}}}
\newcommand{\thetajnninetyninepercent}[1]{\IfEqCase{#1}{{GW230529ay_combined_imrphm_high_spin}{3.01}{GW230529ay_combined_imrphm_low_spin}{3.00}{GW230529ay_imrphen_nsbh}{3.02}{GW230529ay_imrseobv4_nsbh}{3.03}{GW230529ay_imrpv2_low_spin}{3.01}}}
\newcommand{\thetajnfivepercent}[1]{\IfEqCase{#1}{{GW230529ay_combined_imrphm_high_spin}{0.31}{GW230529ay_combined_imrphm_low_spin}{0.31}{GW230529ay_imrphen_nsbh}{0.26}{GW230529ay_imrseobv4_nsbh}{0.26}{GW230529ay_imrpv2_low_spin}{0.28}}}
\newcommand{\thetajnninetyfivepercent}[1]{\IfEqCase{#1}{{GW230529ay_combined_imrphm_high_spin}{2.85}{GW230529ay_combined_imrphm_low_spin}{2.83}{GW230529ay_imrphen_nsbh}{2.88}{GW230529ay_imrseobv4_nsbh}{2.89}{GW230529ay_imrpv2_low_spin}{2.84}}}
\newcommand{\thetajnninetypercent}[1]{\IfEqCase{#1}{{GW230529ay_combined_imrphm_high_spin}{2.74}{GW230529ay_combined_imrphm_low_spin}{2.72}{GW230529ay_imrphen_nsbh}{2.75}{GW230529ay_imrseobv4_nsbh}{2.77}{GW230529ay_imrpv2_low_spin}{2.71}}}
\newcommand{\tidaldisruptionfrequencyminus}[1]{\IfEqCase{#1}{{GW230529ay_combined_imrphm_high_spin}{-}{GW230529ay_combined_imrphm_low_spin}{-}{GW230529ay_imrphen_nsbh}{383}{GW230529ay_imrseobv4_nsbh}{-}{GW230529ay_imrpv2_low_spin}{-}}}
\newcommand{\tidaldisruptionfrequencymed}[1]{\IfEqCase{#1}{{GW230529ay_combined_imrphm_high_spin}{-}{GW230529ay_combined_imrphm_low_spin}{-}{GW230529ay_imrphen_nsbh}{1542}{GW230529ay_imrseobv4_nsbh}{-}{GW230529ay_imrpv2_low_spin}{-}}}
\newcommand{\tidaldisruptionfrequencyplus}[1]{\IfEqCase{#1}{{GW230529ay_combined_imrphm_high_spin}{-}{GW230529ay_combined_imrphm_low_spin}{-}{GW230529ay_imrphen_nsbh}{1530}{GW230529ay_imrseobv4_nsbh}{-}{GW230529ay_imrpv2_low_spin}{-}}}
\newcommand{\tidaldisruptionfrequencyonepercent}[1]{\IfEqCase{#1}{{GW230529ay_combined_imrphm_high_spin}{-}{GW230529ay_combined_imrphm_low_spin}{-}{GW230529ay_imrphen_nsbh}{1064}{GW230529ay_imrseobv4_nsbh}{-}{GW230529ay_imrpv2_low_spin}{-}}}
\newcommand{\tidaldisruptionfrequencyninetyninepercent}[1]{\IfEqCase{#1}{{GW230529ay_combined_imrphm_high_spin}{-}{GW230529ay_combined_imrphm_low_spin}{-}{GW230529ay_imrphen_nsbh}{4759}{GW230529ay_imrseobv4_nsbh}{-}{GW230529ay_imrpv2_low_spin}{-}}}
\newcommand{\tidaldisruptionfrequencyfivepercent}[1]{\IfEqCase{#1}{{GW230529ay_combined_imrphm_high_spin}{-}{GW230529ay_combined_imrphm_low_spin}{-}{GW230529ay_imrphen_nsbh}{1159}{GW230529ay_imrseobv4_nsbh}{-}{GW230529ay_imrpv2_low_spin}{-}}}
\newcommand{\tidaldisruptionfrequencyninetyfivepercent}[1]{\IfEqCase{#1}{{GW230529ay_combined_imrphm_high_spin}{-}{GW230529ay_combined_imrphm_low_spin}{-}{GW230529ay_imrphen_nsbh}{3072}{GW230529ay_imrseobv4_nsbh}{-}{GW230529ay_imrpv2_low_spin}{-}}}
\newcommand{\tidaldisruptionfrequencyninetypercent}[1]{\IfEqCase{#1}{{GW230529ay_combined_imrphm_high_spin}{-}{GW230529ay_combined_imrphm_low_spin}{-}{GW230529ay_imrphen_nsbh}{2482}{GW230529ay_imrseobv4_nsbh}{-}{GW230529ay_imrpv2_low_spin}{-}}}
\newcommand{\tidaldisruptionfrequencyratiominus}[1]{\IfEqCase{#1}{{GW230529ay_combined_imrphm_high_spin}{-}{GW230529ay_combined_imrphm_low_spin}{-}{GW230529ay_imrphen_nsbh}{0.2}{GW230529ay_imrseobv4_nsbh}{-}{GW230529ay_imrpv2_low_spin}{-}}}
\newcommand{\tidaldisruptionfrequencyratiomed}[1]{\IfEqCase{#1}{{GW230529ay_combined_imrphm_high_spin}{-}{GW230529ay_combined_imrphm_low_spin}{-}{GW230529ay_imrphen_nsbh}{0.5}{GW230529ay_imrseobv4_nsbh}{-}{GW230529ay_imrpv2_low_spin}{-}}}
\newcommand{\tidaldisruptionfrequencyratioplus}[1]{\IfEqCase{#1}{{GW230529ay_combined_imrphm_high_spin}{-}{GW230529ay_combined_imrphm_low_spin}{-}{GW230529ay_imrphen_nsbh}{0.5}{GW230529ay_imrseobv4_nsbh}{-}{GW230529ay_imrpv2_low_spin}{-}}}
\newcommand{\tidaldisruptionfrequencyratioonepercent}[1]{\IfEqCase{#1}{{GW230529ay_combined_imrphm_high_spin}{-}{GW230529ay_combined_imrphm_low_spin}{-}{GW230529ay_imrphen_nsbh}{0.3}{GW230529ay_imrseobv4_nsbh}{-}{GW230529ay_imrpv2_low_spin}{-}}}
\newcommand{\tidaldisruptionfrequencyrationinetyninepercent}[1]{\IfEqCase{#1}{{GW230529ay_combined_imrphm_high_spin}{-}{GW230529ay_combined_imrphm_low_spin}{-}{GW230529ay_imrphen_nsbh}{1.6}{GW230529ay_imrseobv4_nsbh}{-}{GW230529ay_imrpv2_low_spin}{-}}}
\newcommand{\tidaldisruptionfrequencyratiofivepercent}[1]{\IfEqCase{#1}{{GW230529ay_combined_imrphm_high_spin}{-}{GW230529ay_combined_imrphm_low_spin}{-}{GW230529ay_imrphen_nsbh}{0.3}{GW230529ay_imrseobv4_nsbh}{-}{GW230529ay_imrpv2_low_spin}{-}}}
\newcommand{\tidaldisruptionfrequencyrationinetyfivepercent}[1]{\IfEqCase{#1}{{GW230529ay_combined_imrphm_high_spin}{-}{GW230529ay_combined_imrphm_low_spin}{-}{GW230529ay_imrphen_nsbh}{1.1}{GW230529ay_imrseobv4_nsbh}{-}{GW230529ay_imrpv2_low_spin}{-}}}
\newcommand{\tidaldisruptionfrequencyrationinetypercent}[1]{\IfEqCase{#1}{{GW230529ay_combined_imrphm_high_spin}{-}{GW230529ay_combined_imrphm_low_spin}{-}{GW230529ay_imrphen_nsbh}{0.9}{GW230529ay_imrseobv4_nsbh}{-}{GW230529ay_imrpv2_low_spin}{-}}}
\newcommand{\tiltoneminus}[1]{\IfEqCase{#1}{{GW230529ay_combined_imrphm_high_spin}{0.65}{GW230529ay_combined_imrphm_low_spin}{0.38}{GW230529ay_imrphen_nsbh}{3.14}{GW230529ay_imrseobv4_nsbh}{3.14}{GW230529ay_imrpv2_low_spin}{0.53}}}
\newcommand{\tiltonemed}[1]{\IfEqCase{#1}{{GW230529ay_combined_imrphm_high_spin}{1.87}{GW230529ay_combined_imrphm_low_spin}{1.93}{GW230529ay_imrphen_nsbh}{3.14}{GW230529ay_imrseobv4_nsbh}{3.14}{GW230529ay_imrpv2_low_spin}{2.00}}}
\newcommand{\tiltoneplus}[1]{\IfEqCase{#1}{{GW230529ay_combined_imrphm_high_spin}{0.75}{GW230529ay_combined_imrphm_low_spin}{0.72}{GW230529ay_imrphen_nsbh}{0.00}{GW230529ay_imrseobv4_nsbh}{0.00}{GW230529ay_imrpv2_low_spin}{0.72}}}
\newcommand{\tiltoneonepercent}[1]{\IfEqCase{#1}{{GW230529ay_combined_imrphm_high_spin}{0.75}{GW230529ay_combined_imrphm_low_spin}{0.95}{GW230529ay_imrphen_nsbh}{0.00}{GW230529ay_imrseobv4_nsbh}{0.00}{GW230529ay_imrpv2_low_spin}{1.15}}}
\newcommand{\tiltoneninetyninepercent}[1]{\IfEqCase{#1}{{GW230529ay_combined_imrphm_high_spin}{2.89}{GW230529ay_combined_imrphm_low_spin}{2.92}{GW230529ay_imrphen_nsbh}{3.14}{GW230529ay_imrseobv4_nsbh}{3.14}{GW230529ay_imrpv2_low_spin}{2.95}}}
\newcommand{\tiltonefivepercent}[1]{\IfEqCase{#1}{{GW230529ay_combined_imrphm_high_spin}{1.23}{GW230529ay_combined_imrphm_low_spin}{1.55}{GW230529ay_imrphen_nsbh}{0.00}{GW230529ay_imrseobv4_nsbh}{0.00}{GW230529ay_imrpv2_low_spin}{1.46}}}
\newcommand{\tiltoneninetyfivepercent}[1]{\IfEqCase{#1}{{GW230529ay_combined_imrphm_high_spin}{2.62}{GW230529ay_combined_imrphm_low_spin}{2.65}{GW230529ay_imrphen_nsbh}{3.14}{GW230529ay_imrseobv4_nsbh}{3.14}{GW230529ay_imrpv2_low_spin}{2.72}}}
\newcommand{\tiltoneninetypercent}[1]{\IfEqCase{#1}{{GW230529ay_combined_imrphm_high_spin}{2.44}{GW230529ay_combined_imrphm_low_spin}{2.48}{GW230529ay_imrphen_nsbh}{3.14}{GW230529ay_imrseobv4_nsbh}{3.14}{GW230529ay_imrpv2_low_spin}{2.56}}}
\newcommand{\tilttwominus}[1]{\IfEqCase{#1}{{GW230529ay_combined_imrphm_high_spin}{1.13}{GW230529ay_combined_imrphm_low_spin}{1.12}{GW230529ay_imrphen_nsbh}{3.14}{GW230529ay_imrseobv4_nsbh}{3.14}{GW230529ay_imrpv2_low_spin}{1.14}}}
\newcommand{\tilttwomed}[1]{\IfEqCase{#1}{{GW230529ay_combined_imrphm_high_spin}{1.71}{GW230529ay_combined_imrphm_low_spin}{1.57}{GW230529ay_imrphen_nsbh}{3.14}{GW230529ay_imrseobv4_nsbh}{3.14}{GW230529ay_imrpv2_low_spin}{1.61}}}
\newcommand{\tilttwoplus}[1]{\IfEqCase{#1}{{GW230529ay_combined_imrphm_high_spin}{0.98}{GW230529ay_combined_imrphm_low_spin}{1.12}{GW230529ay_imrphen_nsbh}{0.00}{GW230529ay_imrseobv4_nsbh}{0.00}{GW230529ay_imrpv2_low_spin}{1.10}}}
\newcommand{\tilttwoonepercent}[1]{\IfEqCase{#1}{{GW230529ay_combined_imrphm_high_spin}{0.27}{GW230529ay_combined_imrphm_low_spin}{0.21}{GW230529ay_imrphen_nsbh}{0.00}{GW230529ay_imrseobv4_nsbh}{0.00}{GW230529ay_imrpv2_low_spin}{0.22}}}
\newcommand{\tilttwoninetyninepercent}[1]{\IfEqCase{#1}{{GW230529ay_combined_imrphm_high_spin}{2.94}{GW230529ay_combined_imrphm_low_spin}{2.95}{GW230529ay_imrphen_nsbh}{3.14}{GW230529ay_imrseobv4_nsbh}{3.14}{GW230529ay_imrpv2_low_spin}{2.92}}}
\newcommand{\tilttwofivepercent}[1]{\IfEqCase{#1}{{GW230529ay_combined_imrphm_high_spin}{0.59}{GW230529ay_combined_imrphm_low_spin}{0.45}{GW230529ay_imrphen_nsbh}{0.00}{GW230529ay_imrseobv4_nsbh}{0.00}{GW230529ay_imrpv2_low_spin}{0.48}}}
\newcommand{\tilttwoninetyfivepercent}[1]{\IfEqCase{#1}{{GW230529ay_combined_imrphm_high_spin}{2.70}{GW230529ay_combined_imrphm_low_spin}{2.69}{GW230529ay_imrphen_nsbh}{3.14}{GW230529ay_imrseobv4_nsbh}{3.14}{GW230529ay_imrpv2_low_spin}{2.71}}}
\newcommand{\tilttwoninetypercent}[1]{\IfEqCase{#1}{{GW230529ay_combined_imrphm_high_spin}{2.52}{GW230529ay_combined_imrphm_low_spin}{2.50}{GW230529ay_imrphen_nsbh}{3.14}{GW230529ay_imrseobv4_nsbh}{3.14}{GW230529ay_imrpv2_low_spin}{2.52}}}
\newcommand{\totalmassminus}[1]{\IfEqCase{#1}{{GW230529ay_combined_imrphm_high_spin}{0.6}{GW230529ay_combined_imrphm_low_spin}{0.6}{GW230529ay_imrphen_nsbh}{0.6}{GW230529ay_imrseobv4_nsbh}{0.6}{GW230529ay_imrpv2_low_spin}{0.4}}}
\newcommand{\totalmassmed}[1]{\IfEqCase{#1}{{GW230529ay_combined_imrphm_high_spin}{5.3}{GW230529ay_combined_imrphm_low_spin}{5.3}{GW230529ay_imrphen_nsbh}{5.4}{GW230529ay_imrseobv4_nsbh}{5.3}{GW230529ay_imrpv2_low_spin}{5.1}}}
\newcommand{\totalmassplus}[1]{\IfEqCase{#1}{{GW230529ay_combined_imrphm_high_spin}{0.6}{GW230529ay_combined_imrphm_low_spin}{0.5}{GW230529ay_imrphen_nsbh}{0.8}{GW230529ay_imrseobv4_nsbh}{0.9}{GW230529ay_imrpv2_low_spin}{0.7}}}
\newcommand{\totalmassonepercent}[1]{\IfEqCase{#1}{{GW230529ay_combined_imrphm_high_spin}{4.7}{GW230529ay_combined_imrphm_low_spin}{4.7}{GW230529ay_imrphen_nsbh}{4.7}{GW230529ay_imrseobv4_nsbh}{4.7}{GW230529ay_imrpv2_low_spin}{4.7}}}
\newcommand{\totalmassninetyninepercent}[1]{\IfEqCase{#1}{{GW230529ay_combined_imrphm_high_spin}{6.3}{GW230529ay_combined_imrphm_low_spin}{6.2}{GW230529ay_imrphen_nsbh}{6.8}{GW230529ay_imrseobv4_nsbh}{6.7}{GW230529ay_imrpv2_low_spin}{6.1}}}
\newcommand{\totalmassfivepercent}[1]{\IfEqCase{#1}{{GW230529ay_combined_imrphm_high_spin}{4.7}{GW230529ay_combined_imrphm_low_spin}{4.7}{GW230529ay_imrphen_nsbh}{4.8}{GW230529ay_imrseobv4_nsbh}{4.7}{GW230529ay_imrpv2_low_spin}{4.7}}}
\newcommand{\totalmassninetyfivepercent}[1]{\IfEqCase{#1}{{GW230529ay_combined_imrphm_high_spin}{5.9}{GW230529ay_combined_imrphm_low_spin}{5.8}{GW230529ay_imrphen_nsbh}{6.2}{GW230529ay_imrseobv4_nsbh}{6.2}{GW230529ay_imrpv2_low_spin}{5.8}}}
\newcommand{\totalmassninetypercent}[1]{\IfEqCase{#1}{{GW230529ay_combined_imrphm_high_spin}{5.8}{GW230529ay_combined_imrphm_low_spin}{5.7}{GW230529ay_imrphen_nsbh}{6.0}{GW230529ay_imrseobv4_nsbh}{5.9}{GW230529ay_imrpv2_low_spin}{5.7}}}
\newcommand{\totalmasssourceminus}[1]{\IfEqCase{#1}{{GW230529ay_combined_imrphm_high_spin}{0.6}{GW230529ay_combined_imrphm_low_spin}{0.6}{GW230529ay_imrphen_nsbh}{0.6}{GW230529ay_imrseobv4_nsbh}{0.6}{GW230529ay_imrpv2_low_spin}{0.4}}}
\newcommand{\totalmasssourcemed}[1]{\IfEqCase{#1}{{GW230529ay_combined_imrphm_high_spin}{5.1}{GW230529ay_combined_imrphm_low_spin}{5.1}{GW230529ay_imrphen_nsbh}{5.2}{GW230529ay_imrseobv4_nsbh}{5.1}{GW230529ay_imrpv2_low_spin}{4.9}}}
\newcommand{\totalmasssourceplus}[1]{\IfEqCase{#1}{{GW230529ay_combined_imrphm_high_spin}{0.6}{GW230529ay_combined_imrphm_low_spin}{0.5}{GW230529ay_imrphen_nsbh}{0.8}{GW230529ay_imrseobv4_nsbh}{0.8}{GW230529ay_imrpv2_low_spin}{0.7}}}
\newcommand{\totalmasssourceonepercent}[1]{\IfEqCase{#1}{{GW230529ay_combined_imrphm_high_spin}{4.4}{GW230529ay_combined_imrphm_low_spin}{4.4}{GW230529ay_imrphen_nsbh}{4.5}{GW230529ay_imrseobv4_nsbh}{4.4}{GW230529ay_imrpv2_low_spin}{4.4}}}
\newcommand{\totalmasssourceninetyninepercent}[1]{\IfEqCase{#1}{{GW230529ay_combined_imrphm_high_spin}{6.1}{GW230529ay_combined_imrphm_low_spin}{5.9}{GW230529ay_imrphen_nsbh}{6.5}{GW230529ay_imrseobv4_nsbh}{6.4}{GW230529ay_imrpv2_low_spin}{5.8}}}
\newcommand{\totalmasssourcefivepercent}[1]{\IfEqCase{#1}{{GW230529ay_combined_imrphm_high_spin}{4.5}{GW230529ay_combined_imrphm_low_spin}{4.5}{GW230529ay_imrphen_nsbh}{4.6}{GW230529ay_imrseobv4_nsbh}{4.5}{GW230529ay_imrpv2_low_spin}{4.4}}}
\newcommand{\totalmasssourceninetyfivepercent}[1]{\IfEqCase{#1}{{GW230529ay_combined_imrphm_high_spin}{5.7}{GW230529ay_combined_imrphm_low_spin}{5.6}{GW230529ay_imrphen_nsbh}{6.0}{GW230529ay_imrseobv4_nsbh}{5.9}{GW230529ay_imrpv2_low_spin}{5.6}}}
\newcommand{\totalmasssourceninetypercent}[1]{\IfEqCase{#1}{{GW230529ay_combined_imrphm_high_spin}{5.5}{GW230529ay_combined_imrphm_low_spin}{5.4}{GW230529ay_imrphen_nsbh}{5.8}{GW230529ay_imrseobv4_nsbh}{5.7}{GW230529ay_imrpv2_low_spin}{5.5}}}

\newcommand{\PEpercentMassGap}[1]{\IfEqCase{#1}{{GW230529ay_combined_imrphm_high_spin}{94}{GW230529ay_combined_imrphm_low_spin}{94}}}
\newcommand{\PEpercentchioneznegative}[1]{\IfEqCase{#1}{{GW230529ay_combined_imrphm_high_spin}{83}{GW230529ay_combined_imrphm_low_spin}{94}}}
\newcommand{\PEpercentchieffnegative}[1]{\IfEqCase{#1}{{GW230529ay_combined_imrphm_high_spin}{93}{GW230529ay_combined_imrphm_low_spin}{94}}}
\newcommand{\PEpercentMassBelowfive}[1]{\IfEqCase{#1}{{GW230529ay_combined_imrphm_high_spin}{99}{GW230529ay_combined_imrphm_low_spin}{100}}}
\newcommand{\PEpercentMassTwoAbovetwo}[1]{\IfEqCase{#1}{{GW230529ay_combined_imrphm_high_spin}{5}{GW230529ay_combined_imrphm_low_spin}{6}}}

\newcommand{\medianEMbrightFracNICER}{0.13}
\newcommand{\lowerEMbrightFracNICER}{-0.11}
\newcommand{\upperEMbrightFracNICER}{0.19}
\newcommand{\upperLimitEMbrightFrac}{0.18}
\newcommand{\upperLimitEMbrightFracOld}{0.06}
\newcommand{\upperLimitTotalEjectaMass}{1.1}

\newcommand{\upperLimitGRBProgenitors}{23}

\newcommand{\medianBHMinimumMassOld}{6.0}
\newcommand{\lowerBHMinimumMassOld}{-3.2}
\newcommand{\upperBHMinimumMassOld}{1.8}
\newcommand{\medianBHMinimumMass}{3.4}
\newcommand{\lowerBHMinimumMass}{-1.2}
\newcommand{\upperBHMinimumMass}{1.0}
\newcommand{\upperLimitMremHighSpin}{0.052}
\newcommand{\upperLimitMremLowSpin}{0.011}
\newcommand{\disruptionProbHighSpin}{0.1}
\newcommand{\disruptionProbLowSpin}{0.042}

\newcommand{\IMRPhenomXPIMRPhenomXASBBH}{0.22}

\newcommand{\MaxLogTenBayesFactorBetweenPopModels}{1.0}

\newcommand{\MMMSDefaultPsrGwHighSpinPrimaryIsNS}{\ensuremath{2.9 \pm 0.4\%}}
\newcommand{\MMMSDefaultPsrGwHighSpinSecondaryIsNS}{\ensuremath{96.1 \pm 0.4\%}}

\newcommand{\MMMSPDBGWTCPsrGWPDBSpinPrimaryIsNS}{\ensuremath{8.8 \pm 2.8\%}}
\newcommand{\MMMSPDBGWTCPsrGWPDBSpinSecondaryIsNS}{\ensuremath{98.4 \pm 1.3\%}}

\newcommand{\MMMSPDBGWTCPsrGWHighSpinPrimaryIsNS}{\ensuremath{27.3 \pm 3.8\%}}

\newcommand{\MMMSPDBAllEventsPDBSpinPrimaryCR}{\ensuremath{2.7_{-0.4}^{+0.9}}}

\newcommand{\mbtaSNR}{11.4}
\newcommand{\mbtaOnlineIFAR}{1.1}
\newcommand{\mbtaOfflineIFAR}{$>$1000}

\newcommand{\pycbcSNR}{11.6}
\newcommand{\pycbcOnlineIFAR}{160.4}
\newcommand{\pycbcOfflineIFAR}{$>$1000}

\newcommand{\gstlalOnlineSNR}{11.3}
\newcommand{\gstlalOnlineIFAR}{1.1}
\newcommand{\gstlalOfflineIFAR}{60.3}

\newcommand{\BGPFracRgapWithoutEvent}{7.5^{+46.4}_{-6.5}}
\newcommand{\BGPFracRgapWithEvent}{33^{+89}_{-29}}

\newcommand{\PDBFracRgapWithoutEvent}{17^{+25}_{-14}}
\newcommand{\PDBFracRgapWithEvent}{24^{+28}_{-16}}

\newcommand{\FGMCRateWithEvent}{94^{+109}_{-64}}

\newcommand{\GWTwoThreeZeroFiveTwoNineKKLrate}{{55}_{-47}^{+127}}

\newcommand{\TotalNSBHKKLrate}{{85}_{-57}^{+116}}

\newcommand{\chirpMassMin}{2.0214}
\newcommand{\chirpMassMax}{2.0331}
\newcommand{\massRatioMin}{0.125}
\newcommand{\massRatioMax}{1}
\newcommand{\distanceMin}{1}
\newcommand{\distanceMax}{500}

\newcommand{\lambdaUpperLimit}{1462}
\newcommand{\skyArea}{$24,100$}

\newcommand{\bilbyarea}{$25,600$}
\newcommand{\bayestararea}{$24,200$}
\newcommand{\probNS}{$>$ 99.9}
\newcommand{\probNSBilby}{98.5}
\newcommand{\probMassGap}{98.5}
\newcommand{\probMassGapBilby}{69.4}
\newcommand{\probRemnant}{12.1}
\newcommand{\probRemnantBilby}{1.1}

\acrodef{LSC}[LSC]{LIGO Scientific Collaboration}
\acrodef{LVC}[LVC]{LIGO Scientific and Virgo Collaboration}
\acrodef{LVK}[LVK]{LIGO Scientific, Virgo, and KAGRA Collaboration}
\acrodef{aLIGO}{Advanced Laser Interferometer Gravitational-Wave Observatory}
\acrodef{aVirgo}{Advanced Virgo}
\acrodef{LIGO}[LIGO]{Laser Interferometer Gravitational-Wave Observatory}
\acrodef{IFO}[IFO]{interferometer}
\acrodef{LHO}[LHO]{LIGO-Hanford}
\acrodef{LLO}[LLO]{LIGO-Livingston}
\acrodef{O1}[O1]{first observing run}
\acrodef{O2}[O2]{second observing run}
\acrodef{O3}[O3]{third observing run}
\acrodef{O3a}[O3a]{first half of the third observing run}
\acrodef{O3b}[O3b]{second half of the third observing run}
\acrodef{O4}[O4]{fourth observing run}
\acrodef{O4a}[O4a]{first half of the fourth observing run}
\acrodef{O4b}[O4b]{second half of the fourth observing run}

\acrodef{BH}[BH]{black hole}
\acrodef{BBH}[BBH]{binary black hole}
\acrodef{BNS}[BNS]{binary neutron star}
\acrodef{IMBH}[IMBH]{intermediate-mass black hole}
\acrodef{NS}[NS]{neutron star}
\acrodef{BHNS}[BHNS]{black hole--neutron star binaries}
\acrodef{NSBH}[NSBH]{neutron star--black hole}
\acrodef{PBH}[PBH]{primordial black hole binaries}
\acrodef{CBC}[CBC]{compact binary coalescence}
\acrodef{GW}[GW]{gravitational-wave}
\acrodef{GWH}[GW]{gravitational-wave}

\acrodef{CWB}[cWB]{coherent WaveBurst}

\acrodef{SNR}[S/N]{signal-to-noise ratio}
\acrodef{FAR}[FAR]{false alarm rate}
\acrodef{IFAR}[IFAR]{inverse false alarm rate}
\acrodef{FAP}[FAP]{false alarm probability}
\acrodef{PSD}[PSD]{power spectral density}

\acrodef{GR}[GR]{general relativity}
\acrodef{NR}[NR]{numerical relativity}
\acrodef{PN}[PN]{post-Newtonian}
\acrodef{EOB}[EOB]{effective-one-body}
\acrodef{ROM}[ROM]{reduced-order model}
\acrodef{IMR}[IMR]{inspiral--merger--ringdown}

\acrodef{PDF}[PDF]{probability density function}
\acrodef{PE}[PE]{parameter estimation}
\acrodef{CL}[CL]{credible level}
\acrodef{GP}[GP]{Gaussian process}

\acrodef{EOS}[EOS]{equation of state}
\acrodef{ISCO}[ISCO]{innermost stable circular orbit}

\acrodef{LAL}[LAL]{LIGO Algorithm Library}

\acrodef{KLD}[KLD]{Kullback--Leibler divergence}
\acrodef{JSD}[JSD]{Jensen--Shannon divergence}

\acrodef{GCN}[GCN]{General Coordinates Network}
\acrodef{EM}[EM]{electromagnetic}
\acrodef{GRB}[GRB]{gamma-ray burst}

\newcommand{\GSTLAL}{\textsc{GstLAL}\xspace}
\newcommand{\BAYESTAR}{\textsc{BAYESTAR}\xspace}

\newcommand{\PYCBC}{\textsc{PyCBC}\xspace}
\newcommand{\MBTA}{\textsc{MBTA}\xspace}

\newcommand{\BAYESWAVE}{\textsc{BayesWave}\xspace}
\newcommand{\BILBY}{\textsc{Bilby}\xspace}

\newcommand{\PBILBY}{\textsc{Parallel \BILBY{}}\xspace}
\newcommand{\ASIMOV}{\textsc{Asimov}\xspace}
\newcommand{\PESUMMARY}{\textsc{PESummary}\xspace}
\newcommand{\GRACEDB}{\textsc{GraceDB}\xspace}
\newcommand{\IDQ}{iDQ\xspace}

\newcommand{\BILBYTGR}{\textsc{Bilby TGR}\xspace}

\newcommand{\NUMPY}{\textsc{NumPy}\xspace}
\newcommand{\SCIPY}{\textsc{SciPy}\xspace}
\newcommand{\PLT}{\textsc{Matplotlib}\xspace}
\newcommand{\SEABORN}{\textsc{seaborn}\xspace}
\newcommand{\GWPY}{\textsc{GWpy}\xspace}
\newcommand{\DYNESTY}{\textsc{Dynesty}\xspace}
\newcommand{\GWPOPULATION}{\textsc{GWPopulation}\xspace}
\newcommand{\PYMC}{\textsc{PyMC}\xspace}
\newcommand{\DMT}{{DMT}\xspace}
\newcommand{\DQR}{{DQR}\xspace}
\newcommand{\DQSEGDB}{{DQSEGDB}\xspace}
\newcommand{\GWDETCHAR}{{gwdetchar}\xspace}
\newcommand{\HVETO}{{hveto}\xspace}
\newcommand{\PYTHONVIRGOTOOLS}{{PythonVirgoTools}\xspace}
\newcommand{\OMICRONSCAN}{{Omicron}\xspace}

\newcommand{\BGP}{\textsc{Binned Gaussian Process}\xspace}
\newcommand{\PDB}{\textsc{Power law + Dip + Break}\xspace}
\newcommand{\NSBHpop}{\textsc{NSBH-pop}\xspace}

\graphicspath{{./}{figures/}}

\begin{document}

\title{Observation of Gravitational Waves from the Coalescence \\of a $2.5\text{--}4.5~\Msun$ Compact Object and a Neutron Star}



\author[0000-0003-4786-2698]{A.~G.~Abac}
\affiliation{Max Planck Institute for Gravitational Physics (Albert Einstein Institute), D-14476 Potsdam, Germany}
\author{R.~Abbott}
\affiliation{LIGO Laboratory, California Institute of Technology, Pasadena, CA 91125, USA}
\author{I.~Abouelfettouh}
\affiliation{LIGO Hanford Observatory, Richland, WA 99352, USA}
\author{F.~Acernese}
\affiliation{Dipartimento di Farmacia, Universit\`a di Salerno, I-84084 Fisciano, Salerno, Italy}
\affiliation{INFN, Sezione di Napoli, I-80126 Napoli, Italy}
\author[0000-0002-8648-0767]{K.~Ackley}
\affiliation{University of Warwick, Coventry CV4 7AL, United Kingdom}
\author{S.~Adhicary}
\affiliation{The Pennsylvania State University, University Park, PA 16802, USA}
\author[0000-0002-4559-8427]{N.~Adhikari}
\affiliation{University of Wisconsin-Milwaukee, Milwaukee, WI 53201, USA}
\author[0000-0002-5731-5076]{R.~X.~Adhikari}
\affiliation{LIGO Laboratory, California Institute of Technology, Pasadena, CA 91125, USA}
\author{V.~K.~Adkins}
\affiliation{Louisiana State University, Baton Rouge, LA 70803, USA}
\author[0000-0002-8735-5554]{D.~Agarwal}
\affiliation{Universit\'e catholique de Louvain, B-1348 Louvain-la-Neuve, Belgium}
\affiliation{Inter-University Centre for Astronomy and Astrophysics, Pune 411007, India}
% ...




\begin{abstract}
We report the observation of a coalescing compact binary with component masses $\massonesourcefivepercent{GW230529ay_combined_imrphm_high_spin} \text{--} \massonesourceninetyfivepercent{GW230529ay_combined_imrphm_high_spin}~\Msun$ and $\masstwosourcefivepercent{GW230529ay_combined_imrphm_high_spin} \text{--} \masstwosourceninetyfivepercent{GW230529ay_combined_imrphm_high_spin}~\Msun$ (all measurements quoted at the 90\% credible level).
The gravitational-wave signal GW230529\_181500 was observed during the fourth observing run of the LIGO--Virgo--KAGRA detector network on 2023 May 29 by the LIGO Livingston observatory.
The primary component of the source has a mass less than $5~\Msun$ at $\PEpercentMassBelowfive{GW230529ay_combined_imrphm_high_spin}\%$ credibility.
We cannot definitively determine from gravitational-wave data alone whether either component of the source is a neutron star or a black hole.
However, given existing estimates of the maximum neutron star mass, we find the most probable interpretation of the source to be the coalescence of a neutron star with a black hole that has a mass between the most massive neutron stars and the least massive black holes observed in the Galaxy.
We provisionally estimate a merger rate density of $\GWTwoThreeZeroFiveTwoNineKKLrate~\perGpcyr$ for compact binary coalescences with properties similar to the source of GW230529\_181500; assuming that the source is a neutron star--black hole merger, GW230529\_181500-like sources may make up the majority of neutron star–black hole coalescences.
The discovery of this system implies an increase in the expected rate of neutron star--black hole mergers with electromagnetic counterparts and provides further evidence for compact objects existing within the purported lower mass gap.

\end{abstract}


\section{Introduction}\label{sec:intro}

In 2023 May, the \ac{O4} of the Advanced LIGO~\citep{LIGOScientific:2014pky}, Advanced Virgo~\citep{VIRGO:2014yos}, and KAGRA~\citep{Somiya:2011np, Aso:2013eba} observatory network commenced following a series of upgrades to increase the sensitivity of the network.
The prior three observing runs opened the field of observational \ac{GW} astronomy, with 90 probable \ac{CBC} candidates reported by the \ac{LVK} at the conclusion of the \acl{O3}~\citep[\acsu{O3};][]{LIGOScientific:2021vkt} and further candidates found by external analyses~\citep[e.g.,][]{Nitz:2021zwj, Mehta:2023zlk, Wadekar:2023gea}.
These included the first observation of merging stellar-mass black holes~\citep{LIGOScientific:2016aoc}, the first observation of a stellar-mass black hole merging with a neutron star~\citep{LIGOScientific:2021qlt}, and the first observation of two merging neutron stars~\citep{LIGOScientific:2017vwq}.
The first observation of two merging neutron stars was also a multimessenger event accompanied by emission across the \ac{EM} spectrum~\citep{LIGOScientific:2017ync, Margutti:2020xbo}.
The continued discovery of \acp{CBC} by the international \ac{GW} detector network in \ac{O4} and beyond promises to reveal new information about the formation pathways of compact binaries and the physics of their evolution.

While \ac{GW} observations have enabled detailed characterization of the population of stellar-mass compact-object binary mergers overall~\citep[e.g.,][]{LIGOScientific:2021duu}, the small number of observed mergers at the low-mass end of the black hole mass spectrum leads to considerable uncertainty in this region of the mass distribution.
Dynamical mass measurements of X-ray binary systems within the Milky Way suggest a paucity of compact objects with masses between $\sim 3$ and $5~\Msun$, and hence a lower mass gap that divides the population of neutron stars (with masses less than $\sim 3~\Msun$; \citealt{Rhoades:1974fn, Kalogera:1996ci}) and stellar-mass black holes (observed to have masses above $5~\Msun$; \citealt{Bailyn:1997xt, Ozel:2010su, Farr:2010tu}).
Although a number of recent observations of noninteracting binary systems~\citep{Thompson:2018ycv, Jayasinghe:2021uqb}, radio pulsar surveys~\citep{Barr:2024wwl}, and \ac{GW} observations~\citep{LIGOScientific:2020zkf, LIGOScientific:2021duu} have found hints of compact objects residing in the lower mass gap, \ac{GW} observations have yet to quantify the extent and potential occupation of this gap~\citep{Fishbach:2020ryj, Farah:2021qom, LIGOScientific:2021duu}.

Here we report on the compact binary merger signal GW230529\_181500, henceforth abbreviated as GW230529, which was detected by the LIGO Livingston observatory on 2023 May 29 at 18:15:00 UTC; all other observatories either were offline or did not have the sensitivity required to observe this signal.
We find that the compact binary source of GW230529 had component masses of $\massonesourcemed{GW230529ay_combined_imrphm_high_spin}^{+\massonesourceplus{GW230529ay_combined_imrphm_high_spin}}_{-\massonesourceminus{GW230529ay_combined_imrphm_high_spin}}~\Msun$ and $\masstwosourcemed{GW230529ay_combined_imrphm_high_spin}^{+\masstwosourceplus{GW230529ay_combined_imrphm_high_spin}}_{-\masstwosourceminus{GW230529ay_combined_imrphm_high_spin}}~\Msun$ (all measurements are reported as symmetric 90\% credible intervals around the median of the marginalized posterior distribution with default uniform priors, unless otherwise specified).
Although we cannot definitively determine the nature of the higher-mass (primary) compact object in the binary system, if we assume that all compact objects with masses below current constraints on the maximum neutron star mass are indeed neutron stars, the most probable interpretation for the source of GW230529 is the coalescence between a $\massonesourcefivepercent{GW230529ay_combined_imrphm_high_spin}\text{--}\massonesourceninetyfivepercent{GW230529ay_combined_imrphm_high_spin}~\Msun$ black hole and a neutron star.
GW230529 provides further evidence that a population of compact objects exists with masses between the heaviest neutron stars and lightest black holes observed in the Milky Way.
Furthermore, if the source of GW230529 was a merger between a neutron star and a black hole, its masses are significantly more symmetric than the \acp{NSBH} previously observed via \acp{GW}~\citep{LIGOScientific:2021vkt}, which increases the expected rate of \acp{NSBH} that may be accompanied by an \ac{EM} counterpart.

We report on the status of the detector network at the time of GW230529 in Section~\ref{sec:detectors} and provide details of the detection in Section~\ref{sec:detection}.
In Section~\ref{sec:source_properties} we present estimates of the source properties, along with a discussion of the inferred masses, spins, tidal effects, and consistency of the signal with general relativity.
We provide updated constraints on merger rates and the inferred properties of the compact binary and \ac{NSBH} populations, as well as population-informed posteriors on source properties, in Section~\ref{sec:rates_populations}.
Section~\ref{sec:source_classification} provides analysis and interpretation of the physical nature of the source components.
Implications for multimessenger astrophysics and the formation of low-mass black holes are discussed in Sections \ref{sec:mma_implications} and \ref{sec:astro_implications}, respectively.
Section~\ref{sec:summary} summarizes our findings.
Data from the analyses in this work are available on Zenodo~\citep{ligo_scientific_collaboration_2024_10845779}.

\section{Observatory Status and Data Quality}\label{sec:detectors}

At the time of GW230529 (2023 May 29 18:15:00.7 UTC), LIGO Livingston was in observing mode; LIGO Hanford was offline, having gone out of observing mode $1.5~\mathrm{hr}$ prior to the detection.
The Virgo observatory was undergoing upgrades and was not operational at the time of the detection.
The KAGRA observatory was in observing mode, but its sensitivity was insufficient to impact the analysis of GW230529.
Hence, only the data from the LIGO Livingston observatory are used in the analysis of GW230529.

LIGO Livingston was observing with stable sensitivity for $\approx 66~\mathrm{hr}$ up to and including the time of the \ac{GW} signal.
At the time of the detection, the sky-averaged \ac{BNS} inspiral range was $\approx 150~\mathrm{Mpc}$.
From the start of \ac{O4} until this event, the LIGO Livingston \ac{BNS} inspiral range~\citep{Chen:2017wpg} varied between $140$--$160~\mathrm{Mpc}$, a $4.5$--$19.4\%$ increase compared to the median \ac{BNS} range in \ac{O3}~\citep{aLIGO:2020wna,LIGOScientific:2021vkt}.
Additional details on upgrades to the LIGO observatories for \ac{O4} can be found in Appendix~\ref{supp:detectors_text}.

The Advanced LIGO observatories are laser interferometers that measure strain~\citep{LIGOScientific:2014pky}.
The observatories are calibrated via photon radiation pressure actuation.
An amplitude-modulated laser beam is directed onto the end test masses, inducing a known change in the arm length from the equilibrium position~\citep{Karki:2016pht,Viets:2017yvy}.
For the strain data used in the GW230529 analysis, the maximum $1\sigma$ bounds on calibration uncertainties at LIGO Livingston were $6\%$ in amplitude and $6.5^\circ$ in phase for the frequency range $20$--$2048~\mathrm{Hz}$.

Detection vetting procedures similar to those of past \ac{GW} candidates~\citep{LIGO:2021ppb,LIGOScientific:2016gtq}, when applied to GW230529, find no evidence that instrumental or environmental artifacts~\citep{Helmling-Cornell:2023wqe,AdvLIGO:2021oxw,Effler:2014zpa} could have caused GW230529.
We find no evidence of transient noise that is likely to impact the recovery of the \ac{GW} signal in the 256~s LIGO Livingston data segment containing the signal.

\section{Detection of GW230529}\label{sec:detection}

GW230529 was initially detected in low latency in data from the LIGO Livingston observatory.
The signal was detected independently by three matched-filter search pipelines: \GSTLAL~\citep{Messick:2016aqy, Sachdev:2019vvd, Hanna:2019ezx, Cannon:2020qnf, ewing2023performance, Tsukada:2023edh}, \MBTA \citep{Adams:2015ulm, Aubin:2020goo} and \PYCBC \citep{Allen:2005fk, Allen:2004gu, DalCanton:2020vpm, Usman:2015kfa, Nitz:2017svb, Davies:2020tsx}.
Although GW230529 occurred when only a single detector was observing, it was detected by all three pipelines with high significance, and stands out from the background distribution of noise triggers.
More details are given in Appendix~\ref{supp:detection_timeline}.

All three pipelines have a similar matched-filter-based approach to identify \ac{GW} candidates but differ in the details of implementation.
Each pipeline begins by performing a matched-filtering analysis on the data from the observatory with a bank of \ac{GW} templates~\citep{sakon2023template, Roy:2017oul, MBTA_technical_note}.
Times of high \ac{SNR} are identified, after which each pipeline performs its own set of signal consistency tests, combines them with the \ac{SNR} to make a single ranking statistic for each candidate \ac{GW} event, and calculates a \ac{FAR} by comparing the ranking statistic of the candidate with the background distribution.
The \ac{FAR} is the expected rate of triggers caused by noise with a ranking statistic greater than or equal to that of the candidate.

For each pipeline, this procedure can be done in two modes: a low-latency or online mode, and an offline mode.
In the online mode, the pipelines only use the background information available at the time of a candidate to estimate its significance, along with low-latency data-quality information.
In the offline mode, pipelines use more background information gathered from subsequent observing times to estimate the significance of candidates and use more refined data-quality information.
The offline mode enables more robust and reproducible results at the cost of greater latency.

In both cases, the background distribution is extrapolated to higher significances.
This enables the inverse \ac{FAR} to be greater than the time for which the background was collected.
Pipelines have differing extrapolation methods and differing methods for calculating the \ac{FAR} of single-detector \ac{GW} candidates, details of which can be found in Appendix~\ref{supp:single_detector_searches}.
These differing methods can cause the \acp{FAR} reported by different pipelines for the same candidate to vary to a significant degree, as seen in the case of GW230529.
However, all three pipelines recover a nearly identical \ac{SNR} for GW230529, as expected for an astrophysical signal with a high match to the search templates.
This, in concert with the fact that the inverse \acp{FAR} from the offline analyses of all pipelines are much greater than the duration of the \ac{O4a}, makes it highly likely that GW230529 is of astrophysical origin.
Table~\ref{table:detection_stats} shows the online \ac{SNR}, online inverse \ac{FAR}, and offline inverse \ac{FAR} for each search pipeline.

\begin{table}
\begin{ruledtabular}
    \caption{
Properties of the detection of GW230529 from each search pipeline.
Significance is measured by the inverse of the \ac{FAR}.
    }
    \label{table:detection_stats}
    \begin{center}
    {\renewcommand{\arraystretch}{1.2}
    \begin{tabular}{lc c c  }
         & GstLAL & MBTA & PyCBC \\
        \hline
        Online \ac{SNR} & \gstlalOnlineSNR & \mbtaSNR & \pycbcSNR \\
        Online inverse FAR (yr) & \gstlalOnlineIFAR & \mbtaOnlineIFAR & \pycbcOnlineIFAR \\
        Offline inverse FAR (yr) & \gstlalOfflineIFAR & \mbtaOfflineIFAR & \pycbcOfflineIFAR \\
    \end{tabular}
    }
    \end{center}
\end{ruledtabular}
\end{table}

\section{Source Properties}\label{sec:source_properties}

The source properties of GW230529 are inferred using a Bayesian analysis of the data from the LIGO Livingston observatory.
We analyze $128~\mathrm{s}$ of data including frequencies in the range of $20\text{--}1792~\mathrm{Hz}$ in the calculation of the likelihood.
To describe the detector noise, we assume a noise \ac{PSD} given by the median estimate provided by \BAYESWAVE~\citep{Littenberg:2014oda, Cornish:2014kda, Cornish:2020dwh}.
We use the \BILBY~\citep{Ashton:2018jfp, Romero-Shaw:2020owr} or \PBILBY~\citep{Smith:2019ucc} inference libraries to generate samples from the posterior distribution of the source parameters using a nested sampling~\citep{Skilling:2006gxv} algorithm, as implemented in the \DYNESTY software package~\citep{Speagle:2019ivv}.

Given the uncertain nature of the compact objects of the source, we analyze GW230529 using a range of waveform models that incorporate a number of key physical effects.
For our primary analysis, we employ \acf{BBH} waveform models that include higher-order multipole moments, the effects of spin-induced orbital precession, and allow for spin magnitudes on both components up to the Kerr limit, but do not include tidal effects on either component.

Systematic errors in inferred source properties due to waveform modeling may be significant for \ac{NSBH} systems~\citep{Huang:2020pba}.
We mitigate these effects by combining the posteriors inferred using two different signal models: the phenomenological frequency-domain model IMRPhenomXPHM~\citep{Pratten:2020fqn, Garcia-Quiros:2020qpx, Pratten:2020ceb}, and a time-domain effective-one-body model SEOBNRv5PHM~\citep{Khalil:2023kep, Pompili:2023tna, Ramos-Buades:2023ehm, vandeMeent:2023ols}.
The posterior samples obtained independently using the two signal models are broadly consistent.
Unless otherwise noted, we present results throughout this work obtained by an equal-weight combination of the posterior samples from both models under the default priors described in Appendix~\ref{supp:waveform_systematics}.
Our measurements of key source parameters for GW230529 are presented in Table~\ref{table:combined_pe}.

\begin{table}
\begin{ruledtabular}
    \caption{
Source properties of GW230529 from the primary combined analysis (\ac{BBH} waveforms, high-spin, default priors).
We report the median values together with the $90\%$ symmetric credible intervals at a reference frequency of $20~\mathrm{Hz}$.
    }
    \label{table:combined_pe}
    \renewcommand{\arraystretch}{1.2}
    \begin{center}
    \begin{tabular}{l c}
        Parameter & Value \\
        \hline
        Primary mass $m_1 / \Msun$ & $\massonesourcemed{GW230529ay_combined_imrphm_high_spin}^{+\massonesourceplus{GW230529ay_combined_imrphm_high_spin}}_{-\massonesourceminus{GW230529ay_combined_imrphm_high_spin}}$ \\
        Secondary mass $m_2 / \Msun $ & $\masstwosourcemed{GW230529ay_combined_imrphm_high_spin}^{+\masstwosourceplus{GW230529ay_combined_imrphm_high_spin}}_{-\masstwosourceminus{GW230529ay_combined_imrphm_high_spin}}$ \\
        Mass ratio $q = m_2 / m_1$ & $\massratiomed{GW230529ay_combined_imrphm_high_spin}^{+\massratioplus{GW230529ay_combined_imrphm_high_spin}}_{-\massratiominus{GW230529ay_combined_imrphm_high_spin}}$ \\
        Total mass $M / \Msun$ & $\totalmasssourcemed{GW230529ay_combined_imrphm_high_spin}^{+\totalmasssourceplus{GW230529ay_combined_imrphm_high_spin}}_{-\totalmasssourceminus{GW230529ay_combined_imrphm_high_spin}}$ \\
        Chirp mass $\mathcal{M} / \Msun$ & $\chirpmasssourcemed{GW230529ay_combined_imrphm_high_spin}^{+\chirpmasssourceplus{GW230529ay_combined_imrphm_high_spin}}_{-\chirpmasssourceminus{GW230529ay_combined_imrphm_high_spin}}$ \\
        Detector-frame chirp mass $(1 + z) \mathcal{M} / \Msun$ & $\chirpmassmed{GW230529ay_combined_imrphm_high_spin}^{+\chirpmassplus{GW230529ay_combined_imrphm_high_spin}}_{-\chirpmassminus{GW230529ay_combined_imrphm_high_spin}}$ \\
        Primary spin magnitude $\chi_1$ & $\aonemed{GW230529ay_combined_imrphm_high_spin}^{+\aoneplus{GW230529ay_combined_imrphm_high_spin}}_{-\aoneminus{GW230529ay_combined_imrphm_high_spin}}$ \\
        Effective inspiral-spin parameter $\chi_{\rm eff}$ & $\chieffmed{GW230529ay_combined_imrphm_high_spin}^{+\chieffplus{GW230529ay_combined_imrphm_high_spin}}_{-\chieffminus{GW230529ay_combined_imrphm_high_spin}}$ \\
        Effective precessing-spin parameter $\chi_{\rm p}$ & $\chipmed{GW230529ay_combined_imrphm_high_spin}^{+\chipplus{GW230529ay_combined_imrphm_high_spin}}_{-\chipminus{GW230529ay_combined_imrphm_high_spin}}$ \\
        Luminosity distance $D_{\rm L} / \rm{Mpc}$ & $\luminositydistancemed{GW230529ay_combined_imrphm_high_spin}^{+\luminositydistanceplus{GW230529ay_combined_imrphm_high_spin}}_{-\luminositydistanceminus{GW230529ay_combined_imrphm_high_spin}}$ \\
        Source redshift $z$ & $\redshiftmed{GW230529ay_combined_imrphm_high_spin}^{+\redshiftplus{GW230529ay_combined_imrphm_high_spin}}_{-\redshiftminus{GW230529ay_combined_imrphm_high_spin}}$ \\
    \end{tabular}
    \end{center}
\end{ruledtabular}
\end{table}

Analysis details and results from other waveform models we consider are reported in Appendix~\ref{supp:waveform_systematics}; we find that the key conclusions of the analyses presented here are not sensitive to the choice of signal model.
In particular, the use of \ac{BBH} models is validated by comparison to waveform models that include tidal effects, finding no evidence that the \ac{BNS} or \ac{NSBH} models are preferred, consistent with previous observations \citep{LIGOScientific:2021qlt}.
This is expected given the moderate \ac{SNR} with which GW230529 was detected~\citep{Huang:2020pba}.

\begin{figure}
    \centering
    \includegraphics[width=\columnwidth]{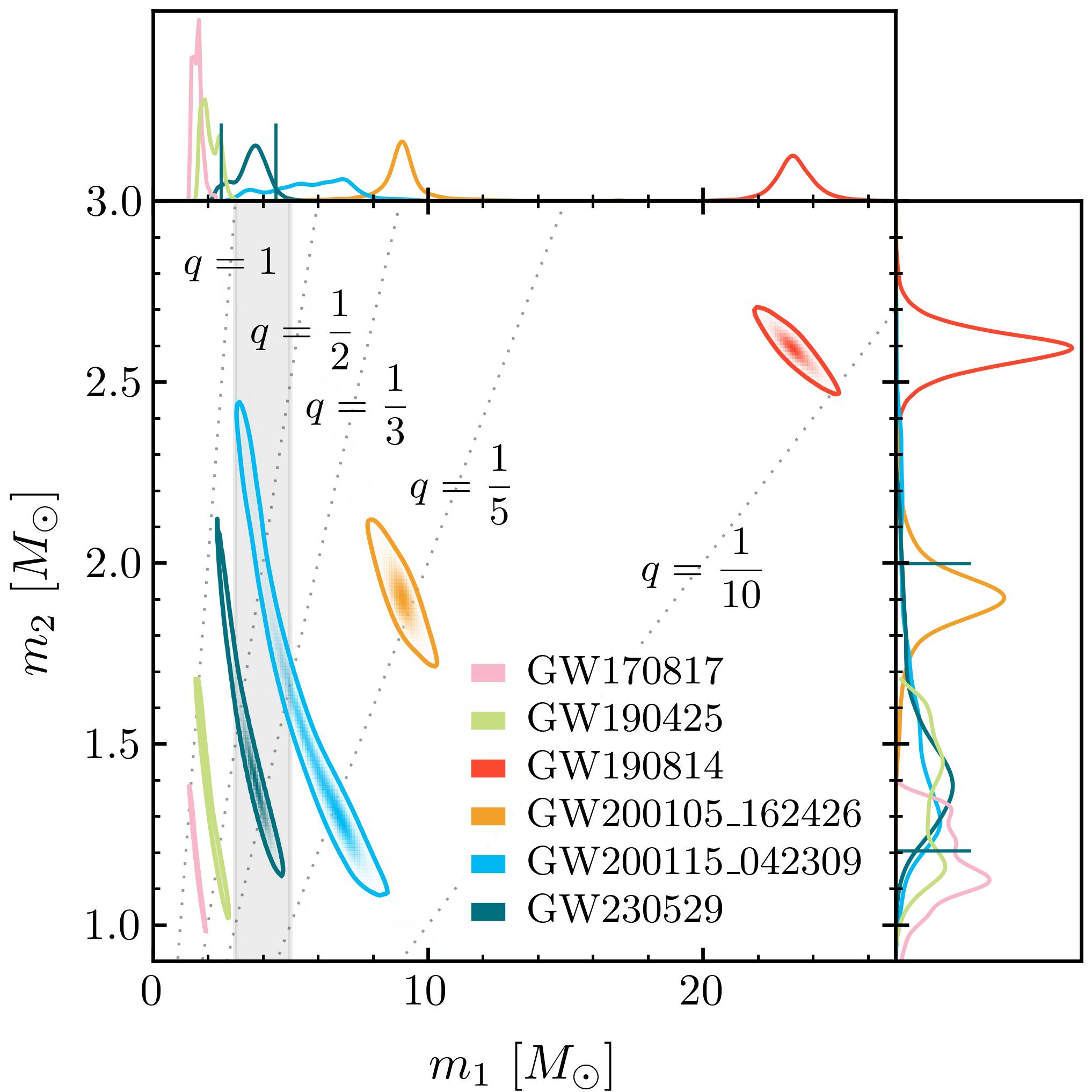}
    \caption{
The one- and two-dimensional posterior probability distributions for the component masses of the source binary of GW230529 (teal).
The contours in the main panel denote the $90\%$ credible regions, with vertical and horizontal lines in the side panels denoting the $90\%$ credible interval for the marginalized one-dimensional posterior distributions.
Also shown are the two \ac{O3} \ac{NSBH} events GW200105\_162426 and GW200115\_042309 (orange and blue, respectively; \citealt{LIGOScientific:2021qlt}) with \ac{FAR} $< 0.25~\text{yr}^{-1}$~\citep{LIGOScientific:2021vkt}, the two confident \ac{BNS} events GW170817 and GW190425 (pink and green, respectively; \citealt{LIGOScientific:2017vwq, LIGOScientific:2018mvr, LIGOScientific:2020aai, LIGOScientific:2021usb}), and GW190814 (red; \citealt{LIGOScientific:2020zkf, LIGOScientific:2021usb}) where the secondary component may be a black hole or a neutron star.
Lines of constant mass ratio are indicated by dotted gray lines.
The gray shaded region marks the $3\text{--}5~\Msun$ range of primary masses.
The \ac{NSBH} events and GW190814 use combined posterior samples assuming a high-spin prior analogous to those presented in this work.
The \ac{BNS} events use high-spin IMRPhenomPv2\_NRTidal \citep{Dietrich:2019kaq} samples.
    }
    \label{fig:ComponentMasses}
\end{figure}

The analysis of GW230529 indicates that it is an asymmetric compact binary with a mass ratio $q = m_2 / m_1 =  \massratiomed{GW230529ay_combined_imrphm_high_spin}^{+ \massratioplus{GW230529ay_combined_imrphm_high_spin}}_{- \massratiominus{GW230529ay_combined_imrphm_high_spin}}$ and source component masses $m_1 = \massonesourcemed{GW230529ay_combined_imrphm_high_spin}^{+\massonesourceplus{GW230529ay_combined_imrphm_high_spin}}_{-\massonesourceminus{GW230529ay_combined_imrphm_high_spin}}~\Msun$ and $m_2 = \masstwosourcemed{GW230529ay_combined_imrphm_high_spin}^{+\masstwosourceplus{GW230529ay_combined_imrphm_high_spin}}_{-\masstwosourceminus{GW230529ay_combined_imrphm_high_spin}}~\Msun$.
The primary is consistent with a black hole that resides in the lower mass gap ($3~\Msun \lesssim m_1 \lesssim 5~\Msun$; \citealt{Ozel:2010su, Farr:2010tu}), with a mass $< 5~\Msun$ at the $\PEpercentMassBelowfive{GW230529ay_combined_imrphm_high_spin}\%$ credible level.
The posterior distribution on the mass of the secondary is peaked around $\sim \masstwosourcemed{GW230529ay_combined_imrphm_high_spin}~\Msun$ with an extended tail beyond $2~\Msun$, such that $P(m_2 > 2 ~\Msun) = \PEpercentMassTwoAbovetwo{GW230529ay_combined_imrphm_high_spin}\%$.
The mass of the secondary is consistent with the distribution of known neutron star masses, including Galactic pulsars~\citep{Antoniadis:2016hxz, Ozel:2016oaf, Alsing:2017bbc, Farrow:2019xnc} and extragalactic \ac{GW} observations~\citep{Landry:2021hvl, LIGOScientific:2021duu}.
Figure~\ref{fig:ComponentMasses} shows the component mass posteriors of GW230529 relative to other \acp{BNS} (GW170817 and GW190425) and \acp{NSBH} (GW200105\_162426 and GW200115\_042309, henceforth abbreviated as GW200105 and GW200115) observed by the \ac{LVK}, as well as GW190814 \citep{LIGOScientific:2017vwq,LIGOScientific:2020aai,LIGOScientific:2020zkf,LIGOScientific:2021qlt}.

\begin{figure}
    \centering
    \includegraphics[width=\columnwidth]{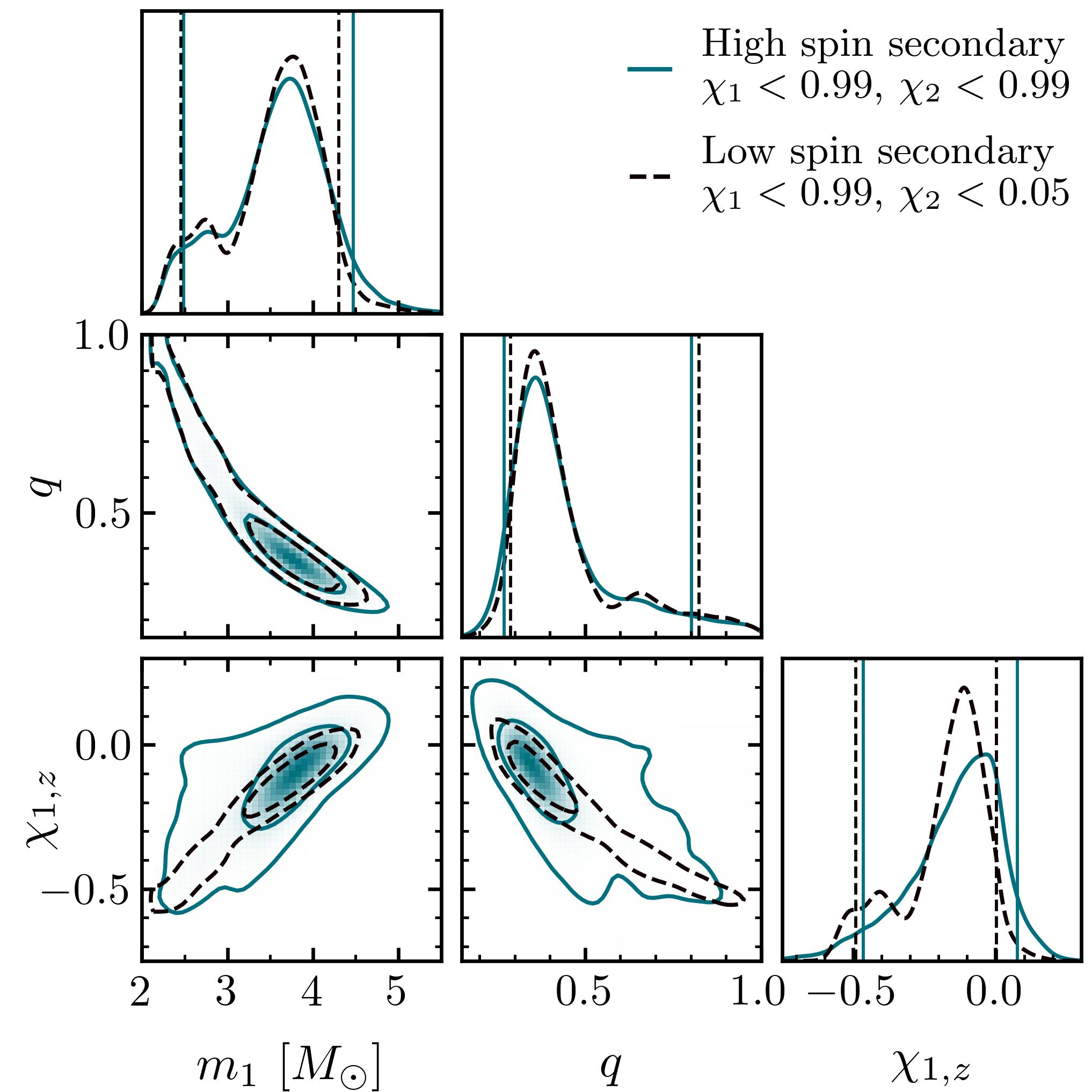}
    \caption{
Selected source properties of GW230529.
The plot shows the one-dimensional (diagonal) and two-dimensional (off-diagonal) marginal posterior distributions for the primary mass $m_1$, the mass ratio $q$, and the spin component parallel to the orbital angular momentum $\chi_{1,z} \equiv {\boldsymbol{\chi}}_{1} \cdot \hat{L}_{\rm N}$.
The shaded regions denote the posterior probability, with the solid (dashed) curves marking the $50\%$ and $90\%$ credible regions for the posteriors determined using a high-spin (low-spin) prior on the secondary of $\chi_2 < 0.99$ ($\chi_2 < 0.05$).
The vertical lines in the one-dimensional marginal posteriors mark the $90\%$ credible intervals.
    }
    \label{fig:SpinProjections}
\end{figure}

To capture dominant spin effects on the \ac{GW} signal, we present constraints on the effective inspiral spin $\chi_{\rm eff}$, which is defined as a mass-weighted projection of the spins along the unit Newtonian orbital angular momentum vector $\hat{L}_{\rm N}$~\citep{Damour:2001tu, Racine:2008qv,  Ajith:2012mn},
\begin{align}
\chi_{\rm eff} &= \left( \frac{m_1}{M} \boldsymbol{\chi}_1 + \frac{m_2}{M} \boldsymbol{\chi}_2 \right) \cdot \hat{L}_{\rm N},
\end{align}
where the dimensionless spin vector $\boldsymbol{\chi}_i$ of each component is related to the spin angular momentum $\boldsymbol{S}_i$ by $\boldsymbol{\chi}_i = c \boldsymbol{S}_i / (G m^2_i)$.
If $\chi_{\rm eff}$ is negative, it indicates that at least one of the spin component projections must be antialigned with respect to the orbital angular momentum, i.e., $\chi_{i,z} \equiv \boldsymbol{\chi}_i \cdot \hat{L}_{\rm N} < 0$.
We measure an effective inspiral spin of $\chi_{\rm eff}$ = $\chieffmed{GW230529ay_combined_imrphm_high_spin}^{+\chieffplus{GW230529ay_combined_imrphm_high_spin}}_{-\chieffminus{GW230529ay_combined_imrphm_high_spin}}$, which is consistent with a binary in which one of the spin components is antialigned or a binary with negligible spins.
The measurement is primarily driven by the spin component of the primary compact object $\chi_{1,z} = \spinonezmed{GW230529ay_combined_imrphm_high_spin}^{+\spinonezplus{GW230529ay_combined_imrphm_high_spin}}_{-\spinonezminus{GW230529ay_combined_imrphm_high_spin}}$, with a probability that $\chi_{1,z} < 0$ of $\PEpercentchioneznegative{GW230529ay_combined_imrphm_high_spin}\%$.
However, there is a degeneracy between the measured masses and spins of the binary components such that more comparable mass ratios correlate to more negative values of $\chi_{\rm eff}$~\citep{Cutler:1994ys} for this system.
We show the correlation between $\chi_{1,z}$ and the mass ratio and primary mass in Figure~\ref{fig:SpinProjections}, with more negative values of $\chi_{1,z}$ corresponding to more symmetric mass ratios and smaller primary masses.
The secondary spin is only weakly constrained and broadly symmetric about $0$, $\chi_{2,z} = \spintwozmed{GW230529ay_combined_imrphm_high_spin}^{+\spintwozplus{GW230529ay_combined_imrphm_high_spin}}_{-\spintwozminus{GW230529ay_combined_imrphm_high_spin}}$.
We find no evidence for precession, with the posteriors on the effective precessing spin $\chi_{\rm p}$ \citep{Schmidt:2014iyl} being uninformative.

The presence of a neutron star in a compact binary imprints tidal effects onto the emitted \ac{GW} signal \citep{Flanagan:2007ix}.
The strength of this interaction is governed by the tidal deformability of the neutron star, which quantifies how easily the star will be deformed in the presence of an external tidal field.
In contrast, the tidal deformability of a black hole is zero~\citep{Binnington:2009bb, Damour:2009vw, Chia:2020yla}, offering a potential avenue for distinguishing between a black hole and a neutron star.
We investigate the tidal constraints for both the primary and secondary components using waveform models that account for tidal effects \citep{Dietrich:2019kaq, Matas:2020wab, Thompson:2020nei}, which do not qualitatively change the mass and spin conclusions discussed above.
Irrespective of whether we analyze GW230529 with a \ac{NSBH} model that assumes only the tidal deformability of the primary compact object to be zero or a \ac{BNS} model that includes the tidal deformability of both objects, we find the tidal deformability of the secondary object to be unconstrained.
The dimensionless tidal deformability of the primary peaks at zero, consistent with a black hole.
The constraints on this parameter are also consistent with dense matter \ac{EOS} predictions for neutron stars in this mass range.

We also perform parameterized tests of the \ac{GW} phase evolution to verify whether GW230529 is consistent with general relativity and find no evidence of inconsistencies.
More detailed information on tidal deformability analyses and testing general relativity can be found in Appendices~\ref{supp:tides} and \ref{supp:tgr}, respectively.

\section{Impact of GW230529 on Merger Rates and Populations}\label{sec:rates_populations}

We provide a provisional update to the \ac{NSBH} merger rate reported in our earlier studies~\citep{LIGOScientific:2021qlt, LIGOScientific:2021duu} by incorporating data from the first 2 weeks of \ac{O4a} using two different methods.
In the first, event-based approach, we consider GW230529 to be representative of a new class of \acp{CBC} and assume its contribution to the total number of \ac{NSBH} detections to be a single Poisson-distributed count~\citep{Kim:2002uw, LIGOScientific:2021qlt} over the span of time from the beginning of the \ac{O1} through the first 2 weeks of \ac{O4a}.
We find the rate of GW230529-like mergers to be $\mathcal{R}_{230529} = \GWTwoThreeZeroFiveTwoNineKKLrate~\perGpcyr$.
When computing the rates of the significant \ac{NSBH} events in \ac{O3} detected with \ac{FAR}$ < 0.25~\mathrm{yr}^{-1}$~\citep{LIGOScientific:2021qlt, LIGOScientific:2021vkt} using the same method, we find a total event-based \ac{NSBH} merger rate of $\mathcal{R}_{\mathrm{NSBH}} = \TotalNSBHKKLrate~\perGpcyr$.
Using this selection criterion, the same as \cite{LIGOScientific:2021duu}, the population of \ac{NSBH} mergers includes GW200105 and GW200115; we do not include GW190814, as its source binary is most probably a \ac{BBH}~\citep{LIGOScientific:2020zkf, LIGOScientific:2021duu, Essick:2020ghc}.

\begin{figure}
    \centering
    \includegraphics[width=\columnwidth]{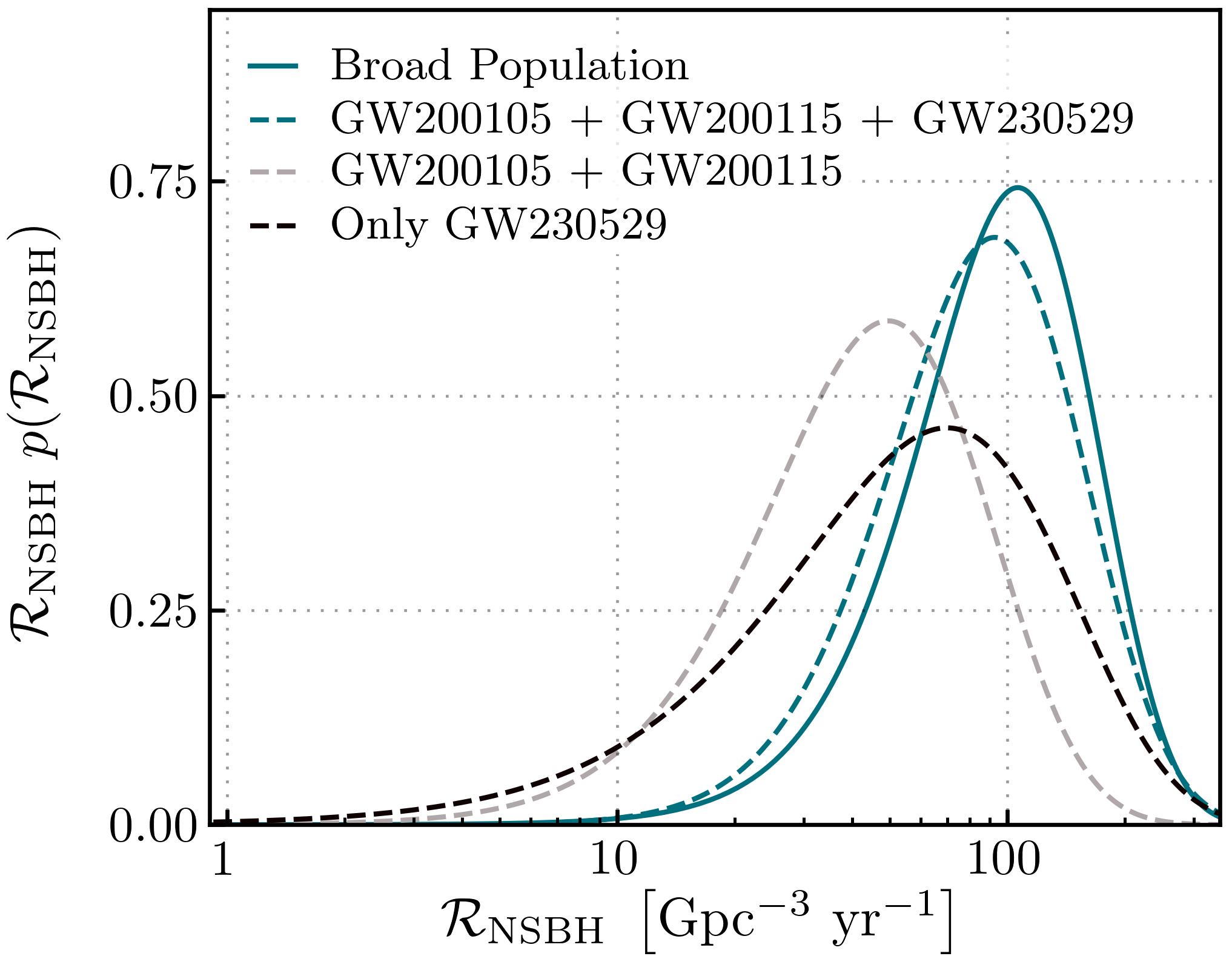}
    \caption{
Posterior on the merger rates of \ac{NSBH} systems.
The solid and dashed lines represent the broad population-based rate calculation and the event-based rate calculation, respectively.
    }
    \label{fig:Rates}
\end{figure}

For the second, broad population-based approach, we consider GW200105, GW200115, and GW230529 to be members of a single \ac{CBC} class, together with an ensemble of less significant candidates.
The definition of this class is determined by a simple cut on masses and spins resulting in an \ac{NSBH}-like region of parameter space.
We aggregate data from all the triggers found by \GSTLAL from the beginning of \ac{O1} through the first 2 weeks of \ac{O4a}, assess their impact on the estimated merger rates of \acp{NSBH} while also accounting for the possibility that some of them are of terrestrial origin~\citep{Farr:2013yna, Kapadia:2019uut}, and update the \ac{NSBH} merger rate obtained at the end of \ac{O3}~\citep{LIGOScientific:2021vkt} to $\mathcal{R}_{\mathrm{NSBH}} = \FGMCRateWithEvent~\perGpcyr$.
Further details regarding the classification of triggers in the population-based rate method can be found in Appendix~\ref{supp:merger_rates}.

\begin{figure*}
    \centering
    \includegraphics[width=\linewidth]{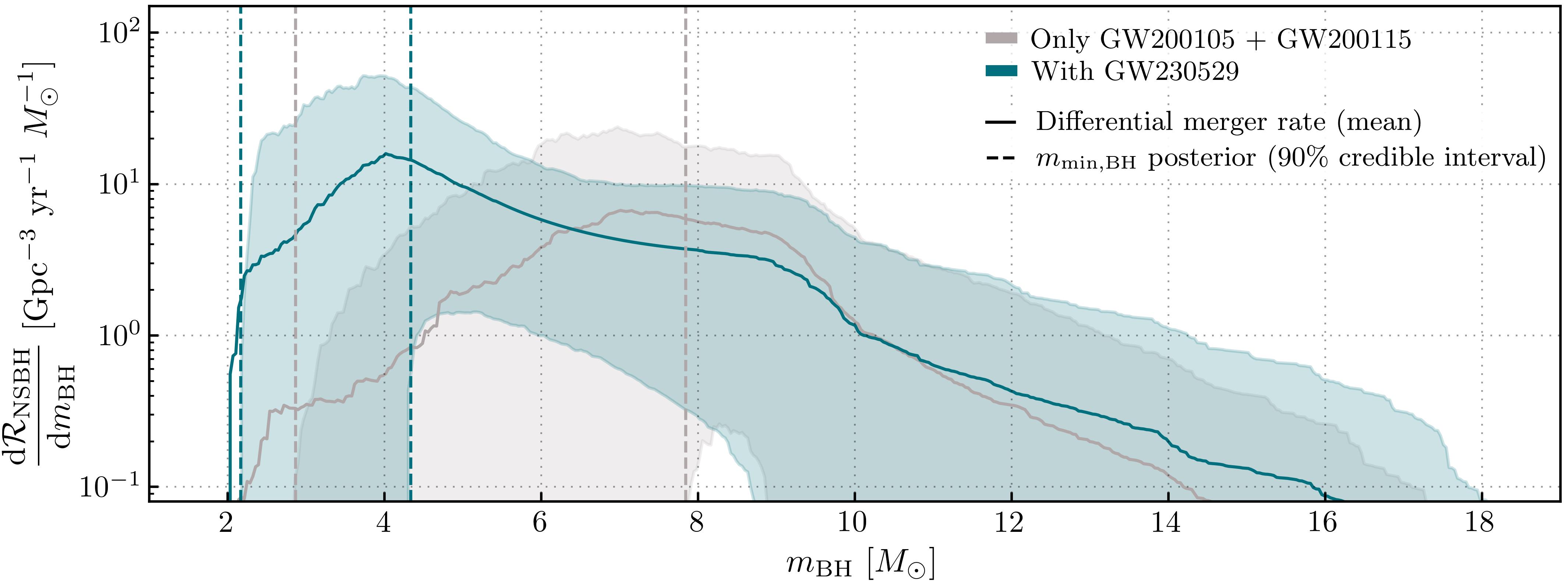}
    \caption{
The differential merger rate of \ac{NSBH} systems as a function of black hole masses (solid curves: mean; shaded regions: 90\% credible interval) and minimum black hole mass (dashed lines: 90\% credible interval) using the \NSBHpop model with (teal) and without (gray) GW230529.
The minimum black hole mass $m_{\mathrm{BH}, \min}$ is a parameter of the \NSBHpop population model; see Table~\ref{table:parameters_nsbhpop}.
While the median rate is always inside the credible region, the mean can be outside the credible region.
    }
    \label{fig:PrimaryMassDistribution}
\end{figure*}

We show updates to both the event-based and population-based \ac{NSBH} rate constraints in Figure~\ref{fig:Rates}.
The event-based rate estimates highlight that GW230529-like systems merge at a similar (or potentially higher) rate to the more asymmetric \acp{NSBH} identified in GWTC-3.
The population-based rate estimate is more representative of the full \ac{NSBH} merger rate, as it includes less significant \ac{GW} events in the data.
Both of our updated estimates are consistent with the findings of \cite{LIGOScientific:2021qlt} within the measurement uncertainties.
Further details about rate estimates can be found in Appendix~\ref{supp:merger_rates}.

In addition to updates to the overall merger rate of \acp{NSBH}, we also study the impact of GW230529 on the mass and spin distributions of the compact binary population as inferred from GWTC-3~\citep{LIGOScientific:2021duu}.
We employ hierarchical Bayesian inference to marginalize over the properties of individual events and infer the parameters of a given population model (e.g., \citealt{Thrane:2018qnx, Mandel:2018mve, Vitale:2020aaz}).
The updates to population model results in this work are provisional because they only include one \ac{GW} signal from \ac{O4a}, although the biases resulting from this selection will not be severe since GW230529 occurred near the start of the observing run.
We quantify this by comparing the event-based rate estimate (which accounts for \ac{O4a} sensitivity to compute its detectable time--volume) with the \ac{NSBH} rates attained by the various population analyses considered in this work, finding that they are consistent with each other.

We use three different population models in our analysis.
The first two models consider the population of compact-object binaries as a whole without distinguishing by source classification, using either the parameterized \PDB model~\citep{Fishbach:2020ryj, Farah:2021qom, LIGOScientific:2021duu} or the nonparametric \BGP model~\citep{Mohite:2022pui, Ray:2023upk, LIGOScientific:2021duu}.
The \PDB model is designed to search for a separation in masses between neutron stars and black holes by explicitly allowing for, but not enforcing, a dip in the component mass distribution.
The \BGP model is designed to capture the structure of the mass distribution with minimal assumptions about the population.
The broad \ac{CBC} population analyses include all candidates reported in GWTC-3 with \ac{FAR} $ < 0.25~\mathrm{yr}^{-1}$, the same selection criterion used in \citet[][see Table 1 therein]{LIGOScientific:2021duu}.
The third model we investigate (\NSBHpop; \citealt{Biscoveanu:2022iue}) considers only the population of \ac{NSBH} mergers with \ac{FAR} $ < 0.25~\mathrm{yr}^{-1}$.
\NSBHpop is a parametric model designed to constrain the population distributions of \ac{NSBH} masses and black hole spin magnitudes.
This model assumes all analyzed events have a black hole primary and neutron star secondary; we do not include GW190814 in this analysis.
Further details regarding the population model parameterizations and priors can be found in Appendix~\ref{supp:population_models}.

The inclusion of GW230529 in our population analyses has several effects on the inferred properties of \ac{NSBH} systems and the \ac{CBC} population as a whole.

\textit{The inferred minimum mass of black holes in the \ac{NSBH} population decreases with the inclusion of GW230529.}
The mass spectrum of black holes in the \ac{NSBH} population with and without GW230529 inferred using the \NSBHpop model is shown in Figure~\ref{fig:PrimaryMassDistribution}.
As the source of GW230529 is the \ac{NSBH} with the smallest black hole mass observed to date, the minimum mass of black holes in \ac{NSBH} mergers shows a significant decrease with the inclusion of this candidate: $m_\mathrm{min, BH} = \medianBHMinimumMass^{+\upperBHMinimumMass}_{\lowerBHMinimumMass}~\Msun$ with GW230529 compared to $m_\mathrm{min, BH} = \medianBHMinimumMassOld^{+\upperBHMinimumMassOld}_{\lowerBHMinimumMassOld}~\Msun$ without.
In contrast, the parameter that governs the lower edge of the dip feature in the \PDB model, which represents the minimum black hole mass in the \ac{CBC} population, does not shift significantly with the inclusion of GW230529.
This difference is because the \PDB model makes no assumptions about the classification of the components and therefore does not enforce sharp features at the edges of each subpopulation, meaning that the source of GW230529 is not necessarily an \ac{NSBH} in this model.
However, the \BGP and \PDB models are designed to capture the structure of the full compact binary mass spectrum; because they do not assign a source classification to either of the binary components, features present in these population models can have differing astrophysical interpretations from the \NSBHpop model.

\begin{figure}
    \centering
    \includegraphics[width=\columnwidth]{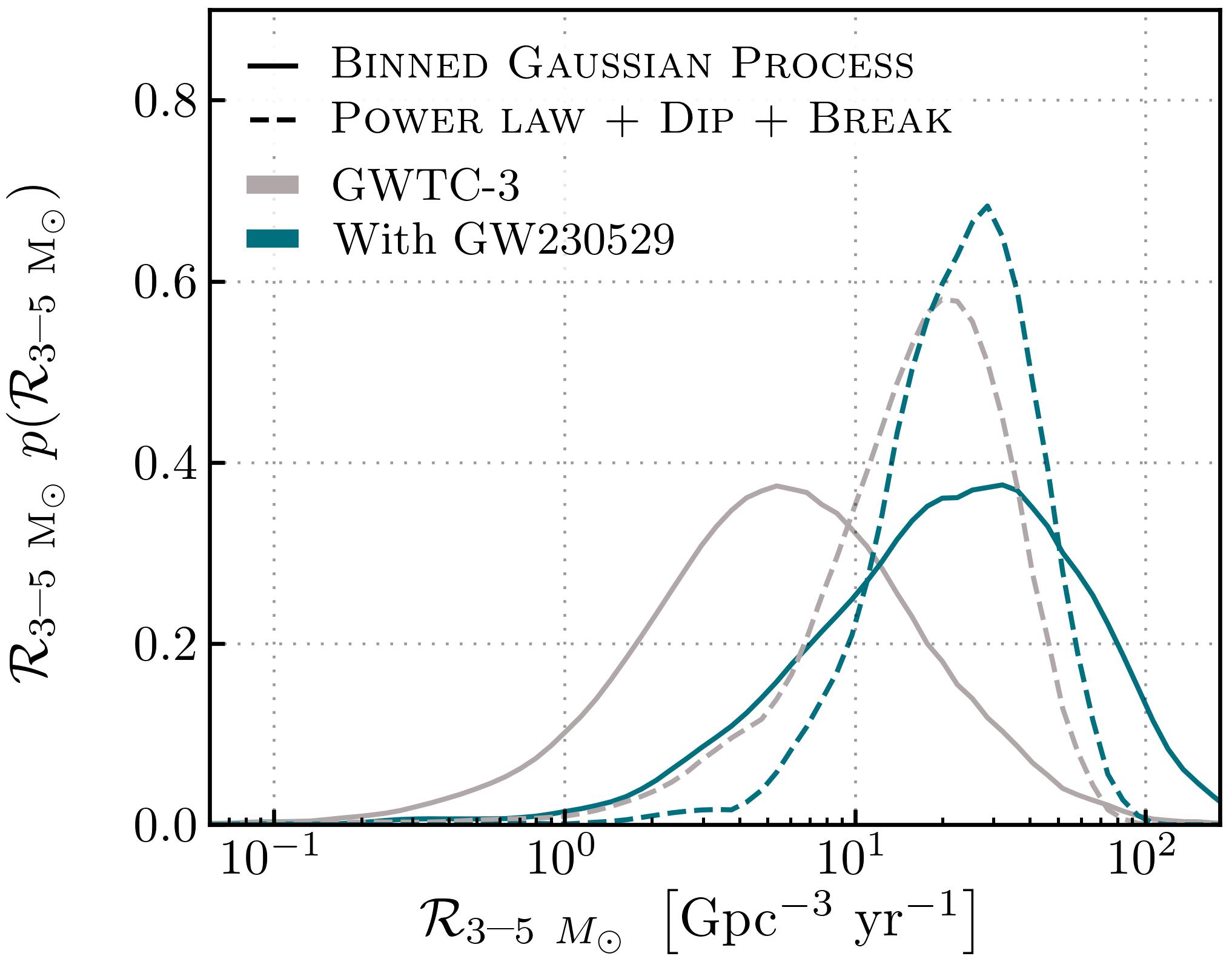}
    \caption{
Posterior on the merger rate of binaries with one or both components between $3$ and $5~\Msun$.
The solid curves show the results from the \BGP analysis, and the dashed curves show the results from the \PDB analysis.
Both models analyze the full black hole and neutron star mass distribution.
The teal and gray curves show the analysis results with and without GW230529, respectively.
    }
    \label{fig:MassGapRate}
\end{figure}

\textit{GW230529 increases the inferred rate of compact binary mergers with a component in the $3\text{--}5~\Msun$ range.}
A region of interest in the mass distribution is the border between the masses of neutron stars and black holes.
We choose $3\text{--}5~\Msun$ to represent the nominal gap between these populations.
In Figure~\ref{fig:MassGapRate} we show the posterior on the rate of mergers with one or both component masses in the $3\text{--}5~\Msun$ range, with and without GW230529.
For the \PDB model there is a small increase in the merger rate within this mass range, $\mathcal{R}_\mathrm{gap} = \PDBFracRgapWithEvent~\perGpcyr$ with GW230529 versus $\mathcal{R}_\mathrm{gap} = \PDBFracRgapWithoutEvent~\perGpcyr$ without.
For the \BGP model there is a larger increase in the merger rate, $\mathcal{R}_\mathrm{gap} = \BGPFracRgapWithEvent~\perGpcyr$ with GW230529 versus $\mathcal{R}_\mathrm{gap} = \BGPFracRgapWithoutEvent~\perGpcyr$ without.
The differing degree of change between the two models is due to different assumptions for the mass ratio distribution of merging compact binaries, as well as the potential dip in the merger rate at low black hole masses built into the \PDB model.
\BGP does not fit for a specific pairing function, while \PDB assumes the same mass ratio distribution throughout the whole population of \acp{CBC}.
As most observed \acp{BBH} favor equal masses, any population model conditioned on those observations should also favor equal masses in the \ac{BBH} mass range.
The implicit assumptions within the \PDB model require this preference to be imposed for all masses, including relatively low mass systems like GW230529.
However, the assumptions in \BGP do not broadcast this preference as strongly across different mass scales and therefore may support more asymmetric mass ratios at lower masses.
Regardless of the mass ratio distribution assumptions made by each model, we find that the inclusion of GW230529 provides further evidence that the $\sim 3\text{--}5~\Msun$ region is not completely empty.

\begin{figure}
    \centering
    \includegraphics[width=\columnwidth]{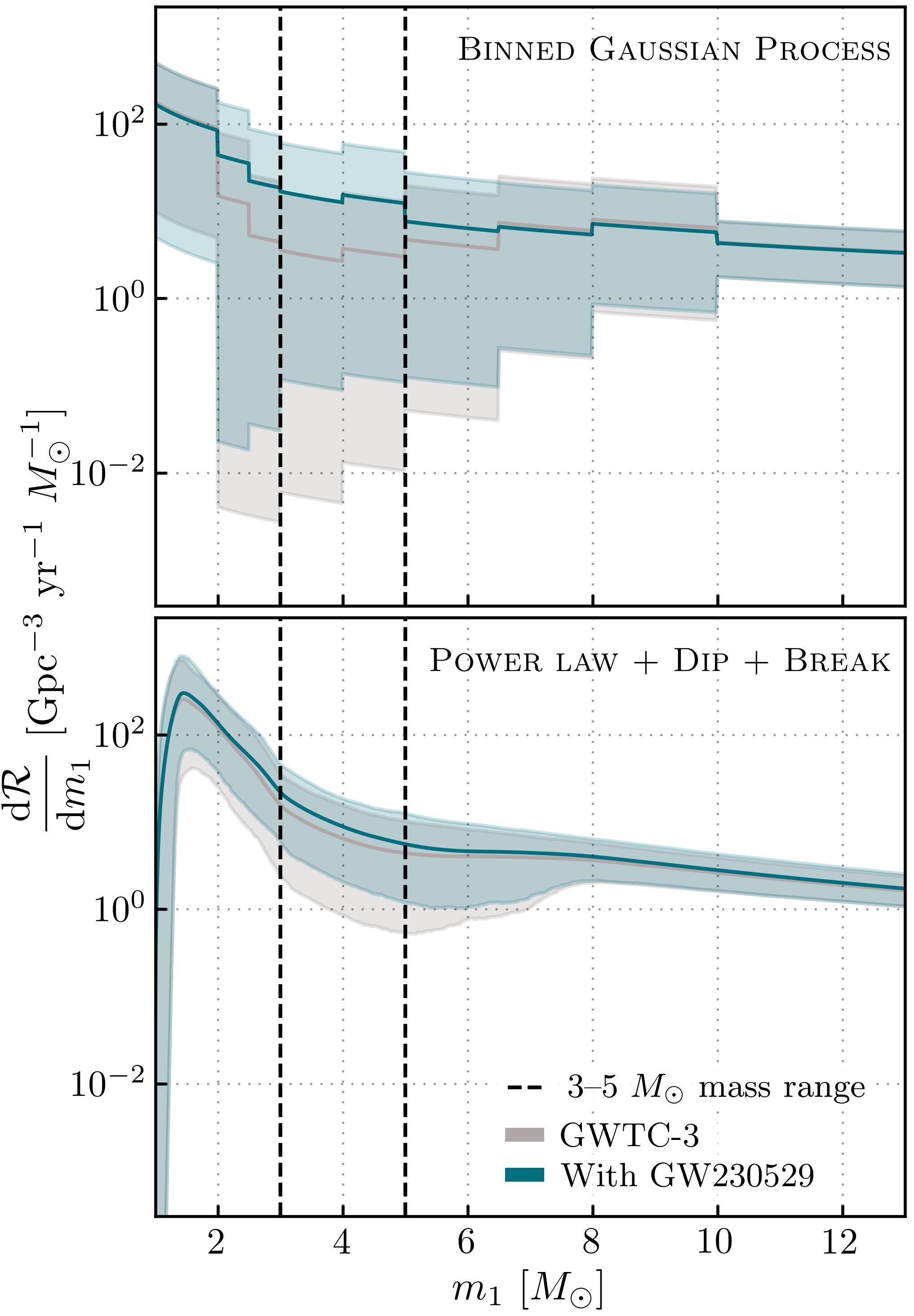}
    \caption{
The differential binary merger rate as a function of the mass of the primary component using the \BGP model (top panel) and the \PDB model (bottom panel) for the full compact binary population (solid curves: mean; shaded regions: 90\% credible interval) with (teal) and without (gray) GW230529.
    }
    \label{fig:PrimaryMassDistributionCBC}
\end{figure}

\textit{GW230529 is consistent with the population inferred from previously observed \ac{CBC} candidates.}
In Figure~\ref{fig:PrimaryMassDistributionCBC}, we show the population distributions of the full black hole and neutron star mass spectrum for the primary component of compact binary mergers using the \BGP (top panel) and \PDB (bottom panel) population models.
We qualitatively see that GW230529 is not an outlier with respect to the masses of previously observed compact-object binaries because the inclusion of GW230529 in the population does not significantly alter the full-population posterior constraints.
This differs from the detection of GW190814, which was an outlier with respect to the rest of the observed \ac{BBH} population at the time due to its small secondary mass~\citep{Essick:2021vlx}.
The observation of GW190814 strongly suggested the region between the masses of neutron stars and black holes was populated~\citep{LIGOScientific:2021duu}, a conclusion that is strengthened with the detection of GW230529 (even though it is not an outlier).

\begin{figure}
    \centering
    \includegraphics[width=\columnwidth]{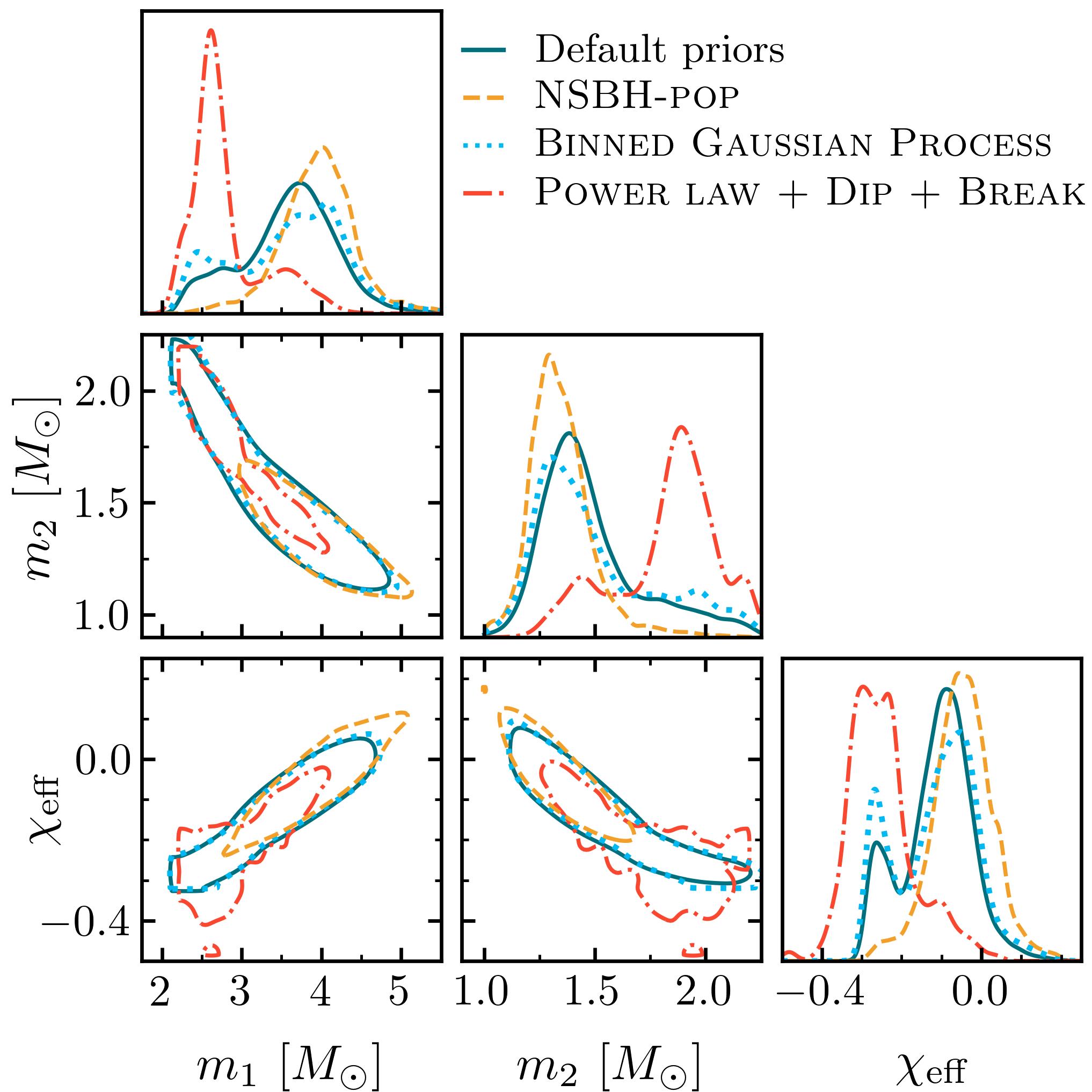}
    \caption{
Posterior distributions of GW230529 under various prior assumptions.
The solid teal curve shows the posterior distributions using the default high-spin priors described in Appendix~\ref{supp:waveform_systematics}.
The various dashed and dotted curves show posteriors obtained using population priors based on the \NSBHpop model (orange), \BGP (blue), and \PDB (red) models.
    }
    \label{fig:PopulationInformedPosteriors}
\end{figure}

\textit{The component masses inferred for GW230529 differ across differing population-informed priors.}
In Figure~\ref{fig:PopulationInformedPosteriors}, we show the mass and spin posteriors obtained using priors informed by each of the population models considered in this work.
The priors informed by the \NSBHpop model prefer unequal mass ratios and small black hole spins; they suppress the extended posterior tail out to equal masses and antialigned spins.
The \BGP model also pulls the posteriors to more asymmetric mass ratios and has less support for antialigned spins.
As shown in the top panel of Figure~\ref{fig:PrimaryMassDistributionCBC}, the merger rate density inferred using the \BGP model is nearly flat across the region of parameter space covered by GW230529, meaning that the priors informed by this population model have less of an impact on the shape of the posterior compared to the parameterized models considered.
Unlike the \NSBHpop model, the \PDB model has a sharp drop in the merger rate above $3~\Msun$.
This sharp feature results in the posterior on the primary component being pulled below $3~\Msun$ and the posterior on the secondary being pulled to higher masses.
The \BGP model has a similar feature, although it is closer to $2~\Msun$ and therefore too low to significantly affect the mass estimates for this system.
The \PDB model also assumes the same preference for equal-mass systems across the entire mass spectrum and thus, given that the majority of the \ac{CBC} population is consistent with symmetric mass ratios, infers that the GW230529 binary components are more similar in mass.
The combination of the drop in the differential merger rate and the preference for symmetric mass ratios increases the support for more extreme spins oriented in the hemisphere opposite the orbital angular momentum.
From these models we find that the binary source of GW230529 either has asymmetric components with low values of $\chi_\mathrm{eff}$ or has components that are similar in mass with $\chi_\mathrm{eff}\sim-0.3$.
These results highlight that the choice of prior has a significant impact on the inferred masses and spins of the GW230529 source and hence the inferred nature of the binary components.
We assess the nature of the components further in Section~\ref{sec:source_classification}.
More details on the population prior reweighting are given in Appendix~\ref{supp:reweighting}.

We assess whether any of our population models are favored as priors for GW230529 by calculating Bayes factors.
Given that the population models consider different sets of \ac{GW} events, Bayesian evidences for each population model are not directly comparable.
However, we may compare the evidence for the single event GW230529 under different models.
In this calculation we only consider the effect of the different shapes of the population models, not their overall normalization over the broader mass parameter space considered by each model.
Thus, we normalize each of the population models over the same range of component masses as the original parameter estimation priors.
We find no significant preference between the population models, with Bayes factors between all three models of $\log_{10}\mathcal{B} \lesssim \MaxLogTenBayesFactorBetweenPopModels$.

\section{Nature of the Compact Objects in the GW230529 Binary}\label{sec:source_classification}

\newcommand{\PrimaryIsNS}{\ensuremath{P(m_1\text{ is NS})}}
\newcommand{\SecondaryIsNS}{\ensuremath{P(m_2\text{ is NS})}}
\newcommand{\BothAreNS}{\ensuremath{P(m_1, m_2\text{ are NSs})}}

\begin{table*}
\begin{ruledtabular}
    \caption{
Probabilistic source classification based on consistency of component masses with the maximum neutron star mass and spin.
All estimates marginalize over uncertainty in the masses, spins, and redshift of the source as well as uncertainty in the astrophysical population and the \ac{EOS}.
We consider three population models: two distributions that use astrophysically agnostic priors and consider either large spins ($\chi_1,\chi_2 \leq 0.99$) or small spins ($\chi_1,\chi_2 \leq 0.05$), and a population prior using the \PDB model fit with only the events from GWTC-3~\citep{LIGOScientific:2021duu}.
We use an \ac{EOS} posterior conditioned on massive pulsars and \ac{GW} observations~\citep{Landry:2020vaw}.
All errors approximate 90\% uncertainty from the finite number of Monte Carlo samples used with the exception of the low-spin results, for which we only place an upper or lower bound.
    }
    \label{table:source_classifications}
    \renewcommand{\arraystretch}{1.4}
    \begin{center}
    \begin{tabular}{c c c c }
        & $\chi_1,\chi_2 \leq 0.99$ & $\chi_1,\chi_2 \leq 0.05$ & \PDB \\
        \hline
        \PrimaryIsNS
          & \MMMSDefaultPsrGwHighSpinPrimaryIsNS
          & $< 0.1\%$ 
          & \MMMSPDBGWTCPsrGWPDBSpinPrimaryIsNS
          \\
        \SecondaryIsNS
          & \MMMSDefaultPsrGwHighSpinSecondaryIsNS
          & $> 99.9\%$ 
          & \MMMSPDBGWTCPsrGWPDBSpinSecondaryIsNS
          \\
    \end{tabular}
    \end{center}
\end{ruledtabular}
\end{table*}

Without clear evidence for or against tidal effects in the signal, the physical nature of the compact objects in the GW230529 source binary can be assessed by comparing the measured masses and spins of each component with the maximum masses and spins of neutron stars allowed by previous observational data.
However, statistical uncertainties in component masses make it especially difficult to determine whether compact objects with masses between $\sim 2.5$ and $5~\Msun$ are consistent with being black holes or neutron stars~\citep{Hannam:2013uu, Littenberg:2015tpa}.
Nevertheless, we assess the nature of the source components by marginalizing over our uncertainties in the masses and spins of GW230529, in the population of merging binaries, and in the supranuclear \ac{EOS}, to compute the posterior probability that each component had a mass and spin less than the maximum mass and spin supported by the \ac{EOS}.
We follow the procedure introduced in the context of GW190814~\citep{LIGOScientific:2020zkf, Essick:2020ghc}, which relied on only the maximum neutron star mass, and has subsequently been extended to include the effects of spin~\citep{LIGOScientific:2020zkf, LIGOScientific:2021qlt, LIGOScientific:2021duu}.
Our analysis provides only an upper limit on the probability that an object is a neutron star, as it assumes that all objects consistent with the maximum mass and spin of a neutron star are indeed neutron stars~\citep{Essick:2020ghc}.

To assess the nature of the component masses of the GW230529 source, we consider two versions of an astrophysically agnostic population model that is uniform in source-frame component masses and spin magnitudes and isotropic in spin orientations: one with component spin magnitudes $\chi_i \equiv |\boldsymbol{\chi_i}|$ that are allowed to be large ($\chi_1,\chi_2 \leq 0.99$) and one where they are restricted \textit{a priori} to be small ($\chi_1,\chi_2 \leq 0.05$).
We also consider the \PDB population model (Appendix~\ref{supp:PDB}) fit to the \ac{GW} candidates from GWTC-3~\citep{LIGOScientific:2021vkt}; including GW230529 in the population model does not significantly affect the results.
For each population, we marginalize over the uncertainty in the maximum neutron star mass conditioned on the existence of massive pulsars and previous \ac{GW} observations~\citep{Landry:2020vaw} using a flexible \ac{GP} representation of the \ac{EOS}~\citep{Landry:2018prl, Essick:2019ldf}.
More information on the \ac{EOS} choices can be found in Appendix~\ref{supp:eos}.

Table~\ref{table:source_classifications} reports the probability that each component of the merger is consistent with a neutron star.
In general, we find that the secondary is almost certainly consistent with a neutron star, and the primary is most probably a black hole.
However, when incorporating information from the \PDB population model, we find that there is a $\sim 1$ in $10$ chance that the primary is consistent with a neutron star.
If we further relax the fixed spin assumptions implicit within the \PDB model for objects with masses $\leq 2.5~\Msun$ from $\chi_i \leq 0.4$ to $\chi_i \leq 0.99$ (see Appendix~\ref{supp:PDB}), we can find probabilities as high as $\PrimaryIsNS = \MMMSPDBGWTCPsrGWHighSpinPrimaryIsNS$.
This ambiguity is similar to the secondary component of GW190814 ($2.50 \leq m_2/\Msun \leq 2.67$), which is consistent with a neutron star if it was rapidly spinning~\citep[e.g.,][]{LIGOScientific:2020zkf, Essick:2020ghc, Most:2020bba}.

The differences observed between population models primarily reflect the uncertainty in the mass ratio and spins of the GW230529 source.
For example, incorporating the \PDB population model as a prior updates the posterior for $m_1$ from $\massonesourcemed{GW230529ay_combined_imrphm_high_spin}^{+\massonesourceplus{GW230529ay_combined_imrphm_high_spin}}_{-\massonesourceminus{GW230529ay_combined_imrphm_high_spin}}~\Msun$ to $\MMMSPDBAllEventsPDBSpinPrimaryCR~\Msun$.
Additional observations of compact objects in or near the lower mass gap may clarify the composition of GW230529 by further constraining the exact shape of the distribution of compact objects below $\sim 5~\Msun$ or the supranuclear \ac{EOS}.

\section{Implications for Multimessenger Astrophysics}\label{sec:mma_implications}

In \ac{NSBH} mergers, the neutron star can either plunge directly into the black hole or be tidally disrupted by its gravitational field.
Tidal disruption would leave some remnant baryonic material outside the black hole that could potentially power a range of \ac{EM} counterparts, including a kilonova~\citep{Lattimer:1974slx, Li:1998bw, Tanaka:2013ana, Tanaka:2013ixa, Fernandez:2016sbf, Kawaguchi:2016ana} or a gamma-ray burst~\citep{Mochkovitch:1993, Janka:1999qu, Paschalidis:2014qra, Shapiro:2017cny, Ruiz:2018wah}.
The conditions for tidal disruption are determined by the mass ratio of the binary, the component of the black hole spin aligned with the orbital angular momentum, and the compactness of the neutron star~\citep{Pannarale:2010vs, Foucart:2012nc, Foucart:2018rjc, Kruger:2020gig}.
While the disruption probability of the neutron star in GW230529 can be inferred based on the binary parameters, we are unlikely to directly observe the disruption in the \ac{GW} signal without next-generation observatories~\citep{Clarke:2023rrm}.

We use the ensemble of fitting formulae collected in \cite{Biscoveanu:2022iue}, including the spin-dependent properties of neutron stars~\citep{Foucart:2018rjc, Cipolletta:2015nga, Breu:2016ufb, Most:2020kyx}, to constrain the remnant baryon mass outside the final black hole following GW230529, assuming it was produced by an \ac{NSBH} merger.
We additionally marginalize over the uncertainty in the \ac{GP}-\ac{EOS} results obtained using the method introduced in Section~\ref{sec:source_classification}~\citep{Legred:2021hdx, legred_isaac_2022_6502467}.
Using the high-spin combined posterior samples obtained with default priors, we find a probability of neutron star tidal disruption of $\disruptionProbHighSpin$, corresponding to an upper limit on the remnant baryon mass produced in the merger of $\upperLimitMremHighSpin~\Msun$ at 99\% credibility.
The low secondary spin priors ($\chi_{2} < 0.05$) yield a tidal disruption probability and remnant baryon mass upper limit of $\disruptionProbLowSpin$ and $\upperLimitMremLowSpin~\Msun$, respectively.
A rapidly spinning neutron star is less compact than a slowly spinning neutron star of the same gravitational mass under the same \ac{EOS}.
This decrease in compactness leads to a larger disruption probability and a larger remnant baryon mass following the merger, explaining the trend we see when comparing the results obtained under the low secondary spin and high-spin priors.
The source binary of GW230529 is the most probable of the confident \acp{NSBH} reported by the \ac{LVK} to have undergone tidal disruption because of the increased symmetry in its component masses.
However, the exact value of the tidal disruption probability and the remnant baryon mass for this system are prior dependent.

\begin{figure}
    \centering
    \includegraphics[width=\columnwidth]{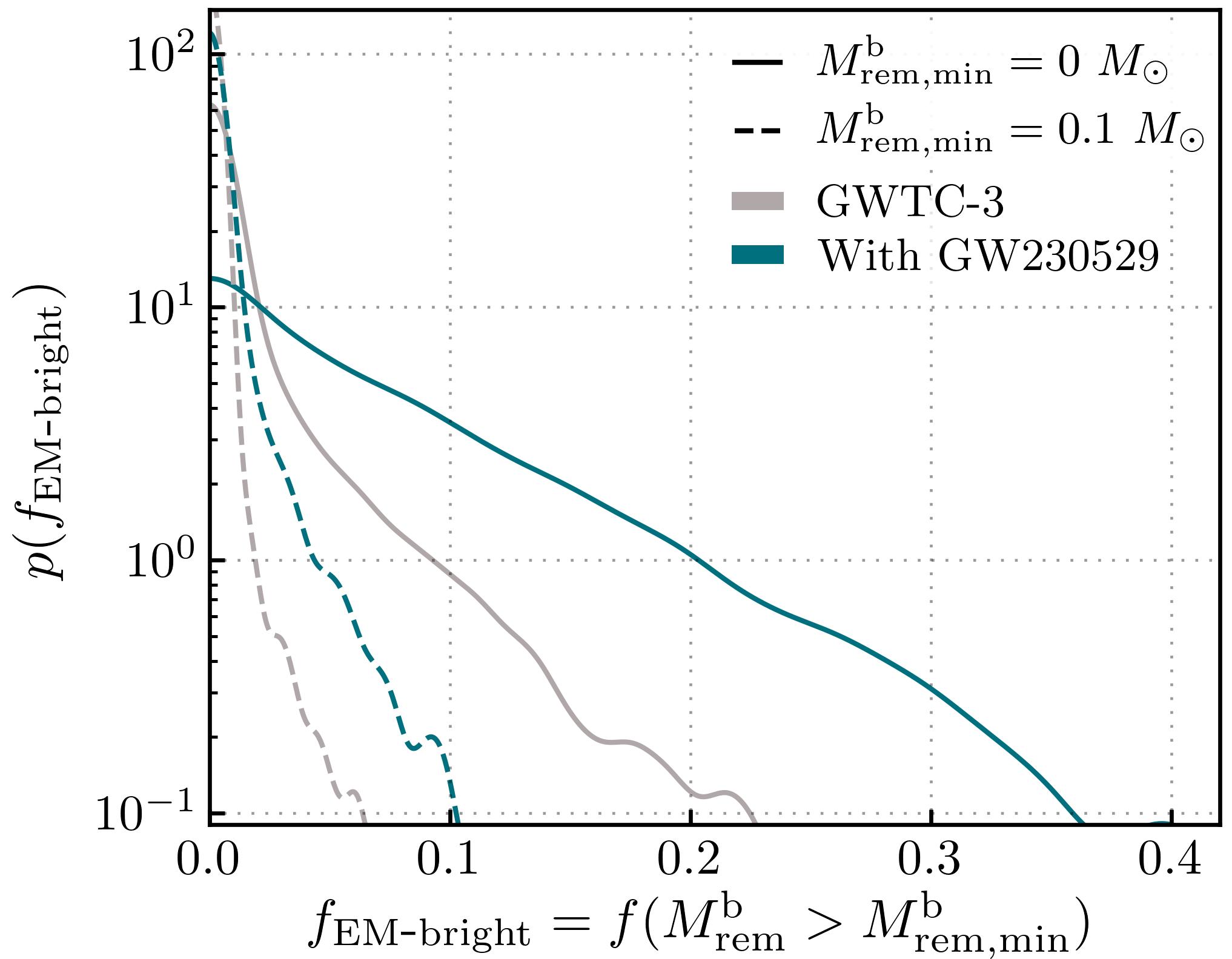}
    \caption{
Posterior on the fraction of \ac{NSBH} systems detected with \acp{GW} that may be \ac{EM} bright, $f_{\mathrm{EM\text{-}bright}}$, depending on the threshold remnant mass required to power a counterpart, $f(M^{\mathrm{b}}_\mathrm{rem} > M^{\mathrm{b}}_\mathrm{rem, min})$.
The solid and dashed curves represent different values of the minimum remnant mass $M^{\mathrm{b}}_\mathrm{rem,min}$.
The teal and gray curves show the analysis results with and without GW230529, respectively.
    }
    \label{fig:EMbrightFraction}
\end{figure}

We can also gauge how the inclusion of GW230529 impacts estimates of the fraction of \ac{NSBH} systems detected in \acp{GW} that may be accompanied by an \ac{EM} counterpart, $f_\mathrm{EM\text{-}bright}$.
Using the mass and spin distributions inferred under the \NSBHpop population model described in Section~\ref{sec:rates_populations}, we find a 90\% credible upper limit on the fraction of \ac{NSBH} mergers that may be \ac{EM}-bright of $f_\mathrm{EM\text{-}bright} \leq \upperLimitEMbrightFrac$ when including GW230529, an increase relative to $f_\mathrm{EM\text{-}bright} \leq \upperLimitEMbrightFracOld$ obtained when excluding GW230529 from the analysis.
When including \ac{GP}-\ac{EOS} constraints additionally conditioned on NICER observations~\citep{Legred:2021hdx, legred_isaac_2022_6502467}, the posterior on the \ac{EM}-bright fraction further increases to peak away from zero, $f_{\mathrm{EM\text{-}bright}} = \medianEMbrightFracNICER^{+\upperEMbrightFracNICER}_{\lowerEMbrightFracNICER}$.
Additional details on analyses with this alternative choice of \ac{EOS} constraint are given in Appendix~\ref{supp:eos}.
These estimates assume that any remnant baryon mass $M^{\mathrm{b}}_\mathrm{rem,min} \geq 0$ could power a counterpart, although the actual threshold value is astrophysically uncertain.
Figure~\ref{fig:EMbrightFraction} shows the posterior for $f_\mathrm{EM\text{--}bright}$ with and without the inclusion of GW230529.
While the exact value of $f_\mathrm{EM\text{-}bright}$ depends on the assumed population model, the increase in $f_\mathrm{EM\text{-}bright}$ for \acp{NSBH} upon the inclusion of GW230529 in the population is robust against modeling assumptions.

Using these updated multimessenger prospects, we can infer the contribution of \ac{NSBH} mergers to the production of heavy elements~\citep{Biscoveanu:2022iue} and the generation of gamma-ray bursts~\citep{Biscoveanu:2023eyp}.
Assuming that all remnant baryon mass produced in \ac{NSBH} mergers is enriched in heavy elements via $r$-process nucleosynthesis~\citep{Lattimer:1974slx, Lattimer:1976kbf}, we infer that \ac{NSBH} mergers contribute at most $\upperLimitTotalEjectaMass~\Msun~\perGpcyr$ to the production of heavy elements and that the rate of gamma-ray bursts with \ac{NSBH} progenitors is at most $\upperLimitGRBProgenitors~\perGpcyr$ at 90\% credibility.
This likely represents a small fraction of all short gamma-ray bursts, for which astrophysical beaming-corrected rate estimates are in the range of $\mathcal{O}(10\text{--}1000)~\perGpcyr$~\citep{Mandel:2021smh}, and of the total $r$-process material produced by compact-object mergers~\citep{Cote:2017evr, Chen:2021fro}.

No significant counterpart candidates have been reported for GW230529~\citep{IceCube:2023, Karambelkar:2023, Lipunov:2023, Longo:2023, Lesage:2023, Savchenko:2023, Sugita:2023a, Sugita:2023b, Waratkar:2023}.
This is unsurprising given that it was only observed by a single detector and hence was poorly localized on the sky.

\section{Astrophysical Implications}\label{sec:astro_implications}

Since the late 1990s, there have been claims about the existence of a mass gap between the maximum neutron star mass and the minimum black hole mass ($\sim 3$--$5~\Msun$) based on dynamical mass measurements of Galactic X-ray binaries~\citep{Bailyn:1997xt, Ozel:2010su, Farr:2010tu}.
The lower edge of this purported mass gap depends on the maximum possible mass with which a neutron star can form in a supernova explosion, which cannot exceed the maximum allowed neutron star mass given by the \ac{EOS}.
Some \acp{EOS} support masses up to $\sim{3}~\Msun$ for nonrotating neutron stars~\citep{Mueller:1996pm, Godzieba:2020tjn} and even larger masses for rotating neutron stars~\citep{1987ApJ...314..594F, Cook:1993qr}, although such large values are disfavored by the tidal deformability inferred for GW170817~\citep{LIGOScientific:2018hze} and Galactic observations~\citep{Alsing:2017bbc, 2020RNAAS...4...65F}.
The upper edge and extent of the mass gap depend on the minimum black hole mass that can form from stellar core collapse.
However, it remains an open question whether observational or evolutionary selection effects inherent to the detection of Galactic X-ray binaries can lead to the observed gap in compact-object masses~\citep{Fryer:1999ht, Kreidberg:2012ud, Siegel:2022gwc}.

In recent years, new \ac{EM} observations have unveiled a few candidates in the $\sim{3}~\Msun$ region, mostly from noninteracting binary systems~\citep{Thompson:2018ycv, 2020Sci...368.3282V, Thompson:2020nbd} and radio surveys for pulsar binary systems~\citep{Barr:2024wwl}.
Microlensing surveys do not support the existence of a mass gap but cannot exclude it either~\citep{Wyrzykowski:2015ppa, Wyrzykowski:2019jyg}.
The \ac{LVK} has already observed one component of a merger whose mass falls between the most massive neutron stars and least massive black holes observed in the Galaxy: the secondary component of GW190814 ($2.5\text{--}2.7~\Msun$ at the 90\% credible level; \citealt{LIGOScientific:2020zkf, LIGOScientific:2021usb}).
The secondary components of the GW200210\_092254 and GW190917\_114630 source binaries also have support in the lower mass gap~\citep{LIGOScientific:2021usb,LIGOScientific:2021vkt}, although their mass estimates are also consistent with a high-mass neutron star.
Unlike GW230529, the primary components of all three of these binaries can be confidently identified as black holes.
Overall, the existence of a mass gap between the most massive neutron stars and least massive black holes still stands as an open question in astrophysics.

GW230529 is the first compact binary with a primary component that has a high probability of residing in the lower mass gap.
Hence, GW230529 reinforces the conclusion that the $3$--$5~\Msun$ range is not completely empty (see Figure~\ref{fig:MassGapRate} in Section~\ref{sec:rates_populations}).
However, the $3$--$5~\Msun$ range may still be less populated than the surrounding regions of the mass spectrum.
This conclusion is consistent with previous population analyses and rate estimates based on GWTC-3~\citep{LIGOScientific:2021duu}.

The formation of GW230529 raises a number of questions.
Given our current understanding of core collapse in massive stars~\citep{OConnor:2010moj, Janka:2012wk, Ertl:2019zks}, it is unlikely that the primary component formed via direct collapse because of its low mass, although stochasticity in the physical mechanisms that determine remnant mass may populate the low-mass end of the black hole mass spectrum~\citep{Mandel:2020qwb, Antoniadis:2021dhe}.
Formation by fallback is a viable scenario: recent numerical models of core-collapse supernovae suggest that the formation of $3$--$6~\Msun$ black holes via substantial fallback is rare but still possible \citep[e.g.,][]{Sukhbold:2015wba, Ertl:2019zks, Vigna-Gomez:2021oqy}.
One-dimensional hydrodynamical simulations of core collapse adopting pure helium star models predict that there is no empty gap, only a less populated region between $3$ and $5~\Msun$, with the lowest mass of black holes produced by prompt implosions starting at $\approx{6}~\Msun$~\citep{Ertl:2019zks}.
Another relevant parameter is the timescale for instability growth and launch of the core-collapse supernova; if this is long enough ($\gtrsim~200~\mathrm{ms}$), the proto-neutron star might accrete enough mass before the explosion to become a mass-gap object~\citep{Fryer:1999ht, Fryer:2011cx}.
Compact binary population synthesis models that allow for longer instability growth timescales naturally produce merging \ac{NSBH} systems with the primary component in the lower mass gap~\citep{Belczynski:2011bn, Chattopadhyay:2020lff, Zevin:2020gma, Broekgaarden:2021iew, Olejak:2022zee}.
It may also be possible to form mass-gap objects through accretion onto a neutron star.
If the first-born neutron star in the binary accretes enough material prior to the formation of the second-born compact object, it can trigger an accretion-induced collapse into a black hole, yielding a mass-gap object~\citep{Siegel:2022gwc}.
This scenario may be aided by super-Eddington accretion onto the first-born neutron star, which has been proposed to explain \acp{NSBH} with mass-gap black holes like the source of GW230529 \citep{Zhu:2023nhy}.
Considering the large number of uncertainties about the outcome of core-collapse supernovae \citep[e.g.,][]{Burrows:2020qrp}, the primary mass of GW230529 provides a piece of evidence to inform and constrain future models.
Overall, astrophysical models in the past have preferentially adopted prescriptions that enforce the presence of a lower mass gap \citep[e.g.,][]{Fryer:2011cx}, and the inferred rate of GW230529-like systems urges a change of paradigm in such model assumptions.

Alternatively, the primary component of GW230529 might be the result of the merger of two neutron stars.
For instance, the 90\% credible intervals for the remnant mass of GW190425 and the primary mass of GW230529 overlap~\citep{LIGOScientific:2020aai}.
This may hint at a scenario where the primary component could be either the member of a former triple or quadruple system~\citep{Fragione:2020aki, Lu:2020gfh, Vynatheya:2021mgl, Gayathri:2023met}, or the result of a dynamical capture in a star cluster~\citep{Clausen:2012zu, Gupta:2019nwj, Rastello:2020sru, ArcaSedda:2020wzl, ArcaSedda:2021zmm} or an active galactic nucleus disk~\citep{Tagawa:2020qll, Yang:2020xyi}.
This scenario was proposed for the formation of a compact object discovered in a binary with a pulsar in the globular cluster NGC 1851, which is measured to have a mass of $2.09\text{--}2.71~\Msun$ at 95\% confidence~\citep{Barr:2024wwl}.
However, dynamically induced \ac{BNS} mergers are predicted to be rare in dense stellar environments ($\mathcal{O}(10^{-2})~\perGpcyr$; e.g., \citealt{Ye:2019xvf, Samsing:2020qqd}).
Thus, the rate of mergers between a \ac{BNS} merger remnant and a neutron star may be at least several orders of magnitude lower than the rate we infer for GW230529-like mergers, making this scenario improbable.

Non-stellar-origin black hole formation scenarios such as primordial black holes \citep[e.g.,][]{Clesse:2020ghq} remain a possibility.
However, there are significant uncertainties in the predicted mass spectrum and merger rate of primordial black hole binaries, thus making it difficult to attribute a primordial origin to compact objects that have masses consistent with predictions from massive-star core collapse.
Furthermore, results from microlensing surveys indicate that primordial black hole mergers cannot be a dominant source of \acp{GW} in the local universe~\citep{Mroz:2024mse}.

It has also been suggested that mergers apparently involving mass-gap objects could instead be gravitationally lensed \acp{BNS}~\citep{Bianconi:2022etr, Magare:2023hgs}, with the lensing magnification making them appear heavier and closer~\citep{Wang:1996as,Dai:2016igl,Hannuksela:2019kle}.
This scenario is difficult to explicitly test in the absence of tidal information~\citep{Pang:2020qow} or \ac{EM} counterparts~\citep{Bianconi:2022etr}, but the expected relative rate of strong lensing is low at current detector sensitivities~\citep{Smith:2022vbp, Magare:2023hgs, LIGOScientific:2023bwz}.

Finally, we also find mild support for the possibility that the primary component is a neutron star rather than a black hole when considering the population-based \PDB prior that incorporates the potential presence of a gap at low black hole masses.
In this case, GW230529 would be the most massive neutron star binary yet observed, with both components $\gtrsim 2.0~\Msun$, and have non zero spins that are anti aligned with the orbital angular momentum.
The effective inspiral spin of the \acp{BNS} in this scenario would differ significantly from \ac{BNS} sources previously observed in \acp{GW}~\citep{LIGOScientific:2017vwq, LIGOScientific:2020aai} as well as those inferred for Galactic \acp{BNS} if they were to merge~\citep{Zhu:2017znf}.
Spins oriented in the hemisphere opposite the binary orbital angular momentum could be the result of supernova natal kicks~\citep{Kalogera:1999tq, Farr:2011gs, OShaughnessy:2017eks, Chan:2020lnd} or spin-axis tossing~\citep{Tauris:2022ggv} at birth.
For example, one of the neutron stars in the binary pulsar system J0737$\text{--}$3039 is significantly misaligned with respect to both the spin of its companion and the orbital angular momentum of the binary system, which may require off-axis kicks~\citep{Farr:2011gs}.
Alternatively, isotropically oriented component spins that result from random pairing in dynamical environments could lead to the significant spin misalignment \citep[e.g.,][]{Rodriguez:2016vmx}.

\section{Summary}\label{sec:summary}

GW230529 is a \ac{GW} signal from the coalescence of a $\massonesourcefivepercent{GW230529ay_combined_imrphm_high_spin}$--$\massonesourceninetyfivepercent{GW230529ay_combined_imrphm_high_spin}~\Msun$ compact object and a compact object consistent with neutron star masses.
The more massive component in the merger provides evidence that compact objects in the hypothesized lower mass gap exist in merging binaries.
Based on mass estimates of the two components in the merger and current constraints on the supranuclear \ac{EOS}, we find the most probable interpretation of the GW230529 source to be an \ac{NSBH} coalescence.
In this scenario, the source binary of GW230529 is the most symmetric-mass \ac{NSBH} merger yet observed, and the primary component of the merger is the lowest-mass primary black hole observed in \acp{GW} to date.
Because \acp{NSBH} with more symmetric masses are more susceptible to tidal disruption, the observation of GW230529 implies that more \acp{NSBH} than previously inferred may produce \ac{EM} counterparts.
However, we cannot rule out the contrasting scenario that the source of GW230529 consisted of two neutron stars rather than a neutron star and a black hole.
In this case, the source of GW230529 would be the only \ac{BNS} coalescence observed to have strong support for nonzero and antialigned spins, as well as the highest-mass \ac{BNS} system observed to date.
Regardless of the true nature of the GW230529 source, it is a novel addition to the growing population of \acp{CBC} observed via their \ac{GW} emission and highlights the importance of continued exploration of the \ac{CBC} parameter space in \ac{O4} and beyond.

Strain data containing the signal from the LIGO Livingston observatory are available from the Gravitational Wave Open Science Center.\footnote{\url{https://doi.org/10.7935/6k89-7q62}}
Specifically, we release the \texttt{L1:GDS-CALIB\_STRAIN\_CLEAN\_AR} channel, where the \texttt{AR} designation means that the strain data are analysis ready; this strain channel was also released at the end of \ac{O3}.
Samples from the posterior distributions of the source parameters, hyperposterior distributions from population analyses, and notebooks for reproducing all results and figures in this paper are available on Zenodo~\citep{ligo_scientific_collaboration_2024_10845779}.
The software packages used in our analyses are open-source.

\vspace{7pt}
%
This material is based on work supported by NSF’s LIGO Laboratory, which is a major facility
fully funded by the National Science Foundation.
The authors also gratefully acknowledge the support of
the Science and Technology Facilities Council (STFC) of the
United Kingdom, the Max-Planck-Society (MPS), and the State of
Niedersachsen/Germany for support of the construction of Advanced LIGO 
and construction and operation of the GEO\,600 detector. 
Additional support for Advanced LIGO was provided by the Australian Research Council.
The authors gratefully acknowledge the Italian Istituto Nazionale di Fisica Nucleare (INFN),  
the French Centre National de la Recherche Scientifique (CNRS), and
the Netherlands Organization for Scientific Research (NWO) 
for the construction and operation of the Virgo detector
and the creation and support  of the EGO consortium. 
The authors also gratefully acknowledge research support from these agencies, as well as by 
the Council of Scientific and Industrial Research of India, 
the Department of Science and Technology, India,
the Science \& Engineering Research Board (SERB), India,
the Ministry of Human Resource Development, India,
the Spanish Agencia Estatal de Investigaci\'on (AEI),
the Spanish Ministerio de Ciencia, Innovaci\'on y Universidades,
the European Union NextGenerationEU/PRTR (PRTR-C17.I1),
the ICSC - CentroNazionale di Ricerca in High Performance Computing, Big Data
and Quantum Computing, funded by the European Union NextGenerationEU,
the Comunitat Auton\`oma de les Illes Balears through the Direcci\'o General de Recerca, Innovaci\'o i Transformaci\'o Digital with funds from the Tourist Stay Tax Law ITS 2017-006,
the Conselleria d'Economia, Hisenda i Innovaci\'o, the FEDER Operational Program 2021-2027 of the Balearic Islands,
the Conselleria d'Innovaci\'o, Universitats, Ci\`encia i Societat Digital de la Generalitat Valenciana and
the CERCA Programme Generalitat de Catalunya, Spain,
the National Science Centre of Poland and the European Union – European Regional Development Fund; Foundation for Polish Science (FNP),
the Polish Ministry of Science and Higher Education,
the Swiss National Science Foundation (SNSF),
the Russian Science Foundation,
the European Commission,
the European Social Funds (ESF),
the European Regional Development Funds (ERDF),
the Royal Society, 
the Scottish Funding Council, 
the Scottish Universities Physics Alliance, 
the Hungarian Scientific Research Fund (OTKA),
the French Lyon Institute of Origins (LIO),
the Belgian Fonds de la Recherche Scientifique (FRS-FNRS), 
Actions de Recherche Concertées (ARC) and
Fonds Wetenschappelijk Onderzoek – Vlaanderen (FWO), Belgium,
the Paris \^{I}le-de-France Region, 
the National Research, Development and Innovation Office Hungary (NKFIH), 
the National Research Foundation of Korea,
the Natural Science and Engineering Research Council Canada,
Canadian Foundation for Innovation (CFI),
the Brazilian Ministry of Science, Technology, and Innovations,
the International Center for Theoretical Physics South American Institute for Fundamental Research (ICTP-SAIFR), 
the Research Grants Council of Hong Kong,
the National Natural Science Foundation of China (NSFC),
the Leverhulme Trust, 
the Research Corporation,
the National Science and Technology Council (NSTC), Taiwan,
the United States Department of Energy,
and
the Kavli Foundation.
The authors gratefully acknowledge the support of the NSF, STFC, INFN, and CNRS for provision of computational resources.
This work was supported by MEXT, JSPS Leading-edge Research Infrastructure Program, JSPS Grant-in-Aid for Specially Promoted Research 26000005, JSPS Grant-in-Aid for Scientific Research on Innovative Areas 2905: JP17H06358, JP17H06361 and JP17H06364, JSPS Core-to-Core Program A. Advanced Research Networks, JSPS Grant-in-Aid for Scientific Research (S) 17H06133 and 20H05639, JSPS Grant-in-Aid for Transformative Research Areas (A) 20A203: JP20H05854, the joint research program of the Institute for Cosmic Ray Research, University of Tokyo, National Research Foundation (NRF), Computing Infrastructure Project of Global Science experimental Data hub Center (GSDC) at KISTI, Korea Astronomy and Space Science Institute (KASI), and Ministry of Science and ICT (MSIT) in Korea, Academia Sinica (AS), AS Grid Center (ASGC) and the National Science and Technology Council (NSTC) in Taiwan under grants including the Rising Star Program and Science Vanguard Research Program, Advanced Technology Center (ATC) of NAOJ, and Mechanical Engineering Center of KEK.
We thank the anonymous journal referee for helpful comments.

\software{
Calibration of the LIGO strain data was performed with a \GSTLAL{}-based calibration software pipeline~\citep{Viets:2017yvy}.
Data-quality products and event-validation results were computed using the \DMT{}~\citep{DMTdocumentation}, \DQR{}~\citep{DQRdocumentation}, \DQSEGDB{}~\citep{Fisher:2020pnr}, \GWDETCHAR{}~\citep{gwdetchar-software}, \HVETO{}~\citep{Smith:2011an}, \IDQ{}~\citep{Essick:2020qpo}, \OMICRONSCAN{}~\citep{Robinet:2020lbf}, and \PYTHONVIRGOTOOLS{}~\citep{pythonvirgotools} software packages and contributing software tools.
Analyses in this catalog relied on software from the LVK Algorithm Library Suite~\citep{lalsuite}.
The detection of the signals and subsequent significance evaluations were performed with the \GSTLAL{}-based inspiral software pipeline~\citep{Messick:2016aqy,Sachdev:2019vvd,Hanna:2019ezx,Cannon:2020qnf}, with the \MBTA{} pipeline~\citep{Adams:2015ulm,Aubin:2020goo}, and with the \PYCBC{}~\citep{Usman:2015kfa,Nitz:2017svb,Davies:2020tsx} packages.
Estimates of the noise spectra and glitch models were obtained using \BAYESWAVE{}~\citep{Cornish:2014kda,Littenberg:2015kpb,Cornish:2020dwh}.
Low-latency source localization was performed using \BAYESTAR{}~\citep{Singer:2015ema}.
Source-parameter estimation was primarily performed with the \BILBY{} and \PBILBY{} libraries~\citep{Ashton:2018jfp,Smith:2019ucc,Romero-Shaw:2020owr} using the \DYNESTY{} nested sampling package~\citep{Speagle:2019ivv}.
SEOBNRv5PHM waveforms used in parameter estimation were generated using pySEOBNR~\citep{Mihaylov:2023bkc}.
FTI and TIGER waveforms used for testing general relativity were generated using \BILBYTGR~\citep{bilby-tgr}.
\PESUMMARY{} was used to post-process and collate parameter estimation results~\citep{Hoy:2020vys}.
The various stages of the parameter estimation analysis were managed with the \ASIMOV{} library~\citep{Williams:2022pgn}.
Plots were prepared with \PLT{}~\citep{Hunter:2007ouj}, \SEABORN{}~\citep{Waskom:2021psk} and \GWPY{}~\citep{gwpy-software}.
\NUMPY{}~\citep{Harris:2020xlr} and \SCIPY{}~\citep{Virtanen:2019joe} were used for analyses in the manuscript.

}

\vspace{10pt}
\appendix

\section{Upgrades to the Detector Network for O4}\label{supp:detectors_text}

The Advanced LIGO, Advanced Virgo, and KAGRA detectors have all undergone a series of upgrades to improve the network sensitivity in preparation for \ac{O4}.
In this appendix we focus on upgrades made to the LIGO Livingston observatory, as it was the only detector to observe GW230529; similar upgrades were made at LIGO Hanford.
Some of these improvements are a part of the Advanced LIGO Plus (A+) detector upgrades~\citep{KAGRA:2013rdx}.

The principal commissioning work done in preparation for \ac{O4} was to enhance the optics systems in the detector.
The pre-stabilized laser was redesigned for \ac{O4} with amplifiers allowing input power up to 125~W~\citep{Bode:2020dge, Cahillane:2022pqm}.
LIGO Livingston operated at 63~W in \ac{O4a}.
Both end test masses at LIGO Livingston were replaced because their mirror coatings contained small, pointlike absorbers that limited detector sensitivity by degrading the power-recycling gain~\citep{LIGOScientific:2021kro}.
For \ac{O4}, frequency-dependent squeezing was implemented with the addition of a 300~m filter cavity to rotate the squeezed vacuum quadrature across the bandwidth of the detector~\citep{Dwyer:2022vbh, McCuller:2020yhw}.
With frequency-dependent squeezing, uncertainty is reduced in both the amplitude and phase quadratures, allowing for a broadband reduction in quantum noise~\citep{LIGOO4Detector:2023wmz}.
Squeezing was implemented independently of frequency in \ac{O3}, which only reduced the quantum noise at high frequencies~\citep{Tse:2019wcy}.

Other commissioning work was done to improve data quality and reduce noise at low frequencies.
For \ac{O4}, scattered-light noise was mitigated by removing some problematic scattering surfaces and improving the resonant damping of other scatterers~\citep{LIGO:2020zwl, LIGO:2021ppb}.
Low-frequency technical noise (i.e., noise that is not intrinsic to the detector design but can limit the detector sensitivity and performance) was reduced with the commissioning of feedback control loops, noise subtraction, and better electronics.
Overall, the upgrades made during the commissioning period for \ac{O4} led to a broadband reduction in noise and improvement in detector sensitivity and performance.

\section{Detection Time Line and Circulars}\label{supp:detection_timeline}

The \ac{LVK} issues low-latency alerts to facilitate prompt follow-up of \ac{GW} candidates~\citep{Chaudhary:2023vec}.
GW230529 was initially identified in a low-latency search using data from the LIGO Livingston observatory.
The candidate was given the name S230529ay in the Gravitational-Wave Candidate Event Database (\GRACEDB).\footnote{\url{https://gracedb.ligo.org/}}

After the detection of GW230529, an initial notice was sent out to astronomers through NASA's \acl{GCN}~\citep[\acsu{GCN};][]{2023GCN_notice}.
The sky localization computed in low latency by \BAYESTAR~\citep{Singer:2015ema, Singer:2016eax} had a $90\%$ credible area of $\approx \bayestararea~{\rm{deg}^2}$; the large credible area was due to GW230529 only being observed by a single detector.
The initial alert also included low-latency estimates of the mass-based source classification (\ac{BNS}, \ac{BBH}, and \ac{NSBH}) produced by the \PYCBC pipeline~\citep{Villa-Ortega:2022qdo}.
Additional estimates that the source binary of GW230529 contained at least one neutron star (\probNS$\%$), contained at least one compact object in the lower mass gap $3\text{--}5~\Msun$ (\probMassGap$\%$), and had neutron star matter ejected outside the final compact object (\probRemnant$\%$) were produced by a machine-learning method trained to infer source properties from the search pipeline results~\citep{Chatterjee:2019avs}.

Low-latency parameter estimation was performed with \BILBY~\citep{Ashton:2018jfp, Romero-Shaw:2020owr} and used to produce an updated sky localization and estimates of source properties sent out in another alert~\citep{2023GCN.33891....1L}.
The updated sky localization had a $90\%$ credible area of $\approx \bilbyarea~{\rm{deg}^2}$.
The updated estimate of source properties led to decreases in the probabilities that the source binary included at least one neutron star (\probNSBilby$\%$), one compact object in the mass gap (\probMassGapBilby$\%$), and ejected matter outside the remnant object (\probRemnantBilby$\%$).

\section{Quantifying Significance for a Single-detector Candidate Event}\label{supp:single_detector_searches}

Each search pipeline has its own method for calculating the \ac{FAR} of a single-detector \ac{GW} candidate.
In this appendix we will summarize these methods.

\subsection{\GSTLAL}

\begin{figure*}
    \centering
    \includegraphics[width=0.48\linewidth]{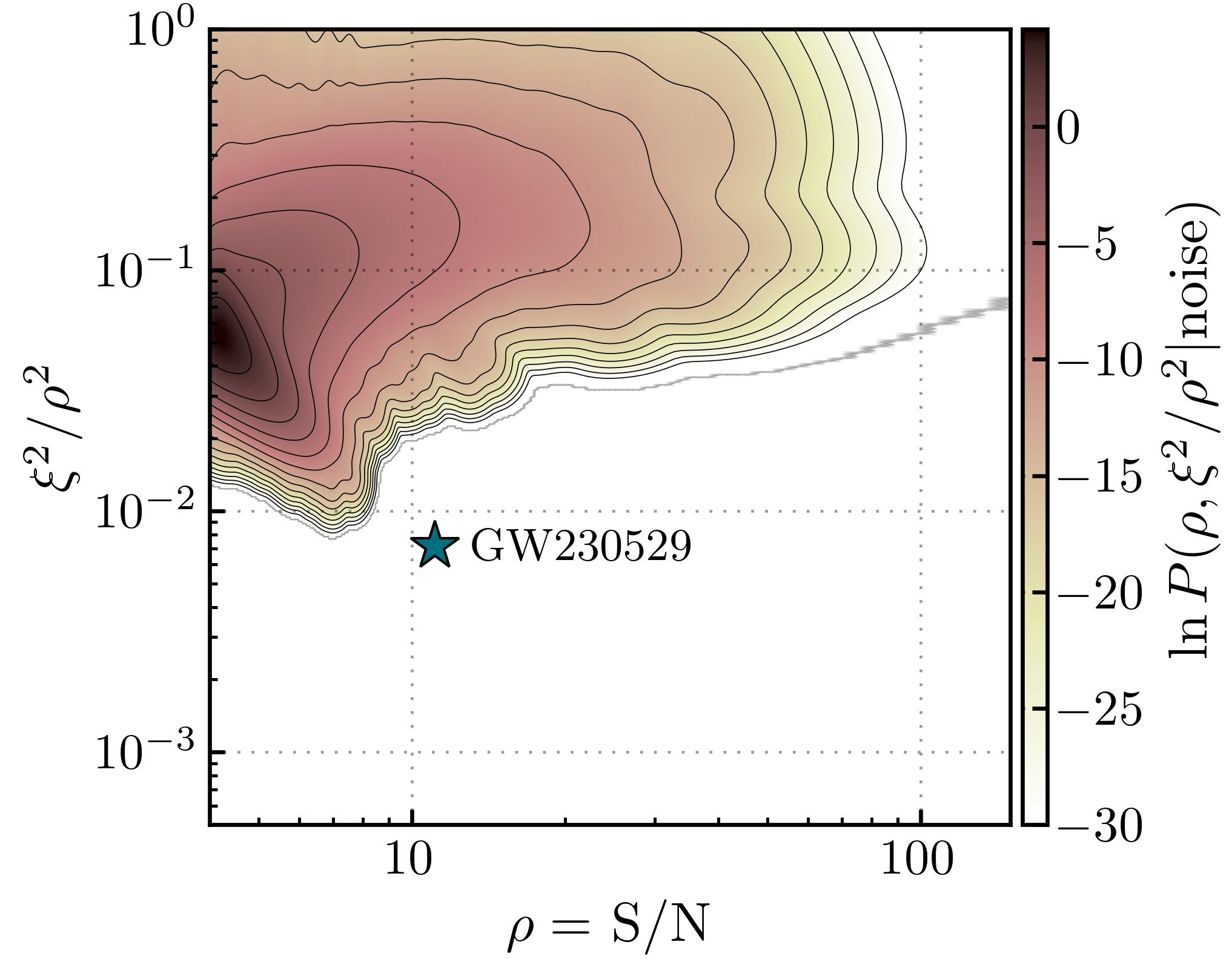}
    \caption{
The noise background for the LIGO Livingston observatory collected by the \GSTLAL pipeline in offline mode, with the \ac{SNR}--$\xi^2$ values of GW230529 overlaid.
The test statistic $\xi^2$ measures signal consistency.
The background is collected from templates that are part of the same bin as the template that recovered GW230529 and is consequently the background used to rank GW230529 and calculate its significance.
The background distribution has effectively no support at the position of GW230529.
The color bar represents the probability of producing a certain \ac{SNR} and $\xi^2$ from noise triggers.
    }
    \label{fig:SignificanceGSTLAL}
\end{figure*}

The \GSTLAL pipeline assigns a likelihood ratio to all of its candidate signals as the ranking statistic to estimate their significance.
The likelihood ratio is the ratio of the probability of obtaining the candidate under the signal hypothesis to the probability of obtaining the candidate under the noise hypothesis.
The latter probability is largely calculated by collecting \ac{SNR} and $\xi^{2}$ statistics of noise triggers during the analysis, where $\xi^{2}$ is a signal consistency test statistic~\citep{Messick:2016aqy}.
Similar templates in the template bank are grouped together on the basis of the \ac{PN} expansion of their waveforms~\citep{Morisaki:2020oqk, sakon2023template}.
Each template bin collects its own \ac{SNR} and $\xi^2$ statistics, which get used to rank candidates recovered by templates in that bin.
Further details about the calculation of the probability of obtaining the candidate under the noise hypothesis, as well as the \GSTLAL likelihood ratio in general, can be found in~\cite{Tsukada:2023edh}.
The \ac{SNR}--$\xi^2$ statistics collected from noise triggers from the same template bin as GW230529 with GW230529 superimposed are shown in Figure~\ref{fig:SignificanceGSTLAL}.
GW230529 clearly stands out from the background, and there are no noise triggers at its position in \ac{SNR}--$\xi^{2}$ space.

Since nonstationary noise transients known as glitches~\citep{Nuttall:2018xhi} are not expected to be correlated across detectors, single-detector \ac{GW} candidates, whether during times when a single or multiple detectors are observing, are more likely to be glitches than coincident \ac{GW} candidates.
To account for this, the \GSTLAL pipeline downweights the logarithm of the likelihood ratio of single-detector candidates by subtracting an empirically tuned factor of 13, which is optimized to maximize the recovery of candidates with an astrophysical origin while minimizing the recovery of glitches~\citep{LIGOScientific:2020aai}.
The \ac{FAR} of the \GSTLAL pipeline for GW230529 quoted in Table~\ref{table:detection_stats} is calculated after the application of this penalty.

Finally, the \ac{FAR} for a candidate in the search is computed by comparing its likelihood ratio with the likelihood ratios of noise triggers not found in coincidence, after accounting for the live time of the analysis.
These noise triggers not found in coincidence allow the pipeline to extrapolate the background distribution of likelihood ratios to large values, enabling the inverse \ac{FAR} of a candidate to be greater than the duration of the analysis.
While in the low-latency mode, the \GSTLAL \ac{FAR} calculation uses the background from the start of the analysis period until the time of the candidate.
In contrast, for its offline analysis, the \GSTLAL pipeline uses background collected from the full analysis period, including the entirety of \ac{O4a}, to rank candidates.
This differs from the other search pipelines presented in this work, which only use background from the first 2 weeks of \ac{O4a} for their offline search results presented here.
The \GSTLAL offline analysis was performed using the same template bank as the online analysis~\citep{sakon2023template, ewing2023performance}.
Details are liable to change for future offline \GSTLAL analyses in \ac{O4a}.
However, for such a significant candidate as GW230529, we do not expect these changes to impact its interpretation as highly likely being of astrophysical origin.

\subsection{\MBTA}

\begin{figure*}
    \centering
    \includegraphics[width=0.48\linewidth]{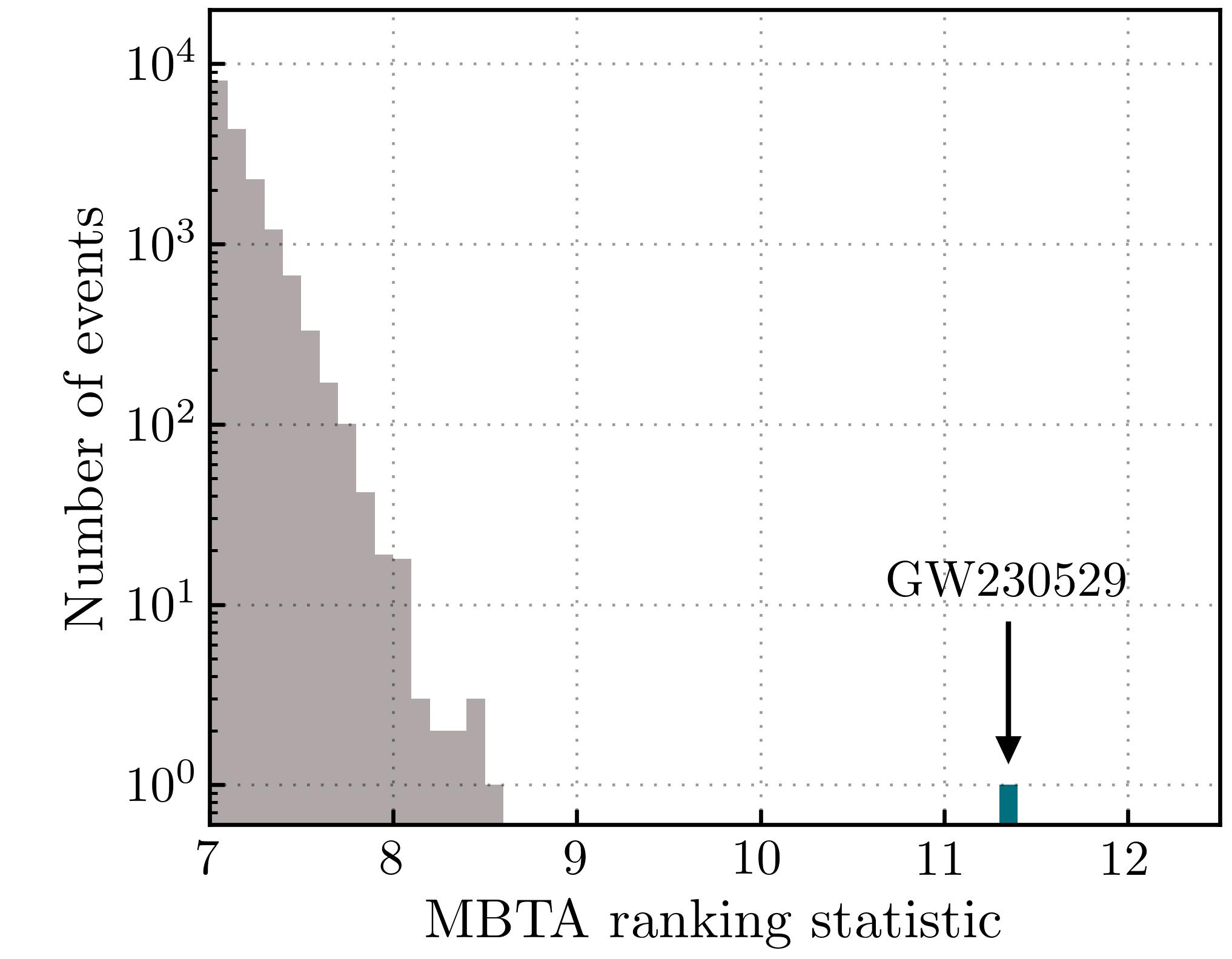}
    \caption{
Ranking statistic distribution of \MBTA single-detector triggers in the LIGO Livingston observatory during the first 2 weeks of \ac{O4}, excluding significant public alerts aside from GW230529.
    }
    \label{fig:SignificanceMBTA}
\end{figure*}

The \MBTA single-detector candidate search for \ac{O4a} focuses on a subset of the full MBTA parameter space.
The goal is to focus on \ac{BNS} and \ac{NSBH} signals, which are most interesting because of their rarity and possible \ac{EM} counterparts.
Their long signal duration also makes them easier to identify and allows for a better sensitivity to be reached when compared to a single-detector candidate search using the whole \MBTA parameter space.
The parameter space for the \ac{O4a} online single-detector search was defined based on the detector-frame (redshifted) primary mass and chirp mass of the templates and motivated by the computation of the probability for whether a nonzero amount of neutron star material remained outside the final remnant compact object~\citep{Chatterjee:2019avs} introduced in Appendix~\ref{supp:detection_timeline}.
The considered space is $(1+z)\,m_1 < 50~\Msun$ and $(1+z)\,\Mc < 5~\Msun$~\citep{Juste:2023xdk}.
For the offline analysis, this space was adapted and we consider only $(1+z)\,\Mc < 7~\Msun$.
\MBTA online also applied selection criteria based on data-quality tests to its single-detector candidates~\citep{Juste:2023xdk}.
This was changed to a reweighting of the ranking statistic for the offline analysis.

The ranking statistic for \MBTA single-detector triggers is a reweighted \ac{SNR}.
The reweighting is based on the computation of a quantity called $\texttt{auto}~\chi^2$~\citep{Aubin:2020goo}, which tests the consistency of the time evolution of the \ac{SNR} time series.
The additional reweighting used in the offline analysis relies on the computation of a quantity that identifies an excess of \ac{SNR} for single-detector triggers.
The excess of \ac{SNR} is larger for triggers that have a noise origin compared to astrophysical signals or injections and therefore allows for discrimination between astrophysical signals and noise.
The ranking statistic distribution of the \MBTA offline analysis for single-detector triggers during the first 2 weeks of \ac{O4a} is shown in Figure~\ref{fig:SignificanceMBTA}.
Other triggers that produced significant public alerts have been removed from the plotted distribution.
GW230529 stands out from the background with a high ranking statistic that reflects the $\texttt{auto}~\chi^2$ value being completely consistent with a signal origin.

The \ac{FAR} in \MBTA is a function of $p_\mathrm{astro}$~\citep{Andres:2021vew}, the probability of astrophysical origin of the candidate.
It is derived from the combined parameterizations of the \ac{FAR} as a function of the ranking statistic and of $p_\mathrm{astro}$ as a function of the ranking statistic.
Inverting the latter gives the ranking statistic as a function of $p_\mathrm{astro}$ and eventually the \ac{FAR} as a function of $p_\mathrm{astro}$.
The background estimation for \MBTA single-detector triggers was computed differently during the online and offline analyses.
The online analysis relies on the computation of simulated single-detector triggers obtained through random combinations of single frequency band data~\citep{Juste:2023xdk}.
\ac{O4a} online analysis and the method of randomly combining single frequency band data showed that the background for \MBTA single-detector triggers follows an exponential distribution and is stable in time beyond statistical fluctuations.
This prompted us to update the model we use for the offline (and future) analyses to reach greater sensitivity.
This new model involves extrapolating the observed distribution of single-detector triggers and removing the significant triggers.
A safety margin is used in the extrapolation such that the \ac{FAR} is overestimated relative to the best-fit extrapolated distribution.
This change in methods, in addition to the difference in handling of single-detector triggers online and offline, explains the difference in inverse \ac{FAR} for the online ($\mbtaOnlineIFAR~\text{yr}$) and offline (\mbtaOfflineIFAR~\text{yr}) analyses.

\subsection{\PYCBC}

\begin{figure*}
    \centering
    \includegraphics[width=0.48\linewidth]{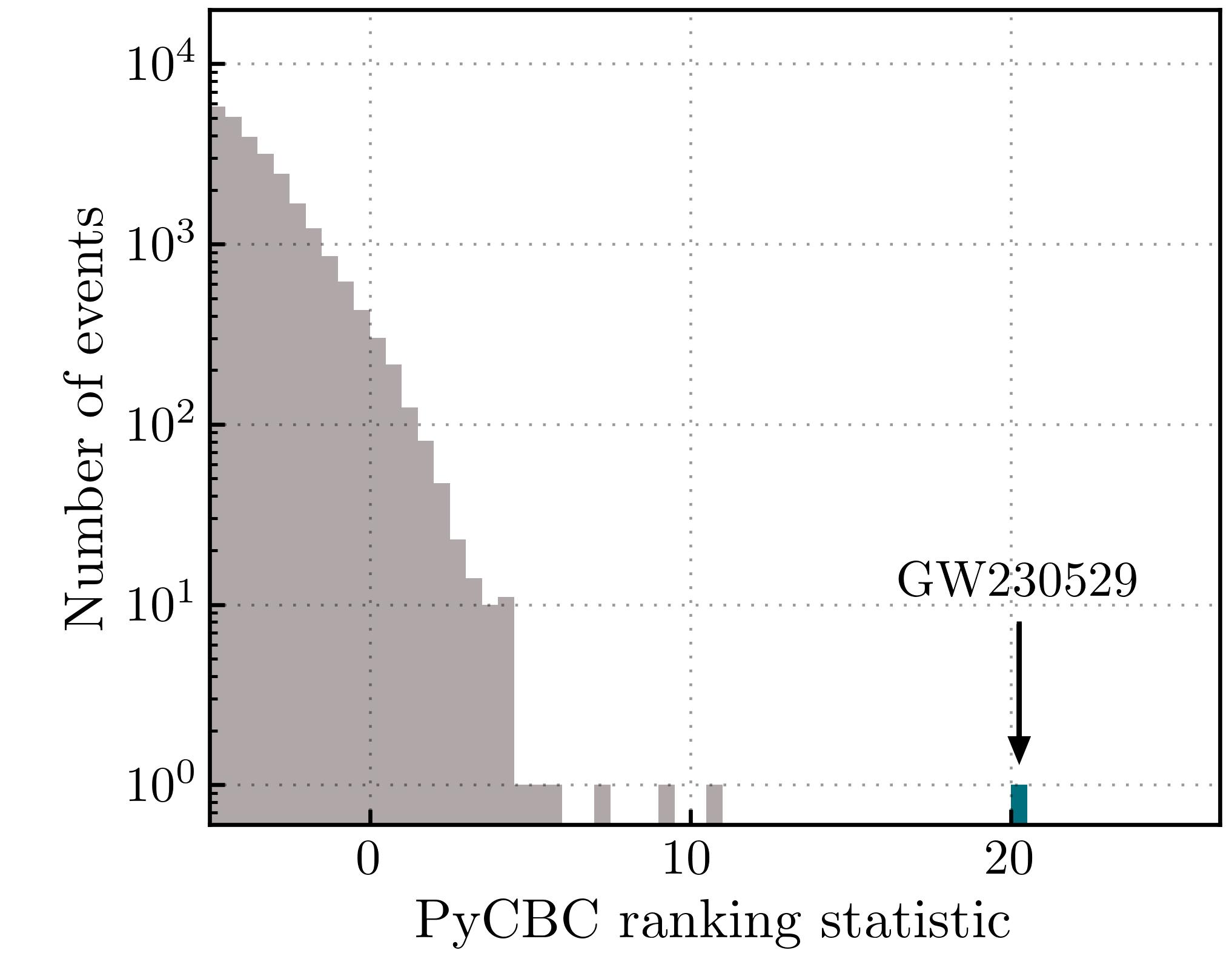}
    \caption{
Ranking statistic distribution of \PYCBC offline single-detector triggers in the LIGO Livingston observatory during the first 2 weeks of \ac{O4}.
The plotted distribution may include triggers from other less significant signals arriving during the period analyzed.
    }
    \label{fig:SignificancePYCBC}
\end{figure*}

In \PYCBC, each \ac{GW} candidate is assigned a ranking statistic, and then a \ac{FAR} is calculated by comparing the ranking statistic of the candidate with the background distribution.
The details of this procedure differ between the online and offline versions of the \PYCBC search.

In the online pipeline, we consider single-detector candidates only from templates with duration greater than $7~\mathrm{s}$ above a starting frequency of $17~\mathrm{Hz}$, corresponding to a range of masses and spins for which \ac{EM} emission due to neutron star ejecta might be expected.
Low-latency data-quality time series produced by \IDQ, a machine-learning framework for autonomous detection of noise artifacts using only auxiliary data channels insensitive to GWs~\citep{Essick:2020qpo}, are used to veto candidates at times when a glitch is likely to be present in the data.
The remaining single-detector candidates are ranked by the reweighted \ac{SNR}~\citep{Nitz:2017lco}.
Only candidates with a $\chi^{2}$ statistic~\citep{Allen:2004gu} $< 2.0$ and reweighted \ac{SNR} $> 6.75$ are kept.
The rate density of single-detector triggers above the reweighted \ac{SNR} threshold is fit with a decreasing exponential.

In offline \PYCBC, we only consider single-detector candidates from templates with duration greater than $0.3~\mathrm{s}$ above a starting frequency of $15~\mathrm{Hz}$.
We also require that candidates have a $\chi^{2}$ statistic $< 10.0$, a reweighted \ac{SNR} $> 5.5$, and a \ac{PSD} variation statistic~\citep{Mozzon:2020gwa} $< 10.0$, in order to exclude high-amplitude noise transients.
Each candidate is assigned a ranking statistic equal to the logarithm of the ratio of astrophysical signal likelihood to detector noise likelihood.
The \PYCBC \ac{O4} offline search introduced two changes to the ranking statistic relative to the \ac{O3} calculation.
First, an explicit model of the signal distribution covering the space of binary masses and spins is included~\citep{Kumar:2024bfe}.
The model is designed to maximize the number of detected signals from the known compact binary distribution, while also maintaining sensitivity to signals in previously unpopulated regions.
Second, the model of the rate density of events caused by detector noise now includes a term describing the variation in rate during times of heightened detector noise~\citep{Davis:2022cmw}.
The rate density of single-detector candidates above a ranking statistic threshold of $0$ is fit with a decreasing exponential.

In both the online and offline versions of \PYCBC, the exponential fit of trigger rate density above the relevant ranking statistic threshold is used to extrapolate the \ac{FAR} of single-detector candidates beyond the observing time of the search~\citep{Davies:2022thw}.
The \ac{FAR} calculations for GW230529 used triggers from the first 2 weeks of \ac{O4} at the LIGO Livingston observatory.
The distribution of ranking statistics for offline single-detector candidates at the LIGO Livingston observatory is shown in Figure~\ref{fig:SignificancePYCBC}.
Similar to the other searches, GW230529 clearly stands out from the background distribution.

\section{Priors, Waveform Systematics, and Bayes Factors}\label{supp:waveform_systematics}

\begin{table}
\begin{ruledtabular}
    \caption{
Summary of parameter estimation analysis choices for GW230529, and the physical content of each waveform model used.
Here tides refers to modeling of neutron star tidal deformability and disruption when tidal forces overcome the self-gravity of the neutron star.
The spin prior denotes any restrictions on the spin magnitude of the binary components.
    }
    \label{table:waveform_table}
    \renewcommand{\arraystretch}{1.4}
    \begin{center}
    \begin{tabular}{ l c c c c c  }
        Waveform Model & Precession & Higher Multipoles & Tides &  Disruption & Spin Prior \\
        \hline
        IMRPhenomNSBH & $-$ & $-$ & \checkmark & \checkmark & $ \chi_1  < 0.50,  \chi_2  < 0.05$ \\
        IMRPhenomPv2\_NRTidalv2 & $\checkmark$ & $-$ & \checkmark & $-$ & $ \chi_1  < 0.99,  \chi_2  < 0.05$ \\
        IMRPhenomXPHM & $\checkmark$ & $\checkmark$ & $-$ & $-$ & $ \chi_1  < 0.99,  \chi_2  < 0.99$ \\
        SEOBNRv5PHM & $\checkmark$ & $\checkmark$ & $-$ & $-$ & $ \chi_1  < 0.99,  \chi_2  < 0.99$ \\
        SEOBNRv4\_ROM\_NRTidalv2\_NSBH & $-$ & $-$ & \checkmark & \checkmark & $ \chi_1  < 0.90,  \chi_2  < 0.05$ \\
        IMRPhenomXPHM & $\checkmark$ & $\checkmark$ & $-$ & $-$ & $ \chi_1  < 0.99,  \chi_2  < 0.05$ \\
        IMRPhenomXP & $\checkmark$ & $-$ & $-$ & $-$ & $ \chi_1  < 0.99,  \chi_2  < 0.99$ \\
        IMRPhenomXHM & $-$ & $\checkmark$ & $-$ & $-$ & $ \chi_1  < 0.99,  \chi_2  < 0.99$ \\
        IMRPhenomXAS & $-$ & $-$ & $-$ & $-$ & $ \chi_1  < 0.99,  \chi_2  < 0.99$ \\
        IMRPhenomXAS & $-$ & $-$ & $-$ & $-$ & $ \chi_1  < 0.50,  \chi_2  < 0.05$ \\
        IMRPhenomPv2\_NRTidalv2 & $\checkmark$ & $-$ & $\checkmark$ & $-$ & $ \chi_1  < 0.05,  \chi_2  < 0.05$ \\
        IMRPhenomXPHM & $\checkmark$ & $\checkmark$ & $-$ & $-$ & $ \chi_1  < 0.05,  \chi_2  < 0.05$ \\
        SEOBNRv5PHM & $\checkmark$ & $\checkmark$ & $-$ & $-$ & $ \chi_1  < 0.99,  \chi_2  < 0.05$ \\
    \end{tabular}
    \end{center}
\end{ruledtabular}
\end{table}

Given the uncertain nature of the compact objects, we analyze GW230529 with a suite of models that incorporate a number of key physical effects.
The choices of data duration ($128~\mathrm{s}$) and frequency bandwidth ($20\text{--}1792~\mathrm{Hz}$) analyzed were informed by comparing waveforms spanning the mass range recovered in preliminary analyses with the detector noise \ac{PSD} at the time of the event.
The high-frequency cutoff is chosen to avoid loss of power at high frequencies due to low-pass filtering of the data.
For a subset of the signal models below, we employ a range of techniques to speed up the evaluation of the likelihood, including heterodyning (also known as relative binning; \citealt{Cornish:2010kf, Cornish:2021lje, Zackay:2018qdy, Krishna:2023bug}), multibanding~\citep{Garcia-Quiros:2020qlt,Morisaki:2021ngj}, and reduced-order quadratures~\citep{Canizares:2014fya,Smith:2016qas,Morisaki:2023kuq}.

The physical effects included and the spin prior ranges for each waveform model considered in this work are shown in Table~\ref{table:waveform_table}.
All analyses use mass priors that are flat in the redshifted component masses with chirp masses $\mathcal{M} = (m_{1}m_{2})^{3/5}/(m_{1}+m_{2})^{1/5} \in [\chirpMassMin, \chirpMassMax]~\Msun$ and mass ratios $q \in [\massRatioMin, \massRatioMax]$.
The luminosity distance prior is uniform in comoving volume and source-frame time in the range $D_\mathrm{L} \in [\distanceMin, \distanceMax]~\mathrm{Mpc}$.
The priors on the tidal deformability parameters are chosen to be uniform in the component tidal deformabilities over $\Lambda \in [0, 5000]$.
Standard priors~\citep[e.g.,][]{Romero-Shaw:2020owr, LIGOScientific:2021usb} are used for all the other extrinsic binary parameters.
Below we provide further motivation for considering this suite of waveform models.

The primary analysis in this work uses the IMRPhenomXPHM and SEOBNRv5PHM \ac{BBH} waveform models (corresponding to the third and fourth rows in Table~\ref{table:waveform_table}), which provide an accurate description of \acp{BBH} but do not incorporate tidal effects or the potential tidal disruption of a companion neutron star.
We did not consider an extension of IMRPhenomXPHM that incorporates additional physics beyond the two precessing higher-order multipole moment \ac{BBH} models used here, as these effects only enter at high frequencies and are not relevant for this event~\citep{Thompson:2023ase}.
In order to quantify the impact of neglecting tidal physics, we analyze GW230529 with \ac{NSBH} and \ac{BNS} waveform models that incorporate different tidal information.
The \ac{NSBH} models only incorporate tidal information from the less massive compact object, are restricted to spins aligned with the orbital angular momentum, and only model the dominant $\ell = m = 2$ multipole moment.
However, they do model the possible tidal disruption of the neutron star, which occurs when tidal forces of the black hole dominate over the self-gravity of the neutron star.

We use two \ac{NSBH} models, a frequency-domain phenomenological model IMRPhenomNSBH~\citep{Thompson:2020nei} and a frequency-domain \ac{EOB} surrogate model SEOBNRv4\_ROM\_NRTidalv2\_NSBH~\citep{Matas:2020wab}.
IMRPhenomNSBH uses the \ac{BBH} models IMRPhenomC~\citep{Santamaria:2010yb} and IMRPhenomD~\citep{Husa:2015iqa, Khan:2015jqa} as baselines for the amplitude and phase, respectively, incorporating corrections to the phase due to tidal effects following the NRTidal model~\citep{Dietrich:2017aum, Dietrich:2018uni, Dietrich:2019kaq} and to the amplitude following~\cite{Pannarale:2013uoa, Pannarale:2015jka}.
SEOBNRv4\_ROM\_NRTidalv2\_NSBH uses the SEOBNRv4 \ac{BBH} model as a baseline~\citep{Bohe:2016gbl} and applies the same corrections to account for tidal deformability and disruption as IMRPhenomNSBH but is additionally calibrated against \ac{NR} simulations of \ac{NSBH} mergers~\citep{Foucart:2013psa, Foucart:2014nda, Foucart:2018lhe, Kyutoku:2010zd, Kyutoku:2011vz}.
As the \ac{NSBH} waveform models do not capture spin-induced orbital precession, we analyze GW230529 with the aligned-spin \ac{BBH} waveform models IMRPhenomXAS~\citep{Pratten:2020fqn}, which only contains the dominant harmonic, and IMRPhenomXHM~\citep{Garcia-Quiros:2020qpx}, which includes higher-order multipole moments.
Models for \ac{NSBH} binaries that contain both higher-order multipole moments and precession are under active development \citep{Gonzalez:2022prs} and could potentially allow for a more consistent model selection between the different source categories.

\begin{figure}
    \centering
    \includegraphics[width=\linewidth]{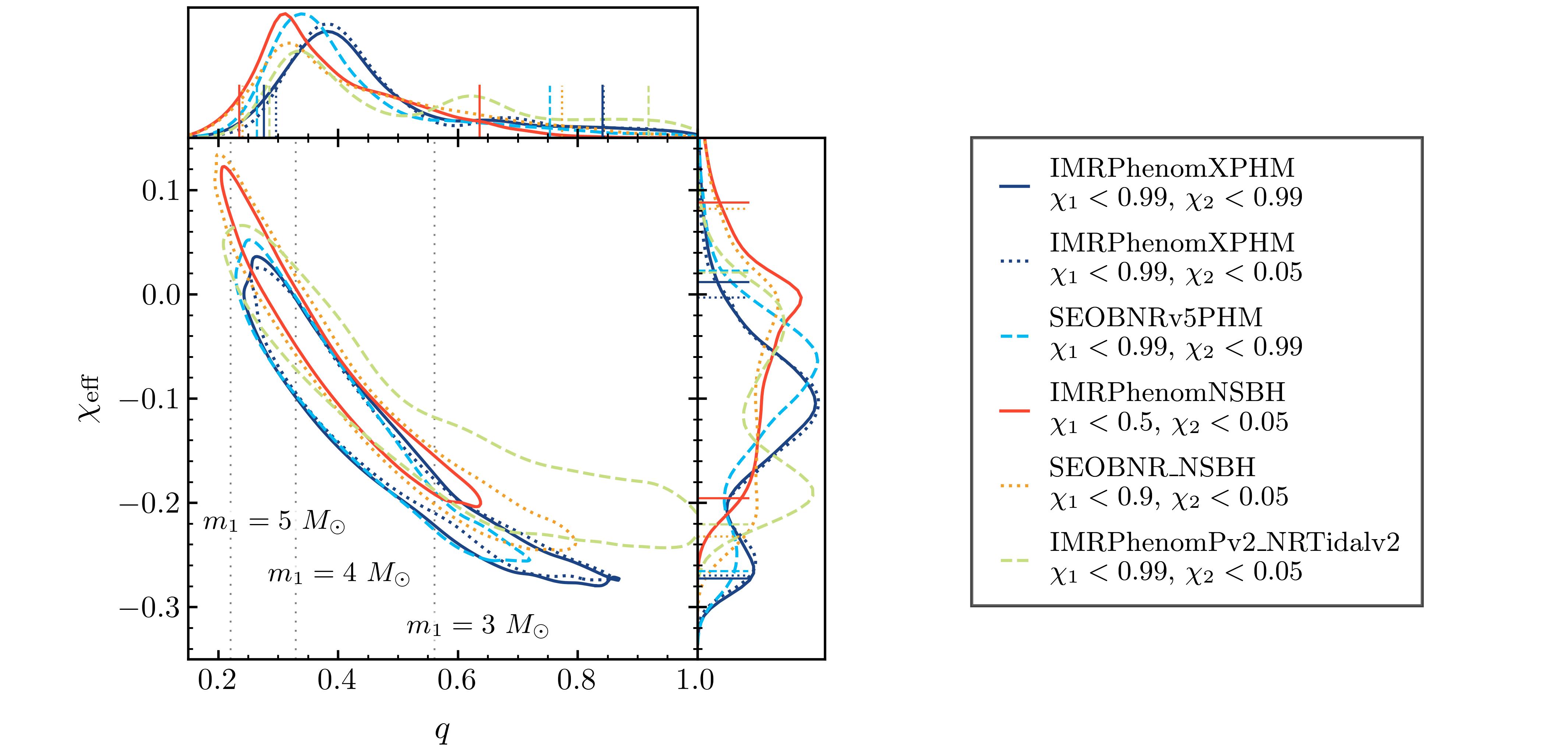}
    \caption{
The two-dimensional $q$--$\chi_{\rm eff}$ posterior probability distributions for GW230529 using various \ac{BBH}, \ac{NSBH}, and \ac{BNS} signal models.
The vertical dotted lines indicate primary masses that have been mapped to the mass ratio given the median chirp mass estimated from the IMRPhenomXPHM high-spin analysis.
    }
    \label{fig:qChieff}
\end{figure}

Given that the mass of the primary overlaps with current estimates of the maximum known neutron star mass~\citep{Landry:2021hvl,Romani:2022jhd}, we also analyze GW230529 with a \ac{BNS} waveform model that allows for tidal interactions on both the primary and the secondary components of the binary.
We use IMRPhenomPv2\_NRTidalv2~\citep{Dietrich:2019kaq}, which adds a model for the tidal phase that is calibrated against both \ac{EOB} and \ac{NR} simulations to the underlying precessing-spin point-particle waveform model IMRPhenomPv2~\citep{Hannam:2013oca}.
A limitation is that IMRPhenomPv2\_NRTidalv2 is calibrated against a suite of equal-mass, nonspinning \ac{BNS} \ac{NR} simulations, where the maximum neutron star mass is only $1.372~\Msun$.
Nevertheless, the model has been validated against a catalog of \ac{EOB}--\ac{NR} hybrid waveforms that includes asymmetric configurations and heavier neutron star masses~\citep{Dietrich:2019kaq,Abac:2023ujg}.
Because of large uncertainties in \ac{BNS} postmerger waveform modeling, IMRPhenomPv2\_NRTidalv2 includes an amplitude taper from $1$ to $1.2$ times the estimated merger frequency~\citep{Dietrich:2018uni}.
The resulting suppression of \ac{SNR} at these frequencies may be responsible for the posterior differences between IMRPhenomPv2\_NRTidalv2 and the \ac{BBH} and \ac{NSBH} waveform models we consider (Figure~\ref{fig:qChieff}).

From the mass estimates alone, the secondary object is consistent with being a neutron star.
As such, we follow earlier analyses and analyze GW230529 with two spin priors~\citep{LIGOScientific:2021qlt}: an agnostic high-spin prior and an astrophysically motivated low-spin prior.
The high-spin prior assumes that the spins on both objects are isotropically oriented and have dimensionless spin magnitudes that are uniformly distributed up to $\chi_1, \chi_2 \leq 0.99$.
The low-spin prior, on the other hand, is inspired by the extrapolated maximum spin observed in Galactic \acp{BNS} that will merge within a Hubble time~\citep{Burgay:2003jj, Stovall:2018ouw} and restricts the spin magnitude of the secondary to be $\chi_2 \leq 0.05$ while keeping $\chi_1 \leq 0.99$.
As the nature of the primary as a black hole or neutron star is uncertain, we also use a third choice of prior where both spins are restricted to $\chi_{1}, \chi_{2} \leq 0.05$ for the IMRPhenomPv2\_NRTidalv2 and IMRPhenomXPHM waveform models (rows 11 and 12 of Table~\ref{table:waveform_table}, respectively).
The purpose of using multiple priors is to help gauge whether astrophysical assumptions on the neutron star spin impact any of the statements on the probability that the primary object lies in the lower mass gap.
In Figure~\ref{fig:qChieff}, we show the impact of waveform systematics and prior assumptions on the strongly correlated mass ratio--effective inspiral spin distribution.

The degree to which a given waveform model under certain assumptions matches the data can be gauged using the Bayes factor.
However, Bayes factors penalize extra degrees of freedom in models that do not improve the fit when those extra degrees of freedom are constrained by the data~\citep{2003itil.book.....M}.
We find no significant preference ($\log_{10}{\mathcal{B}} \lesssim0.5$) between the high-spin and low-spin prior on the secondary; we similarly do not find a statistical preference for or against the effects of precession, higher-order multipole moments, or tidal deformation or disruption of the secondary object.

\section{Additional source properties}\label{supp:source-pars}
\subsection{Component Spins and Precession}\label{supp:spins}

\begin{figure}
    \centering
    \includegraphics[width=0.48\linewidth]{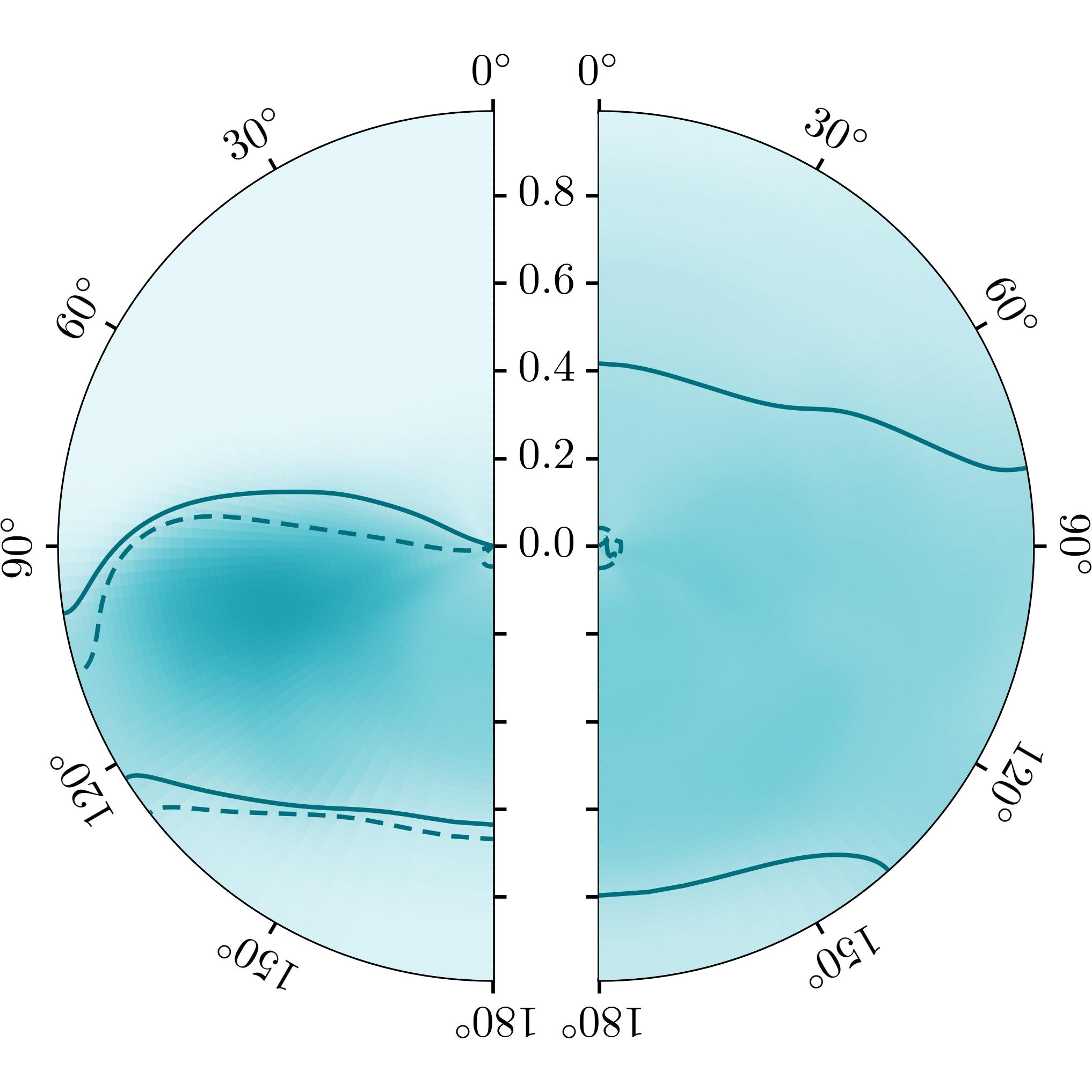}
    \caption{
Two-dimensional posterior probability distributions for the spin magnitude $\chi_i$ and the spin-tilt angle $\theta_i$ for the primary (left) and secondary (right) compact objects at a reference frequency of $20~\mathrm{Hz}$.
A spin-tilt angle of $0^{\circ}$ ($180^{\circ}$) indicates a spin that is perfectly aligned (antialigned) with the orbital angular momentum $\hat{L}$.
The pixels have equal prior probability, and shading denotes the posterior probability of each pixel of the high-spin prior analysis, after marginalizing over azimuthal angles.
The solid (dashed) contours denote the $90\%$ credible region for the high-spin (low-spin) prior analyses.
The probability distributions are marginalized over the azimuthal spin angles.
    }
    \label{fig:SpinDisk}
\end{figure}

Assuming a high-spin prior, the posteriors for the primary spin magnitude are only weakly informative, disfavoring zero and extremal spins with $\chi_1 = \aonefivepercent{GW230529ay_combined_imrphm_high_spin}\text{--}\aoneninetyfivepercent{GW230529ay_combined_imrphm_high_spin}$.
For the secondary, the posteriors are even less informative.
Under the low-spin prior, the primary spin magnitude posterior is peaked at slightly higher values, with zero and extremal spins being disfavored to a larger degree, $\chi_1 = \aonefivepercent{GW230529ay_combined_imrphm_low_spin}\text{--}\aoneninetyfivepercent{GW230529ay_combined_imrphm_low_spin}$.
The secondary spin is completely uninformative over the restricted range of spin magnitudes.
The joint two-dimensional posterior probability distribution of the dimensionless spin magnitude and the spin tilt are shown in Figure~\ref{fig:SpinDisk}.
Regions of high (low) probability are denoted by a darker (lighter) shade.

When the spins are misaligned with the orbital angular momentum, relativistic spin--orbit and spin--spin couplings drive the evolution of the orbital plane and the spins themselves~\citep{Apostolatos:1994mx, Kidder:1995zr}.
The leading-order effect can be captured by an effective precession spin parameter, $0 \leq \chi_{\rm p} \leq 1$, which approximately measures the degree of in-plane spin and can be used to parameterize the rate of precession of the orbital plane~\citep{Schmidt:2014iyl}
\begin{align}
\chi_p &= \max \left[ \chi_1 \, \sin \theta_1 , \left( \frac{3 + 4 q}{4 + 3 q} \right) \, q \, \chi_2 \, \sin \theta_2 \right] .
\end{align}
We find the constraints on $\chi_\mathrm{p}$ to be uninformative, and we are not able to make any significant statements on precession.
The uninformative nature of these results is corroborated by the Bayes factor between the precessing and nonprecessing phenomenological waveform models, $\log_{10} \mathcal{B}^{\mathrm{XP}}_{\mathrm{XAS}} = \IMRPhenomXPIMRPhenomXASBBH$.

\subsection{Source Location and Distance}\label{supp:extrinsic}

As GW230529 was a single-detector observation, the sky localization is poor and covers a sky area of $\approx \skyArea~{\rm{deg}^2}$ at the $90\%$ credible level.
The luminosity distance is inferred to be $D_\mathrm{L} = \luminositydistancemed{GW230529ay_combined_imrphm_high_spin}^{+\luminositydistanceplus{GW230529ay_combined_imrphm_high_spin}}_{-\luminositydistanceminus{GW230529ay_combined_imrphm_high_spin}}~\mathrm{Mpc}$, corresponding to a redshift of $z = \redshiftmed{GW230529ay_combined_imrphm_high_spin}^{+\redshiftplus{GW230529ay_combined_imrphm_high_spin}}_{-\redshiftminus{GW230529ay_combined_imrphm_high_spin}}$ computed using the Planck 2015 cosmological parameters~\citep{Planck:2015fie}.
The luminosity distance has a degeneracy with the inclination angle of the binary's total angular momentum with respect to the line of sight, $\theta_{\rm JN}$~\citep[e.g.,][]{Cutler:1994ys, Nissanke:2009kt, LIGOScientific:2013yzb, Vitale:2018wlg}.
The posterior distribution on $\theta_\mathrm{JN}$ is broadly unconstrained, showing no preference for a total angular momentum vector that is pointed toward or away from the line of sight.

\subsection{Tidal Deformability}\label{supp:tides}

The tidal deformability of a neutron star is defined by
\begin{align}
\lambda &= \frac{2}{3} k_{2} R^5,
\end{align}
where $k_2$ is the gravitational Love number of the object and $R$ is its radius~\citep{Damour:1992sxu, Mora:2003wt, Flanagan:2007ix, Damour:2009vw}.
However, it is often convenient to introduce a dimensionless tidal deformability
\begin{align}
\Lambda &= \frac{\lambda}{m^5} = \frac{2}{3} k_{2} C^{-5},
\end{align}
where $C = m / R$ is the compactness of the neutron star in geometrized units.

\begin{figure}
    \centering
    \includegraphics[width=0.48\linewidth]{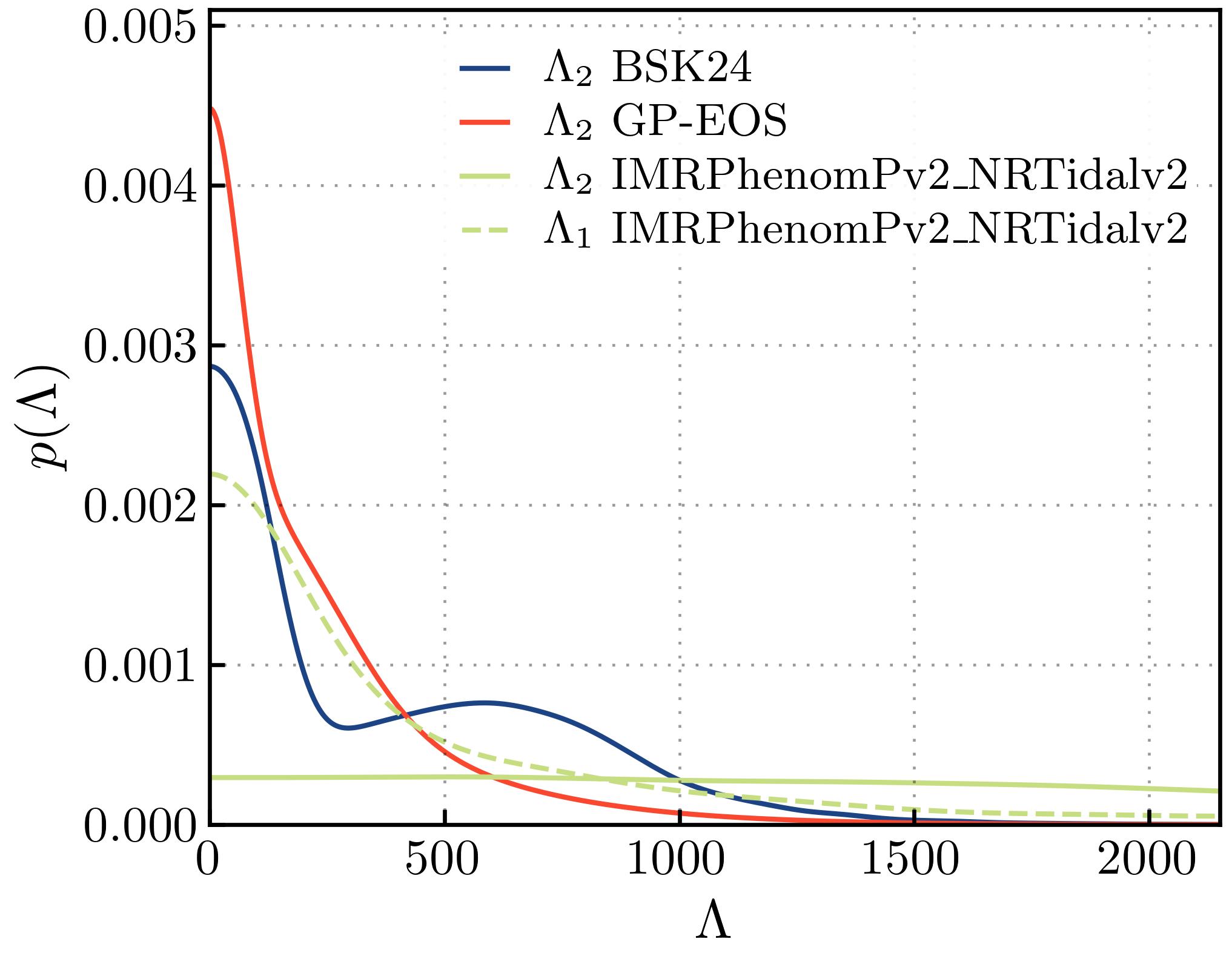}
    \caption{
Constraints on dimensionless tidal deformability $\Lambda$ for the primary (green dashed) and secondary (green solid) components of GW230529.
We also show tidal deformabilities predicted for the secondary object under the BSK24 neutron star \ac{EOS} (navy; \citealt{Goriely:2013nxa, Pearson:2018tkr, Perot:2019gwl}) and constraints obtained from \ac{BNS} and pulsar observations using the \ac{GP} model (\ac{GP}-\ac{EOS}; red).
    }
    \label{fig:TidalDeformability}
\end{figure}

Using the IMRPhenomPv2\_NRTidalv2 model, which allows for tidal deformability on both objects but does not model tidal disruption, we infer $\Lambda_{1} \leq \lambdaUpperLimit$ at 90\% credibility with the $\chi_{1} < 0.99, \chi_{2} < 0.05$ spin prior; the posterior peaks at $\Lambda_1 = 0$.
We do not constrain $\Lambda_2$ relative to the prior.
Constraints on the tidal deformability $\Lambda$ of both components of GW230529 are shown in Figure~\ref{fig:TidalDeformability}.
We also show the tidal deformability predicted by the $\Lambda(m)$ relation of the specific neutron star \ac{EOS} model BSK24~\citep{Goriely:2013nxa, Pearson:2018tkr} weighted by the $m_2$ posterior distribution.
BSK24 is chosen as an illustrative \ac{EOS} that is thermodynamically consistent and falls within the range of support of constraints from nuclear physics and astrophysics, including GW170817~\citep{Perot:2019gwl}.
Even under the assumption that the \ac{EOS} is known (BSK24), $\Lambda_2$ is not well constrained, due to the relatively wide $m_2$ posterior.
The $\Lambda_2$ prediction of the set of \ac{GP}-\ac{EOS} constraints (red; Section~\ref{sec:source_classification}) is consistent with both the inferred value of $\Lambda_2$ (green solid) and the BSK24 value of $\Lambda_2$ (navy).
In addition to the tidal deformability posteriors, we use two distinct likelihood interpolation schemes~\citep{Ray:2022hzg, Landry:2018prl} to directly constrain the neutron star \ac{EOS} using both spectral~\citep{Lindblom:2010bb, Lindblom:2012zi, Lindblom:2013kra} and \ac{GP}~\citep{Landry:2018prl, Essick:2019ldf} representations.
Both methods incorporate consistency with the observed heavy pulsars PSR J0740+6620~\citep{Fonseca:2021wxt} and PSR J0348+0432~\citep{Antoniadis:2013pzd} as priors on the \ac{EOS} and enforce thermodynamic stability and causality.
Unlike other analyses where the \ac{GP} representations are conditioned on \textit{both} \ac{GW} and pulsar data, here we only include the pulsar constraints in the prior, in order to distinguish the \ac{EOS} information provided by GW230529 alone.
These direct \ac{EOS} inferences are also uninformative, returning their respective priors.

\section{Testing General Relativity}\label{supp:tgr}

We perform several analyses to verify whether GW230529 is consistent with general relativity~\citep{Sanger:2024axs}.
Specifically, we perform parameterized tests searching for deviations in the \ac{PN} coefficients that determine the phase evolution of the \ac{GW} signal~\citep{Blanchet:1994ex, Blanchet:1994ez, Arun:2006hn, Arun:2006yw, Yunes:2009ke, Mishra:2010tp, Li:2011cg, Li:2011vx}.
We search for parametric deviations to the \ac{GW} inspiral phasing applied to the frequency-domain waveform models IMRPhenomXP\_NRTidalv2~\citep{Colleoni:2023czp} and IMRPhenomXPHM~\citep{Pratten:2020fqn, Garcia-Quiros:2020qpx, Pratten:2020ceb} using the TIGER framework~\citep{Agathos:2013upa, Meidam:2017dgf}, and SEOBNRv4\_ROM\_NRTidalv2\_NSBH~\citep{Matas:2020wab} and SEOBNRv4HM\_ROM~\citep{Bohe:2016gbl, Cotesta:2018fcv, Cotesta:2020qhw} using the FTI framework~\citep{Mehta:2022pcn}.
These waveform models each capture different physical effects (precession, higher-order multipole moments, and neutron star tidal deformability) to determine whether their absence in the model leads to inferred inconsistencies with general relativity.

For all waveform models and \ac{PN} orders whose analyses have been completed, we find that GW230529 is consistent with general relativity within the inferred uncertainties on the deviation parameters.
We do not yet include results about 0\ac{PN} deviations, which are being examined separately owing to technical issues related to the degeneracy between the 0\ac{PN} deviation parameter and the chirp mass~\citep{Payne:2023kwj}.
The constraints obtained at $-1$\ac{PN} are an order of magnitude tighter than previously reported bounds for \ac{NSBH} and \ac{BBH}~\citep{LIGOScientific:2021sio}.
The previously reported bounds using GW170817 remain the tightest constraints on $-1$PN deviations obtained with \acp{GW}~\citep{LIGOScientific:2018dkp}.

\section{Compact Binary Merger Rate Methods}\label{supp:merger_rates}

A key component of the event-based rate estimation described in Section~\ref{sec:rates_populations} is the sensitive time--volume of our \ac{GW} searches.
The sensitivities to GW200105-, GW200115-, and GW230529-like events are computed from \ac{O1} through the first 2 weeks of \ac{O4a} according to the corresponding mass and spin posteriors from the IMRPhenomXPHM high-spin analyses of these signals.
The posterior samples chosen for this analysis are consistent with previous event-based rate estimates presented for \ac{NSBH} mergers~\citep{LIGOScientific:2021qlt}.
Simulated signals, whose binary parameters are drawn from the mass and spin posteriors for each signal, are distributed uniformly in comoving volume and following the other extrinsic parameter distributions given in Appendix~\ref{supp:waveform_systematics}, and are added to simulated detector noise characterized by representative \acp{PSD} for each observing run.
The detectability of these injections is then calculated semianalytically to determine the sensitive time--volume by calculating network responses to the simulated signals and applying a threshold on the network optimal \SNR of $> 10$.
This choice of threshold was previously tuned to the results of matched-filter searches~\citep{ligo_scientific_collaboration_and_virgo_2021_5636816} and is comparable to threshold statistics calculated for semianalytic sensitivity estimates~\citep{Essick:2023toz}.

For the population-based approach, we estimate the merger rate of three astrophysical populations (\ac{BBH}, \ac{BNS}, and \ac{NSBH}) by aggregating triggers found by \GSTLAL from \ac{O1} through the first 2 weeks of \ac{O4}, while accounting for the possibility that some of these triggers are of terrestrial origin.
Our population analyses do not, however, include data from the engineering run preceding the start of \ac{O4}, during which another \ac{NSBH} candidate was identified~\citep{2023GCN.33813....1L}.
As outlined and implemented in previous \ac{LVK} results~\citep{LIGOScientific:2021qlt, LIGOScientific:2021duu}, we construct the joint likelihood on the Poisson parameters of the astrophysical populations ($\Lambda_{\mathrm{BBH}},\Lambda_{\mathrm{BNS}}, \Lambda_{\mathrm{NSBH}}$) and of the terrestrial triggers \citep[$\Lambda_{\mathrm{background}}$;][]{Farr:2013yna, Kapadia:2019uut}.
We then extrapolate the sensitive time--volume to each astrophysical population ($\VT_{\mathrm{BBH}},\VT_{\mathrm{BNS}},\VT_{\mathrm{NSBH}}$) obtained at the end of \ac{O3}~\citep{LIGOScientific:2021duu} to account for the additional 2 weeks of observation time during \ac{O4} during which GW230529 was detected.
Using a uniform prior on the merger rates $\mathcal{R}_{\alpha}=\Lambda_{\alpha}/\VT_{\alpha}$, we infer their posterior distributions.
The three astrophysical populations are defined by dividing up the space of compact binary component masses and spins into disjoint regions.
We consider all components with masses between $1$ and $3~\Msun$ to be neutron stars and assume a maximum dimensionless spin of 0.05 for such components; all components that are more massive are assumed to be black holes with a maximum dimensionless spin of 0.99.
The distribution of component masses within each region is assumed to be a power law in primary mass with index $-2.35$ and uniform in secondary mass.

Even though we infer the posterior distributions of the merger rates of all three populations, we only present $\mathcal{R}_\mathrm{NSBH}$ given that GW230529 contributes negligibly to $\mathcal{R}_\mathrm{BBH}$ and $\mathcal{R}_\mathrm{BNS}$.

\section{Additional Details of Population Analyses}\label{supp:population_models}

We use three different models to analyze the population of compact-object binaries with and without GW230529.

\subsection{\BGP}\label{supp:BGP}
\begin{table}
\begin{ruledtabular}
    \caption{
Summary of \BGP model parameters~\citep{Ray:2023upk, Mandel:2016prl, Mohite:2022pui}.
Arguments in the priors specify the mean and standard deviation in a normal (N) or half-normal (HN) distribution.
    }
    \label{table:parameters_bgp}
    \renewcommand{\arraystretch}{1.4}
    \begin{center}
    \begin{tabular}{c c c}
        {\bf Parameter} & \textbf{Description} & \textbf{Prior} \\
        \hline
        $\mu$ & Mean $\log\left(\displaystyle\mathrm{Rate}\over\displaystyle \perGpcyr\,\Msun^{-2}\right)$ in each bin & $\mathrm{N}(0,10)$ \\
        $\sigma$ & Amplitude of the covariance kernel & $\mathrm{HN}(0,10)$ \\
        $\mathrm{log}(l)$ & $\log\left(\displaystyle \mathrm{Length\ scale}\over\displaystyle \log(M / \Msun)\right)$ of the covariance kernel & $\mathrm{N}(-0.085,0.93)$ \\
    \end{tabular}
    \end{center}
\end{ruledtabular}
\end{table}

The \BGP method models the merger rate density per log component masses as a piecewise constant function.
By inferring the merger rate density in each bin, we reconstruct the shape of the \ac{CBC} mass spectrum up to the resolution limit imposed by our choice of binning.
A \ac{GP} prior with an exponential quadratic kernel is imposed on the logarithmic rate densities to smooth out the inferred shapes over sparse regions of the parameter space.
To assess the impact of GW230529 on the shape of the \ac{CBC} mass spectrum, we use the same bin locations and priors on the \ac{GP} hyperparamters as the GWTC-3 analysis~\citep{LIGOScientific:2021duu}.
The means~($\mu$), correlation length~($l$), and covariance amplitude~($\sigma$) of the \ac{GP} are drawn from normal, lognormal, and half-normal priors, respectively.
Further details of these priors are summarized in Table~\ref{table:parameters_bgp}.

Unlike more recent implementations of the \BGP model that also fit for the redshift distribution~\citep{Ray:2023upk}, we assume a redshift distribution such that the overall merger rate of compact binaries is uniform in comoving volume and source-frame time.
This facilitates a direct comparison with the findings of the GWTC-3 analysis~\citep{LIGOScientific:2021duu} and avoids any systematic biases in \BGP-based redshift distribution models originating from the inclusion of low-mass events, which remains an active area of study.
As in previous analyses~\citep{LIGOScientific:2021duu,Ray:2023upk}, we fix the spin distributions for each component to be isotropic in direction and uniform in spin magnitude.

\subsection{\PDB}\label{supp:PDB}

The \PDB model is designed to search for a separation in masses between neutron stars and black holes by employing a broken power law with a dip at the location of the power-law break.
The dip is modeled by a notch filter with depth $A$, which is fit along with other model parameters in order to determine the existence and depth of a potential mass gap~\citep{Farah:2021qom}.
A value $A=0$ corresponds to no gap, whereas $A=1$ corresponds to zero merger rate over the interval of the gap, i.e., a maximally deep gap.
\PDB also employs a low-pass filter at high black hole masses to allow for a tapering of the mass spectrum, which has the effect of adding a smooth second break to the power law.

The component mass distributions in this model are both fit by the same broken power law with exponents $\alpha_1$ between $m_{\rm min}$ and $M^{\mathrm{gap}}_{\rm low}$ and $\alpha_2$ between $M^{\mathrm{gap}}_{\rm low}$ and $m_{\rm max}$.
The model additionally includes a power-law pairing function in mass ratio~\citep{Fishbach:2019bbm}, assumed to be the same for all component masses.
The parameters for this mass model are summarized in Table~\ref{table:parameters_pdb}.
Like the \BGP model, the \PDB model additionally assumes that \acp{CBC} are uniformly distributed in comoving volume and source-frame time and that component spins are isotropically oriented and uniformly distributed in magnitude.
These assumptions are made for simplicity and consistency with GWTC-3 analyses~\citep{LIGOScientific:2021duu}.
Components with $m < 2.5~\Msun$ are limited to spin magnitudes $< 0.4$, while components with $m > 2.5~\Msun$ can have spin magnitudes up to 0.99.

\begin{table}
\begin{ruledtabular}
    \caption{
Summary of \PDB model parameters~\citep{Farah:2021qom}.
Arguments in the Prior column specify the lower and upper bounds of a uniform (U) distribution.
The first several entries describe the mass distribution parameters, and the last two entries describe the spin distribution parameters.
    }
    \label{table:parameters_pdb}
    \renewcommand{\arraystretch}{1.2}
    \begin{center}
    \begin{tabular}{c c c}
        {\bf Parameter} & \textbf{Description} & \textbf{Prior} \\
        \hline
        $\alpha_1$ & Spectral index for the power law of the mass distribution below $M^{\mathrm{gap}}_{\rm low}$ & U($-8$, $2$) \\
        $\alpha_2$ & Spectral index for the power law of the mass distribution above $M^{\mathrm{gap}}_{\rm low}$ & U($-3$, $2$) \\
        $A$ & Lower mass gap depth & U($0,1$) \\
        $M^{\mathrm{gap}}_{\rm low}$ ($\Msun$) & Location of the lower end of the mass gap & U($1.4$, $3$) \\
        $M^{\mathrm{gap}}_{\rm high}$ ($\Msun$) & Location of the upper end of the mass gap & U($3.4$, $9$) \\
        $\eta_{\rm low}$ & Parameter controlling how the rate tapers at the low end of the mass gap & 50 \\
        $\eta_{\rm high}$ & Parameter controlling how the rate tapers at the high end of the mass gap & 50 \\
        $\eta$ & Parameter controlling tapering the power law at high black hole mass & U($-4$, $12$) \\
        $\beta$ & Spectral index for the power law-in-mass-ratio pairing function & U($-2$, $7$) \\
        $m_\mathrm{min}$ ($\Msun$) & Minimum mass of the mass distribution & U($1\,$, $1.4$)\\
        $m_\mathrm{max}$ ($\Msun$) &  Maximum mass of the mass distribution & U($35$, $100$)\\
        $\chi_{\mathrm{max, NS}}$ &  Maximum allowed component spin for objects with mass $< 2.5 \Msun$ & $0.4$\\
        $\chi_{\mathrm{max, BH}}$ &  Maximum allowed component spin for objects with mass $\geq 2.5 \Msun$ & $0.99$\\
    \end{tabular}
    \end{center}
\end{ruledtabular}
\end{table}

\subsection{\NSBHpop}\label{supp:NSBHpop}

The \NSBHpop model~\citep{Biscoveanu:2022iue} is designed to capture the mass and spin distributions of the \ac{NSBH} population.
The black hole mass distribution is fit by a truncated power law, and the conditional mass ratio distribution $p(q | m_{1})$ is fit by a truncated Gaussian between $q_{\min} = 1~M_{\odot}/m_{1}$ and $q_{\max} = \min(m_{\mathrm{NS}, \max}/m_\mathrm{1}, 1)$.
The black hole spin magnitude distribution is modeled as a Beta distribution (including singular distributions that peak at the edges of the $\chi_{\mathrm{1}} \in [0,0.99]$ space), while the neutron star spin magnitude distribution is assumed to be uniform over $\chi_{\mathrm{2}} \in [0, 0.7]$ to encapsulate the effect of the neutron star breakup spin at the mass-shedding limit~\citep{Shao:2019ioq, Most:2020bba}.
We assume the redshift distribution is uniform in comoving volume and source-frame time and that spins are isotropically oriented.
The parameters for this model are summarized in Table~\ref{table:parameters_nsbhpop}.

\subsection{Population reweighting}\label{supp:reweighting}
We use the population-level mass and spin distributions inferred using all three population models to reweight (e.g.,~\citealt{Payne:2019wmy}) the high-spin combined parameter estimation samples from the default priors described in Appendix~\ref{supp:waveform_systematics} to three different astrophysically motivated priors given by the one-left-out posterior predictive distributions~\citep{Galaudage:2019jdx, Callister:2021note, Essick:2021note} inferred for each model.
A comparison of the posteriors under the default and astrophysically motivated priors is shown in Figure~\ref{fig:PopulationInformedPosteriors}.

Because of the $q\text{--}\chieff$ degeneracy, the spin distribution assumptions also impact the inferred component masses, and hence classification, for this source.
However, the fact that the \BGP and \NSBHpop model priors recover similar mass posteriors indicates that the different spin distribution assumptions have a subdominant effect compared to the different mass distribution assumptions.

\subsection{Selection effects}\label{supp:selection_effects}
For our population analyses, we account for search selection bias to measure the underlying astrophysical mass and spin distributions of the \ac{NSBH} and \ac{CBC} populations~\citep{Loredo:2004nn, Mandel:2018mve, Farr:2019rap, Vitale:2020aaz}.
To estimate the sensitivity of the searches over the data recorded by the detector network across the binary parameter space, we use a large suite of injections and impose a significance threshold on the \ac{FAR}.
In contrast to the two methods for calculating merger rates described in Appendix~\ref{supp:merger_rates}, we use the same set of injections used for GWTC-3 population studies recovered using matched-filter searches~\citep{LIGOScientific:2021duu}.
By using the combined injection sets from the first three observing runs~\citep{ligo_scientific_collaboration_and_virgo_2021_5636816}, we are effectively assuming that GW230529 occurred at the end of \ac{O3}.
Without accounting for the extra time--volume provided by the first 2 weeks of \ac{O4}, this introduces a bias in our inferred estimates of the merger rate in the mass gap (Figure~\ref{fig:MassGapRate}), rate of gamma-ray bursts with \ac{NSBH} progenitors, and total contribution of \ac{NSBH} mergers to the production of heavy elements (Section~\ref{sec:mma_implications}).
However, this effect is negligible given that GW230529 occurred in the first 2 weeks of \ac{O4a} and the detector sensitivity during this period was not drastically different from that during \ac{O3} (see Section~\ref{sec:detectors}), meaning that the additional time--volume unaccounted for in the injection sets is small.

To generate samples from the hierarchical likelihood for our population analyses, we use the \DYNESTY nested sampler~\citep{Speagle:2019ivv} as implemented in \GWPOPULATION~\citep{Talbot:2019okv} for the \PDB and \NSBHpop analyses and the Hamiltonian Monte Carlo sampler implemented in the \PYMC package~\citep{Salvatier:2016} for the \BGP analysis.

\begin{table}
\begin{ruledtabular}
    \caption{
Summary of \NSBHpop model parameters~\citep{Biscoveanu:2022iue}.
Arguments in the Prior column specify the lower and upper bounds of a uniform (U) distribution.
    }
    \label{table:parameters_nsbhpop}
    \renewcommand{\arraystretch}{1.2}
    \begin{center}
    \begin{tabular}{c c c}
        {\bf Parameter} & \textbf{Description} & \textbf{Prior} \\
        \hline
        $\alpha$ & Black hole mass power-law index &U$(-4, 12)$\\
        $m_{\mathrm{BH}, \min}$ ($\Msun$) & Minimum black hole mass & U$(2, 10)$\\
        $m_{\mathrm{BH}, \max}$ ($\Msun$) & Maximum black hole mass & U$(8, 20)$\\
        $m_{\mathrm{NS}, \max}$ ($\Msun$) & Maximum neutron star mass & U$(1.97, 2.7)$\\
        $\mu$ & Mass ratio mean & U$(0.1, 0.6)$\\
        $\sigma$ & Mass ratio standard deviation & U$(0.1, 1)$\\
        $\alpha_{\chi}$ & Beta distribution shape parameter ($\alpha$) for black hole spin distribution & U$(0.1, 10)$\\
        $\beta_{\chi}$ & Beta distribution shape parameter ($\beta$) for black hole spin distribution & U$(0.1, 10)$\\
    \end{tabular}
    \end{center}
\end{ruledtabular}
\end{table}

\section{Equation of State}\label{supp:eos}

To assess the effect of the \ac{EOS} constraint used for source classification (Section~\ref{sec:source_classification}) and calculation of $f_{\mathrm{EM\text{-}bright}}$ (Section~\ref{sec:mma_implications}), we repeat these analyses using \ac{GP}-\ac{EOS} constraints additionally conditioned on NICER observations~\citep{Legred:2021hdx, legred_isaac_2022_6502467} of J0030+0451~\citep{Miller:2019cac, Riley:2019yda, Vinciguerra:2023qxq} and J0740+6620~\citep{Miller:2021qha, Riley:2021pdl, Salmi:2022cgy, Salmi:2023sfv}.

We find that including the NICER observations only changes the source classification results by at most a few percent.
This is because the constraint on the maximum neutron star mass used in the source classification analysis is primarily driven by the observation of massive pulsars.
Changing the information included within the \ac{EOS} constraint leads to a smaller effect than the choice of astrophysical mass and spin distribution, as discussed in Section~\ref{sec:source_classification}.

The inference on the fraction of \ac{NSBH} systems that may have \ac{EM} counterparts changes more significantly when NICER observations are included in the \ac{EOS} constraint.
This is because the \ac{GW} and pulsar-only results provide much more support for compact neutron stars, which inhibits the probability that the neutron star is disrupted and suppresses the probability of an \ac{EM} counterpart.
The constraints including the NICER observations favor stiffer \acp{EOS}, enhancing the inferred \ac{EM}-bright fraction, $f_{\mathrm{EM\text{-}bright}} = \medianEMbrightFracNICER^{+\upperEMbrightFracNICER}_{\lowerEMbrightFracNICER}$, and pulling the posterior peak away from zero.

\bibliography{references}{}
\bibliographystyle{aasjournal}


\end{document}